\documentclass[10pt,a4paper]{article}
\usepackage{amsfonts,amsmath,amssymb,cite,lscape}
\newcommand{\Real}{\mathop{\textrm{Re}}}
\newcommand{\sgn}{\mathop{\textrm{sgn}}}

\begin{document}
\title{Static electric multipole susceptibilities of the relativistic
hydrogen-like atom in the ground state: Application of the Sturmian
expansion of the generalized Dirac--Coulomb Green function}
\author{Rados{\l}aw Szmytkowski\footnote{~Corresponding author. 
Email: radek@mif.pg.gda.pl} \mbox{}
and Grzegorz {\L}ukasik \\*[3ex]
Atomic Physics Division,
Department of Atomic, Molecular and Optical Physics, \\
Faculty of Applied Physics and Mathematics,
Gda{\'n}sk University of Technology, \\
Narutowicza 11/12, 80--233 Gda{\'n}sk, Poland}
\date{\today}
\maketitle
\begin{abstract} 
The ground state of the Dirac one-electron atom, placed in a weak,
static electric field of definite $2^{L}$-polarity, is studied within
the framework of the first-order perturbation theory. The Sturmian
expansion of the generalized Dirac--Coulomb Green function [R.\
Szmytkowski, J.\ Phys.\ B 30 (1997) 825, erratum: 30 (1997) 2747] is
used to derive closed-form analytical expressions for various
far-field and near-nucleus static electric multipole susceptibilities
of the atom. The far-field multipole susceptibilities --- the
polarizabilities $\alpha_{L}$, electric-to-magnetic
cross-susceptibilities $\alpha_{\mathrm{E}L\to\mathrm{M}(L\mp1)}$ and
electric-to-toroidal-magnetic cross-susceptibilities
$\alpha_{\mathrm{E}L\to\mathrm{T}L}$ --- are found to be expressible
in terms of one or two non-terminating generalized hypergeometric
functions ${}_{3}F_{2}$ with the unit argument. Counterpart formulas
for the near-nucleus multipole susceptibilities --- the electric
nuclear shielding constants $\sigma_{\mathrm{E}L\to\mathrm{E}L}$,
near-nucleus electric-to-magnetic cross-susceptibilities
$\sigma_{\mathrm{E}L\to\mathrm{M}(L\mp1)}$ and near-nucleus
electric-to-toroidal-magnetic cross-susceptibilities
$\sigma_{\mathrm{E}L\to\mathrm{T}L}$ --- involve terminating
${}_{3}F_{2}(1)$ series and for each $L$ may be rewritten in terms of
elementary functions. Exact numerical values of the far-field dipole,
quadrupole, octupole and hexadecapole susceptibilities are provided
for selected hydrogenic ions. Analytical quasi-relativistic
approximations, valid to the second order in $\alpha Z$, where
$\alpha$ is the fine-structure constant and $Z$ is the nuclear charge
number, are derived for all types of the far-field and near-nucleus
susceptibilities considered in the paper.
\vskip1ex
\noindent
\textbf{PACS 2010:} 31.15.ap, 32.10.Dk, 03.65.Pm
\end{abstract}
%
%\newpage
%
\section{Introduction}
\label{I}
\setcounter{equation}{0}
Relativistic studies on static multipole polarizabilities of a
one-electron atom in its ground state, based on the formalism of the
Dirac equation, may be traced back to early 1970s, when Zon \emph{et
al.\/} \cite{Zon72} presented a closed-form analytical expression for
the dipole polarizability $\alpha_{1}$ of such a system. The formula
given in Ref.\ \cite{Zon72} involved a particular generalized
hypergeometric function ${}_{3}F_{2}$ with the unit argument. In the
following decades, several equivalent expressions for $\alpha_{1}$
were derived, with the use of various alternative analytical
techniques, by Labzowsky \cite{Labz73a,Labz73b}, Shestakov and
Khristenko \cite{Shes74}, Labzowsky \emph{et al.\/} \cite{Labz93}, Le
Anh Thu \emph{et al.\/} \cite{LeAn94}, Szmytkowski \cite{Szmy97},
Yakhontov \cite{Yakh03}, and Szmytkowski and Mielewczyk
\cite{Szmy04}. Quasi-relativistic approximations to $\alpha_{1}$,
correct to the second order in $\alpha Z$, where $\alpha$ is the
fine-structure constant and $Z$ is the nuclear charge number, were
provided by Bartlett and Power \cite{Bart69}, Rutkowski and Schwarz
\cite{Rutk90}, and Turski and Sadlej \cite{Turs01} (in this
connection, see also the work of Baluja \cite{Balu95}).

In 1974, Manakov \emph{et al.\/} published a paper \cite{Mana74}
(cf.\ also the review \cite{Zapr81} and the monograph \cite{Zapr85}),
in which they provided an exact analytical formula for a general
static multipole polarizability $\alpha_{L}$; the formula involved
altogether eight different ${}_{3}F_{2}(1)$ functions. In the
particular case of $L=1$, the result arrived at in that work might be
simplified and the expression from Ref.\ \cite{Zon72} was recovered.
An approximate quasi-relativistic representation for $\alpha_{L}$,
given by Kaneko in Ref.\ \cite{Kane77} and by Drachman in an erratum
to Ref.\ \cite{Drac85}, coincided with the corresponding limit
deduced in Ref.\ \cite{Mana74}.\footnote{~It should be mentioned that
a quasi-relativistic expression for $\alpha_{2}$ given in Ref.\
\cite{Turs01} is incorrect and that the criticism of the work
\cite{Mana74} presented in an appendix to Ref.\ \cite{Turs01} was
mostly unjustified.}

In addition to the analytical works listed above, we have tracked
down four papers in which results of purely numerical relativistic
calculations of the multipole polarizabilities were reported for
selected hydrogenic ions. Goldman \cite{Gold89} carried out
variational calculations of the dipole polarizability $\alpha_{1}$,
employing the Slater-type functions used as a variational basis set.
Zhang \emph{et al.\/} \cite{Zhan12} presented results for the
quadrupole polarizability $\alpha_{2}$ computed with the use of the
$B$-spline Galerkin method. The latter study was pushed further in
Ref.\ \cite{Tang12}, where numerical data for $\alpha_{L}$ with
$1\leqslant L\leqslant4$ were provided. Finally, very recently
Filippin \emph{et al.\/} \cite{Fili14} applied their Lagrange-mesh
method in computations of $\alpha_{L}$, with $L$ in the same range as
mentioned above. The calculations reported in Refs.\
\cite{Zhan12,Tang12,Fili14} used the same value of the fine structure
constant (taken from the CODATA 2010 recommendation). Numerical data
for the four multipole polarizabilities presented by both groups,
although obtained with different methods, appeared to be in a very
good agreement.

The multipole polarizabilities $\alpha_{L}$ are closely related to
the far-field \emph{electric\/} multipole moments induced in an atom
by external weak, static, electric multipole fields. However, a
perturbing electric field may also induce in the atom two kinds of
the far-field \emph{magnetic\/} multipole moments: the plain magnetic
moments and the toroidal magnetic moments. Magnitudes of these
induced moments may be characterized, respectively, by the so-called
electric-to-magnetic and electric-to-toroidal-magnetic multipole
cross-susceptibilities. In Ref.\ \cite{Szmy14}, Szmytkowski and
Stefa{\'n}ska derived an exact closed-form analytical expression for
the electric-dipole-to-magnetic-quadrupole cross-susceptibility
$\alpha_{\mathrm{E}1\to\mathrm{M}2}$ for the Dirac one-electron atom
in the ground state. In turn, analytical expressions for the atomic
ground-state electric-dipole-to-toroidal-magnetic-dipole
cross-susceptibility $\alpha_{\mathrm{E}1\to\mathrm{T1}}$ may be
inferred from the papers of Lewis and Blinder \cite{Lewi95} and
Mielewczyk and Szmytkowski \cite{Miel06}. Calculations carried out in
Ref.\ \cite{Lewi95} were partly approximate, while those reported in
Ref.\ \cite{Miel06} were exact at the Dirac--Coulomb level.

The three sets of the electric multipole susceptibilities mentioned
above characterize, through the moments they are linked to, the
first-order field-induced corrections to electromagnetic scalar and
vector potentials generated by the atom in the region distant from
its nucleus. In analogy, one may consider counterpart
susceptibilities related to multipole moments characterizing the
first-order corrections to the scalar and vector potentials in the
close vicinity of the atomic nucleus. The only fully relativistic
research in that direction that we are aware of was done by
Zapryagaev \emph{et al.\/} \cite{Zapr74} (cf.\ also Refs.\
\cite{Zapr81,Zapr85}), who studied the electric multipole shielding
constants $\sigma_{\mathrm{E}L\to\mathrm{E}L}$. From an exact
analytical expression they derived (its explicit, and quite
complicated, form was given only in the chronologically latest Ref.\
\cite[Sec.\ 4.6]{Zapr85}), the quasi-relativistic estimates for the
shielding constants with $L=2$ and $L=3$ were deduced
\cite{Zapr74,Zapr85}. Moreover, a quasi-relativistic formula for
$\sigma_{\mathrm{E}L\to\mathrm{E}L}$ applicable for any $L$ was
derived, in an entirely different way, by Kaneko \cite{Kane77}.

This brief state-of-the-art overview of research on electric
multipole susceptibilities of the Dirac one-electron atom in the
ground state shows that exact analytical expressions for the far- and
near-field electric-to-magnetic and electric-to-toroidal-magnetic
multipole cross-susceptibilities are still missing. We derive them in
the present paper, with the aid of an analytical technique based on
the Sturmian series representation of the Dirac--Coulomb Green
function found by one of us in Ref.\ \cite{Szmy97}. That technique
proved its effectiveness in calculations of various properties of
hydrogenic ions carried out by our group over the past two decades
\cite{Szmy01,Szmy02,Szmy04,Miel06,Szmy11,Stef12,Szmy12,Szmy14,Stef15a,
Stef15b,Stef15c}. Moreover, in view of the annoying complexity of the
representations for the multipole polarizabilities and the electric
shielding constants presented in Refs.\ \cite{Mana74,Zapr85}, we have
decided to reconsider these two families of atomic susceptibilities,
with the goal to arrive at simpler expressions for them. The attempt
has appeared to be successful, and below we present formulas for
$\alpha_{L}$ and $\sigma_{\mathrm{E}L\to\mathrm{E}L}$, each one
containing only \emph{two\/} (as opposed to \emph{eight\/} in Refs.\
\cite{Mana74,Zapr85}) generalized hypergeometric functions
${}_{3}F_{2}(1)$.

The structure of the paper is as follows. Section \ref{II} provides
some basic notions and facts concerning the ground state of the Dirac
one-electron atom placed in a $2^{L}$-pole electric field. In Secs.\
\ref{III}--\ref{V}, we analyze three kinds of the far-field multipole
moments that characterize charge and current distributions of the
atom in such a field. We show that if the field is weak, it induces
in the atom the electric and toroidal magnetic moments of rank $L$
only, as well as the plain magnetic moments of ranks $L-1$ and $L+1$
(except for the case $L=1$, when only the quadrupole magnetic moment
arises). The knowledge of expressions for the induced moments allows
one to deduce closed-form formulas for related atomic
susceptibilities, and this is subsequently done in each of these
sections. We provide exact and approximate (quasi-relativistic)
expressions for the multipole susceptibilities (the polarizabilities,
the electric-to-magnetic cross-susceptibilities and the
electric-to-toroidal-magnetic cross-susceptibilities), and also
tabulate their numerical values computed from the exact formulas for
selected values of the nuclear charge $Z$. Analogous considerations
concerning the near-field moments and the susceptibilities related to
them are carried out in Secs.\ \ref{VI}--\ref{VIII}. The final Sec.\
\ref{IX} contains a brief summary of the most important results
derived in the paper and also discloses our research plans for the
near future. The text is supplemented by five appendices. A
relationship between a multipole polarizability and the second-order
correction to energy of the atom in a multipole electric field is
revealed in Appendix \ref{A}. In Appendix \ref{B}, we show how the
far- and near-field toroidal magnetic multipole moments arise when
the magnetic vector potential is expanded into multipoles. In
Appendices \ref{B} and \ref{C}, we prove that for each of the two
sets of the toroidal multipole moments that have arisen in Appendix
\ref{B}, there is a one-parameter family of equivalent integral
expressions, which may be used as their definitions; this gives one
the precious freedom to define these moments in forms most suitable
for each particular problem they emerge in. Some properties of the
generalized hypergeometric function ${}_{3}F_{2}$ with the unit
argument, relevant to the material presented in Secs.\
\ref{VI}--\ref{VIII}, are discussed in Appendix \ref{E}.
%
%\newpage
%
\section{Preliminaries}
\label{II}
\setcounter{equation}{0}
Consider a Dirac one-electron atom with a motionless, point-like and
spinless nucleus of charge $+Ze$. A position vector of the atomic
electron relative to the nucleus will be hereafter denoted as
$\boldsymbol{r}$. The atom, assumed to be initially in its ground
state of energy $E^{(0)}$ [cf.\ Eq.\ (\ref{2.6}) below], is perturbed
by a static electric $2^{L}$-pole field
$\boldsymbol{\mathcal{E}}_{L}^{(1)}(\boldsymbol{r})$ derivable from
the scalar potential
\begin{equation}
\varphi_{L}^{(1)}(\boldsymbol{r})
=-\sqrt{\frac{4\pi}{2L+1}}\,r^{L}\boldsymbol{\mathsf{C}}_{L}^{(1)}
\cdot\boldsymbol{\mathsf{Y}}_{L}(\boldsymbol{n}_{r})
\qquad (L\geqslant1),
\label{2.1}
\end{equation}
where $\boldsymbol{\mathsf{C}}_{L}^{(1)}$ and
$\boldsymbol{\mathsf{Y}}_{L}(\boldsymbol{n}_{r})$ are spherical
tensor operators of rank $L$ with components $\mathcal{C}_{LM}^{(1)}$
and $Y_{LM}(\boldsymbol{n}_{r})$, respectively. Here
$Y_{LM}(\boldsymbol{n}_{r})$ is the normalized complex spherical
harmonic (in this work, we adopt the Condon--Shortley phase
convention) and $\boldsymbol{n}_{r}$ is the unit vector along
$\boldsymbol{r}$. The components of
$\boldsymbol{\mathsf{C}}_{L}^{(1)}$, which determine both the
strength of the potential and its angular dependence, are constrained
to obey
\begin{equation}
\mathcal{C}_{LM}^{(1)*}=(-)^{M}\mathcal{C}_{L,-M}^{(1)},
\label{2.2}
\end{equation}
where the asterisk denotes the complex conjugation.\footnote{~The
reader should observe that for $L=1$ the components
$\mathcal{C}_{1M}^{(1)}$ of the tensor (then vector)
$\boldsymbol{\mathsf{C}}_{1}^{(1)}$ are simply the cyclic components
of the perturbing dipole electric field
$\boldsymbol{\mathcal{E}}_{1}^{(1)}(\boldsymbol{r})$.} This ensures
that $\varphi_{L}^{(1)}(\boldsymbol{r})$ is real. In terms of
components of the two tensors, the interaction energy between the
atomic electron and the field reads
\begin{equation}
V_{L}^{(1)}(\boldsymbol{r})=e\sqrt{\frac{4\pi}{2L+1}}\,r^{L}
\sum_{M=-L}^{L}\mathcal{C}_{LM}^{(1)*}Y_{LM}(\boldsymbol{n}_{r})
\qquad (L\geqslant1).
\label{2.3}
\end{equation}
Henceforth, it will be assumed that the electric force
$-\boldsymbol{\nabla}V_{L}^{(1)}(\boldsymbol{r})$ acting on the
electron is so weak that the probability that the field ionizes the
atom may be neglected. Within this approximation, the atomic electron
may be considered to be in a stationary state described by the
time-independent Dirac equation
\begin{equation}
\left[-\mathrm{i}c\hbar\boldsymbol{\alpha}\cdot\boldsymbol{\nabla}
+\beta m_{\mathrm{e}}c^{2}-\frac{Ze^{2}}{(4\pi\epsilon_{0})r}
+V_{L}^{(1)}(\boldsymbol{r})-E\right]\Psi(\boldsymbol{r})=0,
\label{2.4}
\end{equation}
where $\boldsymbol{\alpha}$ and $\beta$ are the standard Dirac
matrices. Since, by the above-made assumption, the external electric
field is weak, in what follows the electron--multipole field
interaction term $V_{L}^{(1)}(\boldsymbol{r})$ will be considered as
a small perturbation of the Dirac--Coulomb Hamiltonian. Then, to the
first order in that perturbation, the energy eigenvalue is
\begin{equation}
E\simeq E^{(0)}+E^{(1)},
\label{2.5}
\end{equation}
with the (doubly-degenerate) unperturbed ground-state energy level
$E^{(0)}$ given by
\begin{equation}
E^{(0)}=m_{\mathrm{e}}c^{2}\gamma_{1},
\label{2.6}
\end{equation}
where
\begin{equation}
\gamma_{\kappa}=\sqrt{\kappa^{2}-(\alpha Z)^{2}}
\label{2.7}
\end{equation}
($\alpha$, not to be confused with the Dirac matrix
$\boldsymbol{\alpha}$ or the multipole polarizability $\alpha_{L}$,
is the Sommerfeld fine-structure constant). To the same order, the
electron wave function is
\begin{equation}
\Psi(\boldsymbol{r})\simeq\Psi^{(0)}(\boldsymbol{r})
+\Psi^{(1)}(\boldsymbol{r}),
\label{2.8}
\end{equation}
with the unperturbed component given by
\begin{equation}
\Psi^{(0)}(\boldsymbol{r})=a_{1/2}\Psi_{1/2}^{(0)}(\boldsymbol{r})
+a_{-1/2}\Psi_{-1/2}^{(0)}(\boldsymbol{r}).
\label{2.9}
\end{equation}
The basis states $\Psi_{m}^{(0)}(\boldsymbol{r})$ appearing in Eq.\
(\ref{2.9}) are chosen to be
\begin{equation}
\Psi_{m}^{(0)}(\boldsymbol{r})
=\frac{1}{r}
\left(
\begin{array}{c}
P^{(0)}(r)\Omega_{-1m}(\boldsymbol{n}_{r}) \\
\mathrm{i}Q^{(0)}(r)\Omega_{1m}(\boldsymbol{n}_{r})
\end{array}
\right)
\qquad (m=\pm{\textstyle\frac{1}{2}}),
\label{2.10}
\end{equation}
where
\begin{subequations}
\begin{equation}
P^{(0)}(r)
=-\sqrt{\frac{Z}{a_{0}}\frac{1+\gamma_{1}}{\Gamma(2\gamma_{1}+1)}}
\left(\frac{2Zr}{a_{0}}\right)^{\gamma_{1}}\mathrm{e}^{-Zr/a_{0}},
\label{2.11a}
\end{equation}
\begin{equation}
Q^{(0)}(r)
=\sqrt{\frac{Z}{a_{0}}\frac{1-\gamma_{1}}{\Gamma(2\gamma_{1}+1)}}
\left(\frac{2Zr}{a_{0}}\right)^{\gamma_{1}}\mathrm{e}^{-Zr/a_{0}},
\label{2.11b}
\end{equation}
\label{2.11}%
\end{subequations}
while $\Omega_{\kappa m_{\kappa}}(\boldsymbol{n}_{r})$ are the
spherical spinors \cite{Szmy07}. It is easy to verify that the
functions (\ref{2.10}) are orthonormal in the sense of
\begin{equation}
\int_{\mathbb{R}^{3}}\mathrm{d}^{3}\boldsymbol{r}\:
\Psi_{m}^{(0)\dag}(\boldsymbol{r})
\Psi_{m'}^{(0)}(\boldsymbol{r})=\delta_{mm'}
\qquad (m,m'=\pm{\textstyle\frac{1}{2}}),
\label{2.12}
\end{equation}
so that if the coefficients $a_{\pm1/2}$ are subjected to the
constraint
\begin{equation}
|a_{1/2}|^{2}+|a_{-1/2}|^{2}=1,
\label{2.13}
\end{equation}
the function $\Psi^{(0)}(\boldsymbol{r})$ is normalized to unity:
\begin{equation}
\int_{\mathbb{R}^{3}}\mathrm{d}^{3}\boldsymbol{r}\:
\Psi^{(0)\dag}(\boldsymbol{r})\Psi^{(0)}(\boldsymbol{r})=1.
\label{2.14}
\end{equation}

It follows from the standard Schr{\"o}dinger--Rayleigh perturbation
theory that the so-far unknown corrections $E^{(1)}$ and
$\Psi^{(1)}(\boldsymbol{r})$ [and also the coefficients $a_{\pm1/2}$
hidden in $\Psi^{(0)}(\boldsymbol{r})$] enter the inhomogeneous
Dirac--Coulomb equation
\begin{equation}
\left[-\mathrm{i}c\hbar\boldsymbol{\alpha}\cdot\boldsymbol{\nabla}
+\beta m_{\mathrm{e}}c^{2}-\frac{Ze^{2}}{(4\pi\epsilon_{0})r}
-E^{(0)}\right]\Psi^{(1)}(\boldsymbol{r})
=-[V_{L}^{(1)}(\boldsymbol{r})-E^{(1)}]\Psi^{(0)}(\boldsymbol{r}),
\label{2.15}
\end{equation}
which is to be solved subject to the usual physical regularity
requirements as well as the orthogonality constraints
\begin{equation}
\int_{\mathbb{R}^{3}}\mathrm{d}^{3}\boldsymbol{r}\:
\Psi_{m}^{(0)\dag}(\boldsymbol{r})\Psi^{(1)}(\boldsymbol{r})=0
\qquad (m=\pm{\textstyle\frac{1}{2}}).
\label{2.16}
\end{equation}
After Eq.\ (\ref{2.15}) is projected onto the unperturbed basis
states $\Psi_{m}^{(0)}(\boldsymbol{r})$, ($m=\pm\frac{1}{2}$), this
yields the homogeneous algebraic system
\begin{equation}
\sum_{m'=-1/2}^{1/2}
\big[V_{L,mm'}^{(1)}-E^{(1)}\delta_{mm'}\big]a_{m'}=0
\qquad (m=\pm{\textstyle\frac{1}{2}}),
\label{2.17}
\end{equation}
where
\begin{equation}
V_{L,mm'}^{(1)}=e\sqrt{\frac{4\pi}{2L+1}}
\sum_{M=-L}^{L}\mathcal{C}_{LM}^{(1)*}
\int_{\mathbb{R}^{3}}\mathrm{d}^{3}\boldsymbol{r}\:
\Psi_{m}^{(0)\dag}(\boldsymbol{r})r^{L}
Y_{LM}(\boldsymbol{n}_{r})\Psi_{m'}^{(0)}(\boldsymbol{r}).
\label{2.18}
\end{equation}
Denoting
\begin{equation}
\langle\Omega_{\kappa m_{\kappa}}
\big|Y_{LM}\Omega_{\kappa'm_{\kappa'}}\rangle
\equiv\oint_{4\pi}\mathrm{d}^{2}\boldsymbol{n}_{r}\:
\Omega_{\kappa m_{\kappa}}^{\dag}(\boldsymbol{n}_{r})
Y_{LM}(\boldsymbol{n}_{r})
\Omega_{\kappa'm_{\kappa'}}(\boldsymbol{n}_{r})
\label{2.19}
\end{equation}
and exploiting the known identity
\begin{equation}
\langle\Omega_{-\kappa m_{\kappa}}
\big|Y_{\lambda\mu}\Omega_{-\kappa'm_{\kappa'}}\rangle
=\langle\Omega_{\kappa m_{\kappa}}
\big|Y_{\lambda\mu}\Omega_{\kappa'm_{\kappa'}}\rangle,
\label{2.20}
\end{equation}
we obtain
\begin{equation}
V_{L,mm'}^{(1)}=e\sqrt{\frac{4\pi}{2L+1}}
\int_{0}^{\infty}\mathrm{d}r\:r^{L}\left\{[P^{(0)}(r)]^{2}
+[Q^{(0)}(r)]^{2}\right\}\sum_{M=-L}^{L}\mathcal{C}_{LM}^{(1)*}
\langle\Omega_{-1m}\big|Y_{LM}\Omega_{-1m'}\rangle.
\label{2.21}
\end{equation}
The angular integral in Eq.\ (\ref{2.21}) may be evaluated with the
aid of the known formula
\begin{eqnarray}
&& \hspace*{-3em}
\sqrt{\frac{4\pi}{2L+1}}
\langle\Omega_{\kappa m_{\kappa}}
\big|Y_{LM}\Omega_{\kappa'm_{\kappa'}}\rangle
\nonumber \\
&& =\,(-)^{m_{\kappa}+1/2}2\sqrt{|\kappa\kappa'|}
\left(
\begin{array}{ccc}
|\kappa|-\frac{1}{2} & L & |\kappa'|-\frac{1}{2} \\
-\frac{1}{2} & 0 & \frac{1}{2}
\end{array}
\right)
\left(
\begin{array}{ccc}
|\kappa|-\frac{1}{2} & L & |\kappa'|-\frac{1}{2} \\
m_{\kappa} & -M & -m_{\kappa'}
\end{array}
\right)
\Pi(l_{\kappa},L,l_{\kappa'}),
\nonumber \\
&&
\label{2.22}
\end{eqnarray}
where {\footnotesize${\textstyle
\left(\begin{array}{ccc}j_{a} & j_{b} & j_{c} \\ 
m_{a} & m_{b} & m_{c}\end{array}\right)}$} denotes Wigner's 3$j$
coefficient, while
\begin{equation}
\Pi(l_{\kappa},L,l_{\kappa'})
=\left\{
\begin{array}{ll}
1 & \textrm{for $l_{\kappa}+L+l_{\kappa'}$ even} \\
0 & \textrm{for $l_{\kappa}+L+l_{\kappa'}$ odd},
\end{array}
\right.
\label{2.23}
\end{equation}
with
\begin{equation}
l_{\kappa}=\left|\kappa+\frac{1}{2}\right|-\frac{1}{2}
\label{2.24}
\end{equation} 
and similarly for $l_{\kappa'}$. One finds
\begin{equation}
\langle\Omega_{-1m}\big|Y_{LM}\Omega_{-1m'}\rangle
=\frac{1}{\sqrt{4\pi}}\,\delta_{L0}\delta_{M0}\delta_{mm'}.
\label{2.25}
\end{equation}
Since we have excluded the case $L=0$ from the very beginning, we
have
\begin{equation}
V_{L,mm'}^{(1)}=0,
\label{2.26}
\end{equation}
which implies immediately [cf.\ Eq.\ (\ref{2.17})] that for any $L$
it holds that
\begin{equation}
E^{(1)}=0
\label{2.27}
\end{equation}
and that the coefficients $a_{\pm1/2}$ are arbitrary save for the
normalization condition (\ref{2.13}).

With the result (\ref{2.27}) in mind, the solution to Eq.\
(\ref{2.15}) may be written as
\begin{equation}
\Psi^{(1)}(\boldsymbol{r})=-e\sqrt{\frac{4\pi}{2L+1}}
\sum_{M=-L}^{L}\mathcal{C}_{LM}^{(1)*}
\int_{\mathbb{R}^{3}}\mathrm{d}^{3}\boldsymbol{r}'\:
\bar{G}\mbox{}^{(0)}(\boldsymbol{r},\boldsymbol{r}')
r^{\prime L}Y_{LM}(\boldsymbol{n}_{r}^{\prime})
\Psi^{(0)}(\boldsymbol{r}'),
\label{2.28}
\end{equation}
where $\bar{G}\mbox{}^{(0)}(\boldsymbol{r},\boldsymbol{r}')$ is the
generalized (or reduced) Dirac--Coulomb Green's function \cite[Sec.\
6]{Szmy97} associated with the ground state of the atom under
investigation.
%
%\newpage
%
\section{Electric multipole moments of the atom in the multipole
electric field and atomic multipole polarizabilities}
\label{III} 
\setcounter{equation}{0}
\subsection{Decomposition of the atomic electric multipole moments
into the permanent and the first-order electric-field-induced
components}
\label{III.1}
Being in the state described by the wave function
$\Psi(\boldsymbol{r})$, the electronic cloud of the atom may be
characterized, among others, by its electric multipole moments
$\boldsymbol{\mathsf{Q}}_{\lambda}$ with the spherical
components\footnote{~The reader should observe that the definition
(\ref{3.1}) of the spherical components of the electric multipole
moments, we adopt here, differs from the one we used in Ref.\
\cite{Szmy12} in that in the latter the spherical harmonic was
complex conjugated.}
\begin{equation}
\mathcal{Q}_{\lambda\mu}=\sqrt{\frac{4\pi}{2\lambda+1}}
\int_{\mathbb{R}^{3}}\mathrm{d}^{3}\boldsymbol{r}\:
r^{\lambda}Y_{\lambda\mu}(\boldsymbol{n}_{r})
\rho(\boldsymbol{r}),
\label{3.1}
\end{equation}
where
\begin{equation}
\rho(\boldsymbol{r})
=\frac{-e\Psi^{\dag}(\boldsymbol{r})\Psi(\boldsymbol{r})}
{\int_{\mathbb{R}^{3}}\mathrm{d}^{3}\boldsymbol{r}'\:
\Psi^{\dag}(\boldsymbol{r}')\Psi(\boldsymbol{r}')}
\label{3.2}
\end{equation}
may be considered as a smeared electronic charge density. In the case
the function $\Psi(\boldsymbol{r})$ may be approximated as in Eq.\
(\ref{2.8}), after Eqs.\ (\ref{2.9}), (\ref{2.14}) and (\ref{2.16})
are taken into account, the density $\rho(\boldsymbol{r})$ may be
approximately written as
\begin{equation}
\rho(\boldsymbol{r})\simeq\rho^{(0)}(\boldsymbol{r})
+\rho^{(1)}(\boldsymbol{r}),
\label{3.3}
\end{equation}
where
\begin{equation}
\rho^{(0)}(\boldsymbol{r})=-e\Psi^{(0)\dag}(\boldsymbol{r})
\Psi^{(0)}(\boldsymbol{r})
\label{3.4}
\end{equation}
and
\begin{equation}
\rho^{(1)}(\boldsymbol{r})=-2e\Real[\Psi^{(0)\dag}(\boldsymbol{r})
\Psi^{(1)}(\boldsymbol{r})].
\label{3.5}
\end{equation}
Accordingly, Eq.\ (\ref{3.3}) implies
\begin{equation}
\mathcal{Q}_{\lambda\mu}\simeq\mathcal{Q}_{\lambda\mu}^{(0)}
+\mathcal{Q}_{\lambda\mu}^{(1)},
\label{3.6}
\end{equation}
where
\begin{equation}
\mathcal{Q}_{\lambda\mu}^{(0)}
=-e\sqrt{\frac{4\pi}{2\lambda+1}}
\int_{\mathbb{R}^{3}}\mathrm{d}^{3}\boldsymbol{r}\:
\Psi^{(0)\dag}(\boldsymbol{r})
r^{\lambda}Y_{\lambda\mu}(\boldsymbol{n}_{r})
\Psi^{(0)}(\boldsymbol{r})
\label{3.7}
\end{equation}
and
\begin{equation}
\mathcal{Q}_{\lambda\mu}^{(1)}=-e\sqrt{\frac{4\pi}{2\lambda+1}}
\int_{\mathbb{R}^{3}}\mathrm{d}^{3}\boldsymbol{r}\:
r^{\lambda}Y_{\lambda\mu}(\boldsymbol{n}_{r})
[\Psi^{(0)\dag}(\boldsymbol{r})\Psi^{(1)}(\boldsymbol{r})
+\Psi^{(1)\dag}(\boldsymbol{r})\Psi^{(0)}(\boldsymbol{r})]
\label{3.8}
\end{equation}
are the permanent and the first-order induced electric multipole
moments of the electronic cloud, respectively. Proceeding along the
route that parallels the evaluation of the energy correction
$E^{(1)}$, presented in Sec.\ \ref{II}, it is easy to show that the
only non-vanishing multipole moment of the atom in the unperturbed
ground state is the monopole one:
\begin{equation}
\mathcal{Q}_{\lambda\mu}^{(0)}=\mathcal{Q}_{\lambda\mu}^{(0)}
\delta_{\lambda0}\delta_{\mu0},
\qquad
\mathcal{Q}_{00}^{(0)}=-e.
\label{3.9}
\end{equation}
Therefore, in what follows we shall be concerned with the evaluation
of the induced moment $\mathcal{Q}_{\lambda\mu}^{(1)}$ only.

With the aid of the identity
\begin{equation}
Y_{\lambda\mu}(\boldsymbol{n}_{r})
=(-)^{\mu}Y_{\lambda,-\mu}^{*}(\boldsymbol{n}_{r})
\label{3.10}
\end{equation}
and of the result in Eq.\ (\ref{2.28}),
$\mathcal{Q}_{\lambda\mu}^{(1)}$ may be written as
\begin{equation}
\mathcal{Q}_{\lambda\mu}^{(1)}
=\widetilde{\mathcal{Q}}_{\lambda\mu}^{(1)}
+(-)^{\mu}\widetilde{\mathcal{Q}}_{\lambda,-\mu}^{(1)*},
\label{3.11}
\end{equation}
with
\begin{eqnarray}
\widetilde{\mathcal{Q}}_{\lambda\mu}^{(1)}
&=& e^{2}\frac{4\pi}{\sqrt{(2\lambda+1)(2L+1)}}
\sum_{M=-L}^{L}\mathcal{C}_{LM}^{(1)*}
\nonumber \\
&& \times\int_{\mathbb{R}^{3}}\mathrm{d}^{3}\boldsymbol{r}
\int_{\mathbb{R}^{3}}\mathrm{d}^{3}\boldsymbol{r}'\:
\Psi^{(0)\dag}(\boldsymbol{r})r^{\lambda}
Y_{\lambda\mu}(\boldsymbol{n}_{r})
\bar{G}^{(0)}(\boldsymbol{r},\boldsymbol{r}')
r^{\prime L}Y_{LM}(\boldsymbol{n}_{r}^{\prime})
\Psi^{(0)}(\boldsymbol{r}').
\label{3.12}
\end{eqnarray}
It is possible to separate out radial and angular integrations in
Eq.\ (\ref{3.12}). To this end, one may exploit the following
multipole expansion of the generalized Green's function
$\bar{G}\mbox{}^{(0)}(\boldsymbol{r},\boldsymbol{r}')$:
\begin{eqnarray}
\bar{G}\mbox{}^{(0)}(\boldsymbol{r},\boldsymbol{r}')
&=& \frac{4\pi\epsilon_{0}}{e^{2}}
\sum_{\substack{\kappa=-\infty \\ (\kappa\neq0)}}^{\infty}
\sum_{m_{\kappa}=-|\kappa|+1/2}^{|\kappa|-1/2}\frac{1}{rr'}
\nonumber \\
&& \hspace*{-5em}
\times\,\left(
\begin{array}{cc}
\bar{g}\mbox{}^{(0)}_{(++)\kappa}(r,r')
\Omega_{\kappa m_{\kappa}}(\boldsymbol{n}_{r})
\Omega_{\kappa m_{\kappa}}^{\dag}(\boldsymbol{n}_{r}^{\prime}) &
-\mathrm{i}\bar{g}\mbox{}^{(0)}_{(+-)\kappa}(r,r')
\Omega_{\kappa m_{\kappa}}(\boldsymbol{n}_{r})
\Omega_{-\kappa m_{\kappa}}^{\dag}(\boldsymbol{n}_{r}^{\prime})
\\*[1ex]
\mathrm{i}\bar{g}\mbox{}^{(0)}_{(-+)\kappa}(r,r')
\Omega_{-\kappa m_{\kappa}}(\boldsymbol{n}_{r})
\Omega_{\kappa m_{\kappa}}^{\dag}(\boldsymbol{n}_{r}^{\prime}) &
\bar{g}\mbox{}^{(0)}_{(--)\kappa}(r,r')
\Omega_{-\kappa m_{\kappa}}(\boldsymbol{n}_{r})
\Omega_{-\kappa m_{\kappa}}^{\dag}(\boldsymbol{n}_{r}^{\prime})
\end{array}
\right).
\nonumber \\
&&
\label{3.13}
\end{eqnarray}
After this expansion is plugged into Eq.\ (\ref{3.12}) and use is
made of Eq.\ (\ref{2.20}), one arrives at
\begin{eqnarray}
\widetilde{\mathcal{Q}}_{\lambda\mu}^{(1)}
&=& (4\pi\epsilon_{0})\frac{4\pi}{\sqrt{(2\lambda+1)(2L+1)}}
\sum_{\substack{\kappa=-\infty \\ (\kappa\neq0)}}^{\infty}
R_{\kappa}^{(\lambda,L)}\big(P^{(0)},Q^{(0)};P^{(0)},Q^{(0)}\big)
\nonumber \\
&& \times\sum_{M=-L}^{L}\sum_{m_{\kappa}=-|\kappa|+1/2}^{|\kappa|-1/2}
\sum_{m=-1/2}^{1/2}\sum_{m'=-1/2}^{1/2}
a_{m}^{*}a_{m'}\mathcal{C}_{LM}^{(1)*}
\langle\Omega_{-1m}|Y_{\lambda\mu}\Omega_{\kappa m_{\kappa}}\rangle
\langle\Omega_{\kappa m_{\kappa}}|Y_{LM}\Omega_{-1m'}\rangle,
\nonumber \\
&&
\label{3.14}
\end{eqnarray}
where $R_{\kappa}^{(\lambda,L)}\big(P^{(0)},Q^{(0)};
P^{(0)},Q^{(0)}\big)$ is a particular case of a general double radial
integral
\begin{equation}
R_{\kappa}^{(L_{1},L_{2})}\big(F_{a},F_{b};F_{c},F_{d}\big)
=\int_{0}^{\infty}\mathrm{d}r\int_{0}^{\infty}\mathrm{d}r'\:
\left(
\begin{array}{cc}
F_{a}(r) & F_{b}(r)
\end{array}
\right)
r^{L_{1}}\bar{\mathsf{G}}\mbox{}^{(0)}_{\kappa}(r,r')r^{\prime L_{2}}
\left(
\begin{array}{c}
F_{c}(r') \\
F_{d}(r')
\end{array}
\right)
\label{3.15}
\end{equation}
(other particular forms of this integral will appear in Secs.\
\ref{IV} to \ref{VIII}), with the matrix
\begin{equation}
\bar{\mathsf{G}}\mbox{}^{(0)}_{\kappa}(r,r')
=\left(
\begin{array}{cc}
\bar{g}\mbox{}^{(0)}_{(++)\kappa}(r,r') &
\bar{g}\mbox{}^{(0)}_{(+-)\kappa}(r,r') \\*[1ex]
\bar{g}\mbox{}^{(0)}_{(-+)\kappa}(r,r') &
\bar{g}\mbox{}^{(0)}_{(--)\kappa}(r,r')
\end{array}
\right)
\label{3.16}
\end{equation}
being the radial generalized Dirac--Coulomb Green's function
associated with the ground-state atomic energy level (\ref{2.6}). The
two angular integrals in Eq.\ (\ref{3.14}) may be taken with the help
of the general formula (\ref{2.22}). Once this is done, it is then
possible to carry out summations over the quantum numbers
$m_{\kappa}$, $m$ and $m'$. After straightforward, though tedious,
calculations, one finds that the only non-vanishing contributions to
$\widetilde{\mathcal{Q}}_{\lambda\mu}^{(1)}$ come from the terms with
$\kappa=L$ and $\kappa=-L-1$; one has
\begin{equation}
\widetilde{\mathcal{Q}}_{\lambda\mu}^{(1)}
=\widetilde{\mathcal{Q}}_{\lambda\mu,L}^{(1)}
+\widetilde{\mathcal{Q}}_{\lambda\mu,-L-1}^{(1)},
\label{3.17}
\end{equation}
with
\begin{eqnarray}
\widetilde{\mathcal{Q}}_{\lambda\mu,\kappa}^{(1)}
&=& \delta_{\lambda L}(4\pi\epsilon_{0})
\frac{\sgn(\kappa)}{(2L+1)^{2}}
R_{\kappa}^{(L,L)}\big(P^{(0)},Q^{(0)};P^{(0)},Q^{(0)}\big)
\left\{\left[\kappa+\mu(|a_{1/2}|^{2}-|a_{-1/2}|^{2})\right]
\mathcal{C}_{L\mu}^{(1)}\right.
\nonumber \\
&& \left.+\,\sqrt{(L-\mu)(L+\mu+1)}\,
a_{1/2}a_{-1/2}^{*}\mathcal{C}_{L,\mu+1}^{(1)}
+\sqrt{(L+\mu)(L-\mu+1)}\,
a_{1/2}^{*}a_{-1/2}\mathcal{C}_{L,\mu-1}^{(1)}\right\}
\nonumber \\
&& \hspace*{25em} (\kappa=L,-L-1).
\label{3.18}
\end{eqnarray}
The asterisks at the components of the tensor
$\boldsymbol{\mathsf{C}}_{L}^{(1)}$ in the above equation have
disappeared in virtue of the identity (\ref{2.2}). With Eqs.\
(\ref{3.17}) and (\ref{3.18}) in hand, we return back to Eq.\
(\ref{3.11}). This eventually yields $\mathcal{Q}_{\lambda\mu}^{(1)}$
in the form
\begin{equation}
\mathcal{Q}_{\lambda\mu}^{(1)}=\mathcal{Q}_{\lambda\mu}^{(1)}
\delta_{\lambda L},
\label{3.19}
\end{equation}
with
\begin{equation}
\mathcal{Q}_{L\mu}^{(1)}=\mathcal{Q}_{L\mu,L}^{(1)}
+\mathcal{Q}_{L\mu,-L-1}^{(1)},
\label{3.20}
\end{equation}
where
\begin{equation}
\mathcal{Q}_{L\mu,\kappa}^{(1)}=(4\pi\epsilon_{0})
\frac{2|\kappa|}{(2L+1)^{2}}\,
R_{\kappa}^{(L,L)}\big(P^{(0)},Q^{(0)};P^{(0)},Q^{(0)}\big)
\mathcal{C}_{L\mu}^{(1)}
\qquad (\kappa=L,-L-1).
\label{3.21}
\end{equation}
Equation (\ref{3.19}) shows that the only electric moment induced in
the atom is that one which is precisely of the same multipole
character as the perturbing electric field. In other words, one has
\begin{equation}
\boldsymbol{\mathsf{Q}}_{\lambda}
\simeq\boldsymbol{\mathsf{Q}}_{\lambda}^{(0)}\delta_{\lambda0}
+\boldsymbol{\mathsf{Q}}_{\lambda}^{(1)}\delta_{\lambda L}.
\label{3.22}
\end{equation}
\subsection{Atomic multipole polarizabilities}
\label{III.2}
The $2^{L}$-pole polarizability of the atom in the ground state,
$\alpha_{\mathrm{E}L\to\mathrm{E}L}$, is defined as a proportionality
factor between the induced electric multipole moment
$\boldsymbol{\mathsf{Q}}_{L}^{(1)}$ and the field tensor
$\boldsymbol{\mathsf{C}}_{L}^{(1)}$ appearing in the expression
(\ref{2.1}) for the perturbing $2^{L}$-pole electric
potential:\footnote{~The multipole polarizability $\alpha_{L}$ may be
equivalently defined through the formula
$E^{(2)}=-\frac{1}{2}(4\pi\epsilon_{0})\alpha_{L}
\boldsymbol{\mathsf{C}}_{L}^{(1)}
\cdot\boldsymbol{\mathsf{C}}_{L}^{(1)}$, where $E^{(2)}$ is the
second-order correction to energy. The reader is referred to Appendix
\ref{A} for the justification of this statement.}
\begin{equation}
\boldsymbol{\mathsf{Q}}_{L}^{(1)}=(4\pi\epsilon_{0})
\alpha_{\mathrm{E}L\to\mathrm{E}L}\boldsymbol{\mathsf{C}}_{L}^{(1)}
\label{3.23}
\end{equation}
(the SI factor $4\pi\epsilon_{0}$ has been separated out in order to
secure that the physical dimension of
$\alpha_{\mathrm{E}L\to\mathrm{E}L}$ is \textsf{L}$^{2L+1}$, where
\textsf{L} stands for length). It follows from Eqs.\ (\ref{3.23}) and
(\ref{3.20})--(\ref{3.22}) that the polarizability
$\alpha_{\mathrm{E}L\to\mathrm{E}L}$, hereafter denoted in the
standard manner as $\alpha_{L}$, may be written as the sum
\begin{equation}
\alpha_{L}=\alpha_{L,L}+\alpha_{L,-L-1},
\label{3.24}
\end{equation}
with the constituents being given by
\begin{equation}
\alpha_{L,\kappa}=\frac{2|\kappa|}{(2L+1)^{2}}\,
R_{\kappa}^{(L,L)}\big(P^{(0)},Q^{(0)};P^{(0)},Q^{(0)}\big)
\qquad (\kappa=L,-L-1).
\label{3.25}
\end{equation}

The remaining task is to evaluate the double radial integral
$R_{\kappa}^{(L,L)}\big(P^{(0)},Q^{(0)};P^{(0)},Q^{(0)}\big)$.
This will be done below with the aid of the Sturmian expansion of the
radial generalized Dirac--Coulomb Green's function \cite{Szmy97}:
\begin{equation}
\bar{\mathsf{G}}\mbox{}^{(0)}_{\kappa}(r,r')
=\sum_{n_{r}=-\infty}^{\infty}
\frac{1}{\mu_{n_{r}\kappa}^{(0)}-1}
\left(
\begin{array}{c}
S_{n_{r}\kappa}^{(0)}(r) \\*[1ex]
T_{n_{r}\kappa}^{(0)}(r)
\end{array}
\right)
\left(
\begin{array}{cc}
\mu_{n_{r}\kappa}^{(0)}S_{n_{r}\kappa}^{(0)}(r') 
& T_{n_{r}\kappa}^{(0)}(r')
\end{array}
\right)
\qquad (\kappa\neq-1),
\label{3.26}
\end{equation}
where 
\begin{subequations}
\begin{eqnarray}
\hspace*{-5em}
S_{n_{r}\kappa}^{(0)}(r)
&=& \sqrt{\frac{(1+\gamma_{1})(|n_{r}|+2\gamma_{\kappa})|n_{r}|!}
{2ZN_{n_{r}\kappa}(N_{n_{r}\kappa}-\kappa)
\Gamma(|n_{r}|+2\gamma_{\kappa})}}
\nonumber \\
&& \times\,\left(\frac{2Zr}{a_{0}}\right)^{\gamma_{\kappa}}
\textrm{e}^{-Zr/a_{0}}
\left[L_{|n_{r}|-1}^{(2\gamma_{\kappa})}\left(\frac{2Zr}{a_{0}}\right)
+\frac{\kappa-N_{n_{r}\kappa}}{|n_{r}|+2\gamma_{\kappa}}
L_{|n_{r}|}^{(2\gamma_{\kappa})}\left(\frac{2Zr}{a_{0}}\right)\right]
\label{3.27a}
\end{eqnarray}
and
\begin{eqnarray}
\hspace*{-5em}
T_{n_{r}\kappa}^{(0)}(r)
&=& \sqrt{\frac{(1-\gamma_{1})(|n_{r}|+2\gamma_{\kappa})|n_{r}|!}
{2ZN_{n_{r}\kappa}(N_{n_{r}\kappa}-\kappa)
\Gamma(|n_{r}|+2\gamma_{\kappa})}}
\nonumber \\
&& \times\,\left(\frac{2Zr}{a_{0}}\right)^{\gamma_{\kappa}}
\textrm{e}^{-Zr/a_{0}}
\left[L_{|n_{r}|-1}^{(2\gamma_{\kappa})}
\left(\frac{2Zr}{a_{0}}\right)
-\frac{\kappa-N_{n_{r}\kappa}}{|n_{r}|+2\gamma_{\kappa}}
L_{|n_{r}|}^{(2\gamma_{\kappa})}\left(\frac{2Zr}{a_{0}}\right)\right]
\label{3.27b}
\end{eqnarray}
\label{3.27}%
\end{subequations}
are the radial Dirac--Coulomb Sturmian functions associated with the
atomic ground-state energy level (\ref{2.6}) [here
$L_{n}^{(\alpha)}(\rho)$ denotes the generalized Laguerre polynomial
\cite{Magn66}; we define $L_{-1}^{(\alpha)}(\rho)\equiv0$], while
\begin{equation}
\mu_{n_{r}\kappa}^{(0)}
=\frac{|n_{r}|+\gamma_{\kappa}+N_{n_{r}\kappa}}{\gamma_{1}+1}
\label{3.28}
\end{equation}
is the pertinent Sturmian eigenvalue. The ``apparent principal
quantum number'' appearing in Eqs.\ (\ref{3.27}) and (\ref{3.28}) is
defined as
\begin{equation}
N_{n_{r}\kappa}
=\pm\sqrt{(|n_{r}|+\gamma_{\kappa})^{2}+(\alpha Z)^{2}}
=\pm\sqrt{|n_{r}|^{2}+2|n_{r}|\gamma_{\kappa}+\kappa^{2}},
\label{3.29}
\end{equation}
where the plus sign is to be chosen for $n_{r}>0$ and the minus sign
for $n_{r}<0$; for $n_{r}=0$ one should choose the plus sign if
$\kappa<0$ and the minus sign if $\kappa>0$, i.e., one has
$N_{0\kappa}=-\kappa$.

Substitution of the expansion (\ref{3.26}) into the definition of the
double radial integral which appears in Eq.\ (\ref{3.25}) yields
$\alpha_{L,\kappa}$ in the form
\begin{eqnarray}
\hspace*{-5em}
\alpha_{L,\kappa}
&=& \frac{2|\kappa|}{(2L+1)^{2}}
\sum_{n_{r}=-\infty}^{\infty}\frac{1}{\mu_{n_{r}\kappa}^{(0)}-1}
\int_{0}^{\infty}\mathrm{d}r\:r^{L}
\left[P^{(0)}(r)S_{n_{r}\kappa}^{(0)}(r)
+Q^{(0)}(r)T_{n_{r}\kappa}^{(0)}(r)\right]
\nonumber \\
&& \times\int_{0}^{\infty}\mathrm{d}r'\:r^{\prime L}
\left[\mu_{n_{r}\kappa}^{(0)}P^{(0)}(r')S_{n_{r}\kappa}^{(0)}(r')
+Q^{(0)}(r')T_{n_{r}\kappa}^{(0)}(r')\right]
\qquad (\kappa=L,-L-1).
\label{3.30}
\end{eqnarray}
The evident advantage of the use of the Sturmian expansion
(\ref{3.26}) is that in the resulting Eq.\ (\ref{3.30}) the
integrations over $r$ and $r'$ may be carried out separately. On
exploiting Eqs.\ (\ref{2.11}), (\ref{3.27}), (\ref{3.28}), the
integral formula \cite[Eq.\ (7.414.11)]{Grad07}
\begin{equation} 
\int_{0}^{\infty}\mathrm{d}\rho\:
\rho^{\gamma}\mathrm{e}^{-\rho}L_{n}^{(\alpha)}(\rho)
=\frac{\Gamma(\gamma+1)\Gamma(n+\alpha-\gamma)}
{n!\Gamma(\alpha-\gamma)}
\qquad (\Real\gamma>-1)
\label{3.31}
\end{equation}
and the identity
\begin{equation}
\gamma_{\kappa}^{2}=\gamma_{1}^{2}+\kappa^{2}-1,
\label{3.32}
\end{equation}
one finds that
\begin{eqnarray}
&& \hspace*{-5em} 
\int_{0}^{\infty}\mathrm{d}r\:r^{L}
\left[P^{(0)}(r)S_{n_{r}\kappa}^{(0)}(r)
+Q^{(0)}(r)T_{n_{r}\kappa}^{(0)}(r)\right]
\nonumber \\
&=& -\,\left(\frac{a_{0}}{2Z}\right)^{L+1}
\frac{\sqrt{2}(N_{n_{r}\kappa}-\kappa)
[\gamma_{1}(N_{n_{r}\kappa}+\kappa)
-(|n_{r}|+\gamma_{\kappa}-\gamma_{1}-L-1)]}
{\sqrt{a_{0}|n_{r}|!N_{n_{r}\kappa}(N_{n_{r}\kappa}-\kappa)
\Gamma(2\gamma_{1}+1)\Gamma(|n_{r}|+2\gamma_{\kappa}+1)}}
\nonumber \\
&& \times\,\frac{\Gamma(\gamma_{\kappa}+\gamma_{1}+L+1)
\Gamma(|n_{r}|+\gamma_{\kappa}-\gamma_{1}-L-1)}
{\Gamma(\gamma_{\kappa}-\gamma_{1}-L)}
\label{3.33}
\end{eqnarray}
and
\begin{eqnarray}
&& \hspace*{-3em} 
\int_{0}^{\infty}\mathrm{d}r\:r^{L}
\left[\mu_{n_{r}\kappa}^{(0)}P^{(0)}(r)S_{n_{r}\kappa}^{(0)}(r)
+Q^{(0)}(r)T_{n_{r}\kappa}^{(0)}(r)\right]
\nonumber \\
&=& -\,\left(\frac{a_{0}}{2Z}\right)^{L+1}
\frac{(\mu_{n_{r}\kappa}^{(0)}-1)(N_{n_{r}\kappa}-\kappa)}
{\sqrt{2a_{0}|n_{r}|!N_{n_{r}\kappa}(N_{n_{r}\kappa}-\kappa)
\Gamma(2\gamma_{1}+1)\Gamma(|n_{r}|+2\gamma_{\kappa}+1)}}
\nonumber \\
&& \times\,\frac{\Gamma(\gamma_{\kappa}+\gamma_{1}+L+1)
\Gamma(|n_{r}|+\gamma_{\kappa}-\gamma_{1}-L-1)}
{\Gamma(\gamma_{\kappa}-\gamma_{1}-L)}
\bigg\{[(N_{n_{r}\kappa}+\kappa)-\gamma_{1}
(|n_{r}|+\gamma_{\kappa}-\gamma_{1}-L-1)]
\nonumber \\
&& \quad 
+\,\frac{N_{n_{r}\kappa}+1}{|n_{r}|+\gamma_{\kappa}-\gamma_{1}}
[\gamma_{1}(N_{n_{r}\kappa}+\kappa)
-(|n_{r}|+\gamma_{\kappa}-\gamma_{1}-L-1)]\bigg\}.
\label{3.34}
\end{eqnarray}
Once these two equations are plugged into Eq.\ (\ref{3.30}), then use
is made of Eqs.\ (\ref{3.29}) and (\ref{3.32}) and subsequently the
terms in the summand corresponding to the same absolute value of the
summation index $n_{r}$ are collected together, with a good deal of
labor one finds the following infinite-series representation for
$\alpha_{L,\kappa}$:
\begin{eqnarray}
\hspace*{-7em}
\alpha_{L,\kappa} &=& \frac{a_{0}^{2L+1}}{Z^{2L+2}}
\frac{|\kappa|\Gamma^{2}(\gamma_{\kappa}+\gamma_{1}+L+1)}
{2^{2L}(2L+1)^{2}\Gamma(2\gamma_{1}+1)
\Gamma^{2}(\gamma_{\kappa}-\gamma_{1}-L)}
\nonumber \\
&& \times\,\bigg\{\gamma_{1}[\gamma_{1}(\kappa+1)+2(L+1)]
\sum_{n_{r}=0}^{\infty}
\frac{\Gamma^{2}(n_{r}+\gamma_{\kappa}-\gamma_{1}-L)}
{n_{r}!(n_{r}+\gamma_{\kappa}-\gamma_{1}+1)
\Gamma(n_{r}+2\gamma_{\kappa}+1)}
\nonumber \\
&& \quad -\,(\kappa-1)\sum_{n_{r}=0}^{\infty}
\frac{\Gamma^{2}(n_{r}+\gamma_{\kappa}-\gamma_{1}-L)}
{n_{r}!(n_{r}+\gamma_{\kappa}-\gamma_{1})
\Gamma(n_{r}+2\gamma_{\kappa}+1)}\bigg\}
\qquad (\kappa=L,-L-1).
\label{3.35}
\end{eqnarray}
The two series in Eq.\ (\ref{3.35}) may be expressed in terms of the
generalized hypergeometric function ${}_{3}F_{2}$ of the unit
argument. With the aid of the identity
\begin{eqnarray}
&& \sum_{n=0}^{\infty}\frac{\Gamma(n+a_{1})\Gamma(n+a_{2})
\Gamma(n+a_{3})}{n!\Gamma(n+b_{1})\Gamma(n+b_{2})}
=\frac{\Gamma(a_{1})\Gamma(a_{2})\Gamma(a_{3})}
{\Gamma(b_{1})\Gamma(b_{2})}\,
{}_{3}F_{2}
\left(
\begin{array}{c}
a_{1},\:
a_{2},\:
a_{3} \\
b_{1},\:
b_{2}
\end{array}
;1
\right)
\nonumber \\
&& \hspace*{18em} [\Real(b_{1}+b_{2}-a_{1}-a_{2}-a_{3})>0],
\label{3.36}
\end{eqnarray}
one arrives at
\begin{eqnarray}
\alpha_{L,\kappa} &=& \frac{a_{0}^{2L+1}}{Z^{2L+2}}
\frac{|\kappa|\Gamma^{2}(\gamma_{\kappa}+\gamma_{1}+L+1)}
{2^{2L}(2L+1)^{2}\Gamma(2\gamma_{1}+1)\Gamma(2\gamma_{\kappa}+1)}
\nonumber \\
&& \times\,\bigg\{\frac{\gamma_{1}[\gamma_{1}(\kappa+1)+2(L+1)]}
{\gamma_{\kappa}-\gamma_{1}+1}\,
{}_{3}F_{2}
\left(
\begin{array}{c}
\gamma_{\kappa}-\gamma_{1}-L,\:
\gamma_{\kappa}-\gamma_{1}-L,\:
\gamma_{\kappa}-\gamma_{1}+1 \\
\gamma_{\kappa}-\gamma_{1}+2,\:
2\gamma_{\kappa}+1
\end{array}
;1
\right)
\nonumber \\
&& \quad -\,\frac{\gamma_{\kappa}+\gamma_{1}}{\kappa+1}\,
{}_{3}F_{2}
\left(
\begin{array}{c}
\gamma_{\kappa}-\gamma_{1}-L,\:
\gamma_{\kappa}-\gamma_{1}-L,\:
\gamma_{\kappa}-\gamma_{1} \\
\gamma_{\kappa}-\gamma_{1}+1,\:
2\gamma_{\kappa}+1
\end{array}
;1
\right)
\bigg\}
\qquad (\kappa=L,-L-1).
\nonumber \\
&&
\label{3.37}
\end{eqnarray}
A simplification of the above result may be attained with the help of
the relation
\begin{eqnarray}
{}_{3}F_{2}
\left(
\begin{array}{c}
a_{1},\:
a_{2},\:
a_{3} \\
a_{3}+1,\:
b
\end{array}
;1
\right)
&=& \frac{\Gamma(b)\Gamma(b-a_{1}-a_{2}+1)}
{(b-a_{3}-1)\Gamma(b-a_{1})\Gamma(b-a_{2})}
\nonumber \\
&& -\,\frac{(a_{1}-a_{3}-1)(a_{2}-a_{3}-1)}{(a_{3}+1)(b-a_{3}-1)}
\,{}_{3}F_{2}
\left(
\begin{array}{c}
a_{1},\:
a_{2},\:
a_{3}+1 \\
a_{3}+2,\:
b
\end{array}
;1
\right)
\nonumber \\
&& \hspace*{10em} [\Real(b-a_{1}-a_{2})>-1],
\label{3.38}
\end{eqnarray}
which may be used to eliminate one of the two ${}_{3}F_{2}(1)$'s in
favor of the other. It appears that a bit more compact result is
obtained if the first ${}_{3}F_{2}(1)$ is retained:
\begin{eqnarray}
\alpha_{L,\kappa} &=& \frac{a_{0}^{2L+1}}{Z^{2L+2}}
\frac{|\kappa|\Gamma(2\gamma_{1}+2L+2)}
{2^{2L}(\kappa+1)(2L+1)^{2}\Gamma(2\gamma_{1}+1)}
\bigg\{-1+\frac{[\gamma_{1}(\kappa+1)+L+1]^{2}
\Gamma^{2}(\gamma_{\kappa}+\gamma_{1}+L+1)}
{(\gamma_{\kappa}-\gamma_{1}+1)\Gamma(2\gamma_{1}+2L+2)
\Gamma(2\gamma_{\kappa}+1)}
\nonumber \\
&& \times\,
{}_{3}F_{2}
\left(
\begin{array}{c}
\gamma_{\kappa}-\gamma_{1}-L,\:
\gamma_{\kappa}-\gamma_{1}-L,\:
\gamma_{\kappa}-\gamma_{1}+1 \\
\gamma_{\kappa}-\gamma_{1}+2,\:
2\gamma_{\kappa}+1
\end{array}
;1
\right)
\bigg\}
\qquad (\kappa=L,-L-1).
\label{3.39}
\end{eqnarray}
Specializing to the two admitted values of $\kappa$, we finally
arrive at
\begin{subequations}
\begin{eqnarray}
\alpha_{L,L} 
&=& \frac{a_{0}^{2L+1}}{Z^{2L+2}}\frac{L\Gamma(2\gamma_{1}+2L+2)}
{2^{2L}(L+1)(2L+1)^{2}\Gamma(2\gamma_{1}+1)}
\bigg[-1+\frac{(L+1)^{2}(\gamma_{1}+1)^{2}
\Gamma^{2}(\gamma_{L}+\gamma_{1}+L+1)}{(\gamma_{L}-\gamma_{1}+1)
\Gamma(2\gamma_{1}+2L+2)\Gamma(2\gamma_{L}+1)}
\nonumber \\
&& \times\,
{}_{3}F_{2}
\left(
\begin{array}{c}
\gamma_{L}-\gamma_{1}-L,\:
\gamma_{L}-\gamma_{1}-L,\:
\gamma_{L}-\gamma_{1}+1 \\
\gamma_{L}-\gamma_{1}+2,\:
2\gamma_{L}+1
\end{array}
;1
\right)
\bigg]
\label{3.40a}
\end{eqnarray}
and
\begin{eqnarray}
\alpha_{L,-L-1} 
&=& \frac{a_{0}^{2L+1}}{Z^{2L+2}}\frac{(L+1)\Gamma(2\gamma_{1}+2L+2)}
{2^{2L}L(2L+1)^{2}\Gamma(2\gamma_{1}+1)}
\bigg\{1-\frac{(L\gamma_{1}-L-1)^{2}
\Gamma^{2}(\gamma_{L+1}+\gamma_{1}+L+1)}{(\gamma_{L+1}-\gamma_{1}+1)
\Gamma(2\gamma_{1}+2L+2)\Gamma(2\gamma_{L+1}+1)}
\nonumber \\
&& \times\,
{}_{3}F_{2}
\left(
\begin{array}{c}
\gamma_{L+1}-\gamma_{1}-L,\:
\gamma_{L+1}-\gamma_{1}-L,\:
\gamma_{L+1}-\gamma_{1}+1 \\
\gamma_{L+1}-\gamma_{1}+2,\:
2\gamma_{L+1}+1
\end{array}
;1
\right)
\bigg\},
\label{3.40b}
\end{eqnarray}
\label{3.40}%
\end{subequations}
respectively. Hence, the closed-form expression for the $2^{L}$-pole
polarizability of the hydrogen-like in the ground state is
\begin{eqnarray}
\alpha_{L} &=& \frac{a_{0}^{2L+1}}{Z^{2L+2}}
\frac{\Gamma(2\gamma_{1}+2L+2)}
{2^{2L}L(L+1)(2L+1)\Gamma(2\gamma_{1}+1)}
\nonumber \\
&& \times\,\bigg\{1+\frac{L^{2}(L+1)^{2}(\gamma_{1}+1)^{2}
\Gamma^{2}(\gamma_{L}+\gamma_{1}+L+1)}
{(2L+1)(\gamma_{L}-\gamma_{1}+1)\Gamma(2\gamma_{1}+2L+2)
\Gamma(2\gamma_{L}+1)}
\nonumber \\
&& \qquad \times\,{}_{3}F_{2}
\left(
\begin{array}{c}
\gamma_{L}-\gamma_{1}-L,\:
\gamma_{L}-\gamma_{1}-L,\:
\gamma_{L}-\gamma_{1}+1 \\
\gamma_{L}-\gamma_{1}+2,\:
2\gamma_{L}+1
\end{array}
;1
\right) 
\nonumber \\
&& \quad -\,\frac{(L+1)^{2}(L\gamma_{1}-L-1)^{2}
\Gamma^{2}(\gamma_{L+1}+\gamma_{1}+L+1)}
{(2L+1)(\gamma_{L+1}-\gamma_{1}+1)\Gamma(2\gamma_{1}+2L+2)
\Gamma(2\gamma_{L+1}+1)}
\nonumber \\
&& \qquad \times\,{}_{3}F_{2}
\left(
\begin{array}{c}
\gamma_{L+1}-\gamma_{1}-L,\:
\gamma_{L+1}-\gamma_{1}-L,\:
\gamma_{L+1}-\gamma_{1}+1 \\
\gamma_{L+1}-\gamma_{1}+2,\:
2\gamma_{L+1}+1
\end{array}
;1
\right)
\bigg\}.
\label{3.41}
\end{eqnarray}
In the dipole case ($L=1$), Eq.\ (\ref{3.41}) yields
\begin{eqnarray}
\alpha_{1} &=& \frac{a_{0}^{3}}{Z^{4}}
\left[\frac{(\gamma_{1}+1)(2\gamma_{1}+1)
(4\gamma_{1}^{2}+13\gamma_{1}+12)}{36}
-\frac{(\gamma_{1}-2)^{2}\Gamma^{2}(\gamma_{2}+\gamma_{1}+2)}
{18(\gamma_{2}-\gamma_{1}+1)\Gamma(2\gamma_{1}+1)
\Gamma(2\gamma_{2}+1)}\right.
\nonumber \\
&& \left.\times\,
{}_{3}F_{2}
\left(
\begin{array}{c}
\gamma_{2}-\gamma_{1}-1,\:
\gamma_{2}-\gamma_{1}-1,\:
\gamma_{2}-\gamma_{1}+1 \\
\gamma_{2}-\gamma_{1}+2,\:
2\gamma_{2}+1
\end{array}
;1
\right)
\right],
\label{3.42}
\end{eqnarray}
which is in agreement with earlier findings (cf.\ Refs.\ \cite[Eq.\
(16)]{Yakh03} and \cite[Eq.\ (3.24)]{Szmy04}).

Exact numerical values of the dipole to hexadecapole polarizabilities
for the ground state of the hydrogen ($Z=1$) atom, derived directly
from the analytical formula (\ref{3.41}), are presented in Table
\ref{T.1}. Calculations have been done for two values of the inverse
of the fine-structure constant: $\alpha^{-1}=137.035\:999\:139$ (from
CODATA 2014) and $\alpha^{-1}=137.035\:999\:074$ (from CODATA 2010),
in the latter case to enable making comparison with data available in
Refs.\ \cite{Tang12,Fili14}. Table \ref{T.1} confirms almost perfect
numerical accuracy of results obtained computationally by Tang
\emph{et al.\/} \cite{Tang12} using the $B$-spline Galerkin method,
and also high quality of numbers generated by Filippin \emph{et
al.\/} \cite{Fili14} with the use of the Langrange-mesh method.

Tabulation of exact numerical values of the first four multipole
polarizabilities for ground states of selected hydrogenic ions is
presented in Table \ref{T.2}. Again two data sets are displayed, as a
result of the use of the two aforementioned values of the inverse of
the fine-structure constant. In general, the present data generated
with the CODATA 2010 value of $\alpha^{-1}$ validate completely
counterpart numbers from Refs.\ \cite{Zhan12,Tang12} and imply only
minor inaccuracies (the maximal relative error being of the order of
$10^{-12}$) in the data listed in Ref.\ \cite{Fili14}.
\begin{center}
[Place for Tables \ref{T.1} and \ref{T.2}]
\end{center}

In the final step, we shall provide an approximate formula for the
polarizability $\alpha_{L}$ that is correct to the second order in
$\alpha Z$. Using
\begin{equation}
\gamma_{\kappa}\simeq|\kappa|-\frac{(\alpha Z)^{2}}{2|\kappa|}
\label{3.43}
\end{equation}
and
\begin{equation}
\Gamma(a\gamma_{\kappa}+a'\gamma_{\kappa'}+b)
\simeq\Gamma(a|\kappa|+a'|\kappa'|+b)
\left[1-\frac{(\alpha Z)^{2}}{2}
\left(\frac{a}{|\kappa|}+\frac{a'}{|\kappa'|}\right)
\psi(a|\kappa|+a'|\kappa'|+b)\right],
\label{3.44}
\end{equation}
where
\begin{equation}
\psi(z)=\frac{1}{\Gamma(z)}\frac{\mathrm{d}\Gamma(z)}{\mathrm{d}z}
\label{3.45}
\end{equation}
is the digamma function, one finds
\begin{eqnarray}
&& \hspace*{-5em}
{}_{3}F_{2}
\left(
\begin{array}{c}
\gamma_{L}-\gamma_{1}-L,\:
\gamma_{L}-\gamma_{1}-L,\:
\gamma_{L}-\gamma_{1}+1 \\
\gamma_{L}-\gamma_{1}+2,\:
2\gamma_{L}+1
\end{array}
;1
\right)
\nonumber \\
&\simeq& \frac{2L^{2}+4L+1}{(L+1)(2L+1)}
-(\alpha Z)^{2}\frac{4L^{4}+2L^{3}-8L^{2}-3L+1}
{2L(L+1)^{2}(2L+1)^{2}}
\label{3.46}
\end{eqnarray}
and
\begin{equation}
{}_{3}F_{2}
\left(
\begin{array}{c}
\gamma_{L+1}-\gamma_{1}-L,\:
\gamma_{L+1}-\gamma_{1}-L,\:
\gamma_{L+1}-\gamma_{1}+1 \\
\gamma_{L+1}-\gamma_{1}+2,\:
2\gamma_{L+1}+1
\end{array}
;1
\right)
\simeq 1,
\label{3.47}
\end{equation}
and further
\begin{subequations}
\begin{equation}
\alpha_{L,L} \simeq \frac{a_{0}^{2L+1}}{Z^{2L+2}}
\frac{(L+2)(2L)!}{2^{2L}}\left\{1-(\alpha Z)^{2}
\left[\psi(2L+3)-\psi(3)+\frac{2L^{5}+7L^{4}+6L^{3}-L-1}
{L^{2}(L+1)(L+2)(2L+1)}\right]\right\},
\label{3.48a}
\end{equation}
\begin{eqnarray}
\hspace*{-2em}
\alpha_{L,-L-1} &\simeq& \frac{a_{0}^{2L+1}}{Z^{2L+2}}
\frac{(L+1)(L+2)(2L-1)!}{2^{2L-1}}
\nonumber \\
&& \times\,\left\{1-(\alpha Z)^{2}
\left[\psi(2L+3)-\psi(3)+\frac{4L^{3}+10L^{2}+7L+2}
{2(L+1)^{2}(L+2)(2L+1)}\right]\right\}.
\label{3.48b}
\end{eqnarray}
\label{3.48}%
\end{subequations}
Adding Eqs.\ (\ref{3.48a}) and (\ref{3.48b}) yields the following
approximation for $\alpha_{L}$:
\begin{eqnarray}
\alpha_{L} &\simeq& \frac{a_{0}^{2L+1}}{Z^{2L+2}}
\frac{(L+2)(2L+1)!}{2^{2L}L}
\nonumber \\
&& \times\,\left\{1-(\alpha Z)^{2}
\left[\psi(2L+3)-\psi(3)+\frac{4L^{5}+18L^{4}+22L^{3}+7L^{2}-2}
{2L(L+1)(L+2)(2L+1)^{2}}\right]\right\},
\label{3.49}
\end{eqnarray}
which is identical with the one given in Refs.\
\cite{Mana74,Zapr81,Zapr85} and may be also shown to be equivalent to
the counterpart expressions given in Ref.\ \cite{Kane77} and in an
erratum to Ref.\ \cite{Drac85}. If in the above equation one uses
recursively the relation
\begin{equation}
\psi(z+1)=\psi(z)+\frac{1}{z},
\label{3.50}
\end{equation}
a bit more compact approximation to $\alpha_{L}$ is obtained:
\begin{eqnarray}
\alpha_{L} &\simeq& \frac{a_{0}^{2L+1}}{Z^{2L+2}}
\frac{(L+2)(2L+1)!}{2^{2L}L}
\nonumber \\
&& \times\,\left\{1-(\alpha Z)^{2}
\left[\psi(2L)-\psi(2)+\frac{14L^{3}+43L^{2}+40L+15}
{2(L+1)(L+2)(2L+1)^{2}}\right]\right\}.
\label{3.51}
\end{eqnarray}
Explicit expressions for the quasi-relativistic approximations to
$\alpha_{L}$ with $1\leqslant L\leqslant4$, resulting from Eq.\
(\ref{3.51}), are displayed in Table \ref{T.3}.
\begin{center}
[Place for Table \ref{T.3}]
\end{center}
%
%\newpage
%
\section{Magnetic multipole moments of the atom in the multipole
electric field and atomic E$L\to$M$(L\mp1)$ multipole
cross-susceptibilities}
\label{IV}
\setcounter{equation}{0}
\subsection{Decomposition of the atomic magnetic multipole moments
into the permanent and the first-order electric-field-induced
components}
\label{IV.1}
Next, we proceed to the investigation of electric-field induced
magnetic multipole moments of the Dirac one-electron atom in the
ground state. In Appendix \ref{B}, components of the $2^{L}$-pole
magnetic moment $\boldsymbol{\mathsf{M}}_{\lambda}$ for a stationary
sourceless current distribution $\boldsymbol{j}(\boldsymbol{r})$ are
defined as
\begin{equation}
\mathcal{M}_{\lambda\mu}
=-\mathrm{i}\,\sqrt{\frac{4\pi\lambda}{(\lambda+1)(2\lambda+1)}}
\int_{\mathbb{R}^{3}}\mathrm{d}^{3}\boldsymbol{r}\:
r^{\lambda}\boldsymbol{Y}_{\lambda\mu}^{\lambda}(\boldsymbol{n}_{r})
\cdot\boldsymbol{j}(\boldsymbol{r}),
\label{4.1}
\end{equation}
where $\boldsymbol{Y}_{\lambda\mu}^{\lambda}(\boldsymbol{n}_{r})$ is
a particular vector spherical harmonic \cite[Sec.\ 7.3.1]{Vars75}.
The right-hand side of Eq.\ (\ref{4.1}) may be transformed to another
form using the identity \cite[Sec.\ 7.3.1]{Vars75}
\begin{equation}
\boldsymbol{Y}_{\lambda\mu}^{\lambda}(\boldsymbol{n}_{r})
=\frac{\boldsymbol{\Lambda}Y_{\lambda\mu}(\boldsymbol{n}_{r})}
{\sqrt{\lambda(\lambda+1)}},
\label{4.2}
\end{equation}
where
\begin{equation}
\boldsymbol{\Lambda}
=-\mathrm{i}\boldsymbol{r}\times\boldsymbol{\nabla}
\label{4.3}
\end{equation}
is the orbital angular momentum operator. Plugging Eq.\ (\ref{4.2})
into Eq.\ (\ref{4.1}), after exploiting the Hermiticity property of
the operator $\boldsymbol{\Lambda}$, one obtains the formula
\begin{equation}
\mathcal{M}_{\lambda\mu}
=\frac{\mathrm{i}}{\lambda+1}\sqrt{\frac{4\pi}{2\lambda+1}}
\int_{\mathbb{R}^{3}}\mathrm{d}^{3}\boldsymbol{r}\:
r^{\lambda}Y_{\lambda\mu}(\boldsymbol{n}_{r})
\boldsymbol{\Lambda}\cdot\boldsymbol{j}(\boldsymbol{r}),
\label{4.4}
\end{equation}
which appears to be optimal for the use in the subsequent
considerations.

In the weak-perturbing-field case considered in this work, the atomic
wave function may be approximated as in Eq.\ (\ref{2.8}). Hence,
after using Eqs.\ (\ref{2.9}), (\ref{2.14}) and (\ref{2.16}), to the
first order in the perturbing electric multipole field, the electronic
current in the atom
\begin{equation}
\boldsymbol{j}(\boldsymbol{r})
=\frac{-ec\Psi^{\dag}(\boldsymbol{r})
\boldsymbol{\alpha}\Psi(\boldsymbol{r})}
{\int_{\mathbb{R}^{3}}\mathrm{d}^{3}\boldsymbol{r}'\:
\Psi^{\dag}(\boldsymbol{r}')\Psi(\boldsymbol{r}')}
\label{4.5}
\end{equation}
may be approximated as
\begin{equation}
\boldsymbol{j}(\boldsymbol{r})
\simeq\boldsymbol{j}^{(0)}(\boldsymbol{r})
+\boldsymbol{j}^{(1)}(\boldsymbol{r}),
\label{4.6}
\end{equation}
where
\begin{equation}
\boldsymbol{j}^{(0)}(\boldsymbol{r})
=-ec\Psi^{(0)\dag}(\boldsymbol{r})\boldsymbol{\alpha}
\Psi^{(0)}(\boldsymbol{r})
\label{4.7}
\end{equation}
is the electronic current in the unperturbed atomic ground state,
while
\begin{equation}
\boldsymbol{j}^{(1)}(\boldsymbol{r})
=-2ec\Real[\Psi^{(0)\dag}(\boldsymbol{r})\boldsymbol{\alpha}
\Psi^{(1)}(\boldsymbol{r})]
\label{4.8}
\end{equation}
is the leading term in the field-induced electronic current. The
approximation in Eq.\ (\ref{4.6}) implies that
\begin{equation}
\mathcal{M}_{\lambda\mu}
\simeq \mathcal{M}_{\lambda\mu}^{(0)}
+\mathcal{M}_{\lambda\mu}^{(1)},
\label{4.9}
\end{equation}
where
\begin{equation}
\mathcal{M}_{\lambda\mu}^{(0)}
=\frac{\mathrm{i}}{\lambda+1}\sqrt{\frac{4\pi}{2\lambda+1}}
\int_{\mathbb{R}^{3}}\mathrm{d}^{3}\boldsymbol{r}\:
r^{\lambda}Y_{\lambda\mu}(\boldsymbol{n}_{r})
\boldsymbol{\Lambda}\cdot\boldsymbol{j}^{(0)}(\boldsymbol{r})
\label{4.10}
\end{equation}
are components of the magnetic $2^{\lambda}$-pole moment of the atom
in the unperturbed state $\Psi^{(0)}(\boldsymbol{r})$, while
\begin{equation}
\mathcal{M}_{\lambda\mu}^{(1)}
=\frac{\mathrm{i}}{\lambda+1}\sqrt{\frac{4\pi}{2\lambda+1}}
\int_{\mathbb{R}^{3}}\mathrm{d}^{3}\boldsymbol{r}\:
r^{\lambda}Y_{\lambda\mu}(\boldsymbol{n}_{r})
\boldsymbol{\Lambda}\cdot\boldsymbol{j}^{(1)}(\boldsymbol{r})
\label{4.11}
\end{equation}
is the first-order field-induced correction to
$\mathcal{M}_{\lambda\mu}^{(0)}$.

It has been shown by us in Ref.\ \cite{Szmy14} that the only
non-vanishing unperturbed multipole moment
$\boldsymbol{\mathsf{M}}_{\lambda}^{(0)}$ is the dipole one:
\begin{equation}
\mathcal{M}_{\lambda\mu}^{(0)}=\mathcal{M}_{\lambda\mu}^{(0)}
\delta_{\lambda1},
\label{4.12}
\end{equation}
the spherical components of which are 
\begin{subequations}
\begin{equation}
\mathcal{M}_{1,0}^{(0)}=-\frac{1}{3}(2\gamma_{1}+1)\mu_{\mathrm{B}}
(|a_{1/2}|^{2}-|a_{-1/2}|^{2}),
\label{4.13a}
\end{equation}
\begin{equation}
\mathcal{M}_{1,\pm1}^{(0)}=\pm\frac{\sqrt{2}}{3}(2\gamma_{1}+1)
\mu_{\mathrm{B}}a_{\pm1/2}^{*}a_{\mp{1/2}},
\label{4.13b}
\end{equation}
\label{4.13}%
\end{subequations}
where
\begin{equation}
\mu_{\mathrm{B}}=\frac{e\hbar}{2m_{\mathrm{e}}}
\label{4.14}
\end{equation}
is the Bohr magneton. If we introduce the unit vector
$\boldsymbol{\nu}$ with the cyclic components
\begin{equation}
\nu_{0}=|a_{1/2}|^{2}-|a_{-1/2}|^{2},
\qquad
\nu_{\pm1}=\mp\sqrt{2}\,a_{\pm1/2}^{*}a_{\mp{1/2}},
\label{4.15}
\end{equation}
the magnetic dipole moment vector may be compactly written as
\begin{equation}
\boldsymbol{\mathsf{M}}_{1}^{(0)}=-\frac{2\gamma_{1}+1}{3}
\mu_{\mathrm{B}}\boldsymbol{\nu}.
\label{4.16}
\end{equation}
The parametrization
\begin{equation}
a_{1/2}=\mathrm{e}^{-\mathrm{i}(\chi+\phi)/2}\cos(\vartheta/2),
\qquad
a_{-1/2}=\mathrm{e}^{-\mathrm{i}(\chi-\phi)/2}\sin(\vartheta/2)
\qquad (\textrm{$0\leqslant\chi,\phi<2\pi$, 
$0\leqslant\vartheta\leqslant\pi$})
\label{4.17}
\end{equation}
implies that
\begin{equation}
\nu_{0}=\cos\vartheta,
\qquad \nu_{\pm1}=\mp\frac{1}{\sqrt{2}}\,
\mathrm{e}^{\pm\mathrm{i}\phi}\sin\vartheta,
\label{4.18}
\end{equation}
i.e., $\vartheta$ and $\phi$ may be considered as the polar and the
azimuthal angles, respectively, of the vector $\boldsymbol{\nu}$ in
the spherical coordinate system.

Once the nature of the unperturbed moments has been explained, we
proceed to the analysis of the induced moments
$\boldsymbol{\mathsf{M}}_{\lambda}^{(1)}$. Substituting the
expression (\ref{4.8}) for the induced current into the definition
(\ref{4.11}), after using the identity (\ref{3.10}), we obtain
\begin{equation}
\mathcal{M}_{\lambda\mu}^{(1)}
=\widetilde{\mathcal{M}}_{\lambda\mu}^{(1)}
+(-)^{\mu}\widetilde{\mathcal{M}}_{\lambda,-\mu}^{(1)*},
\label{4.19}
\end{equation}
where
\begin{equation}
\widetilde{\mathcal{M}}^{(1)}_{\lambda\mu}
=-\frac{\mathrm{i}ec}{\lambda+1}\sqrt{\frac{4\pi}{2\lambda+1}}
\int_{\mathbb{R}^{3}}\mathrm{d}^{3}\boldsymbol{r}\:
r^{\lambda}Y_{\lambda\mu}(\boldsymbol{n}_{r})
\boldsymbol{\Lambda}\cdot[\Psi^{(0)\dag}(\boldsymbol{r})
\boldsymbol{\alpha}\Psi^{(1)}(\boldsymbol{r})].
\label{4.20}
\end{equation}
As $\Psi^{(1)}(\boldsymbol{r})$ is given by Eq.\ (\ref{2.28}), the
above equation may be rewritten in the form of the double integral
\begin{eqnarray}
\hspace*{-5em}
\widetilde{\mathcal{M}}_{\lambda\mu}^{(1)}
&=& \frac{\mathrm{i}e^{2}c}{\lambda+1}
\frac{4\pi}{\sqrt{(2\lambda+1)(2L+1)}}
\sum_{M=-L}^{L}\mathcal{C}_{LM}^{(1)*}
\nonumber \\
&& \times\int_{\mathbb{R}^{3}}\mathrm{d}^{3}\boldsymbol{r}
\int_{\mathbb{R}^{3}}\mathrm{d}^{3}\boldsymbol{r}'\:
r^{\lambda}Y_{\lambda\mu}(\boldsymbol{n}_{r})
\boldsymbol{\Lambda}\cdot[\Psi^{(0)\dag}(\boldsymbol{r})
\boldsymbol{\alpha}\bar{G}\mbox{}^{(0)}
(\boldsymbol{r},\boldsymbol{r}')]
r^{\prime L}Y_{LM}(\boldsymbol{n}_{r}^{\prime})
\Psi^{(0)}(\boldsymbol{r}').
\label{4.21}
\end{eqnarray}
A simplification occurs if one exploits the obvious identity
\begin{equation}
\boldsymbol{\Lambda}\cdot[\Psi^{(0)\dag}(\boldsymbol{r})
\boldsymbol{\alpha}\bar{G}^{(0)}(\boldsymbol{r},\boldsymbol{r}')]
=-[\boldsymbol{\alpha}\cdot\boldsymbol{\Lambda}
\Psi^{(0)}(\boldsymbol{r})]^{\dag}
\bar{G}^{(0)}(\boldsymbol{r},\boldsymbol{r}')
+\Psi^{(0)\dag}(\boldsymbol{r})
\boldsymbol{\alpha}\cdot\boldsymbol{\Lambda}
\bar{G}^{(0)}(\boldsymbol{r},\boldsymbol{r}'),
\label{4.22}
\end{equation}
the multipole expansion (\ref{3.13}) of the Green's function
$\bar{G}^{(0)}(\boldsymbol{r},\boldsymbol{r}')$ and the relation
\cite[Eq.\ (3.2.3)]{Szmy07}
\begin{equation}
\boldsymbol{\sigma}\cdot\boldsymbol{\Lambda}
\Omega_{\kappa m_{\kappa}}(\boldsymbol{n}_{r})
=-(\kappa+1)\Omega_{\kappa m_{\kappa}}(\boldsymbol{n}_{r}).
\label{4.23}
\end{equation}
This results in a separation of integrations over radial and angular
variables, and one obtains
\begin{eqnarray}
\widetilde{\mathcal{M}}_{\lambda\mu}^{(1)}
&=& -\frac{(4\pi\epsilon_{0})c}{\lambda+1}
\frac{4\pi}{\sqrt{(2\lambda+1)(2L+1)}}
\sum_{\substack{\kappa=-\infty \\ (\kappa\neq0)}}^{\infty}(\kappa-1)
R_{\kappa}^{(\lambda,L)}\big(Q^{(0)},P^{(0)};P^{(0)},Q^{(0)}\big)
\nonumber \\
&& \times\sum_{M=-L}^{L}\sum_{m_{\kappa}=-|\kappa|+1/2}^{|\kappa|-1/2}
\sum_{m=-1/2}^{1/2}\sum_{m'=-1/2}^{1/2}
a_{m}^{*}a_{m'}\mathcal{C}_{LM}^{(1)*}
\langle\Omega_{-1m}|Y_{\lambda\mu}\Omega_{-\kappa m_{\kappa}}\rangle
\langle\Omega_{\kappa m_{\kappa}}|Y_{LM}\Omega_{-1m'}\rangle,
\nonumber \\
&&
\label{4.24}
\end{eqnarray}
where
$R_{\kappa}^{(\lambda,L)}\big(Q^{(0)},P^{(0)};P^{(0)},Q^{(0)}\big)$
is a particular case of the double radial integral defined in Eq.\
(\ref{3.15}) and the bracket notation has been used to denote the
angular integrals [cf.\ Eq.\ (\ref{2.19})]. Evaluating the latter
with the aid of the formula in Eq.\ (\ref{2.22}) and carrying out the
summations, we find that $\widetilde{\mathcal{M}}_{\lambda\mu}^{(1)}$
does not vanish only if $\lambda=L\mp1$, i.e.,
\begin{equation}
\widetilde{\mathcal{M}}_{\lambda\mu}^{(1)}
=\widetilde{\mathcal{M}}_{\lambda\mu}^{(1)}
\delta_{\lambda,L-1}
+\widetilde{\mathcal{M}}_{\lambda\mu}^{(1)}
\delta_{\lambda,L+1}.
\label{4.25}
\end{equation}
In these two cases, one has
\begin{eqnarray}
\widetilde{\mathcal{M}}_{L-1,\mu}^{(1)}
&=& -(4\pi\epsilon_{0})c\,\frac{L-1}{L(4L^{2}-1)}
R_{L}^{(L-1,L)}\big(Q^{(0)},P^{(0)};P^{(0)},Q^{(0)}\big)
\nonumber \\
&& \times\,\bigg[-\sqrt{L^{2}-\mu^{2}}\,
(|a_{1/2}|^{2}-|a_{-1/2}|^{2})\mathcal{C}_{L\mu}^{(1)}
+\sqrt{(L+\mu)(L+\mu+1)}\,a_{1/2}a_{-1/2}^{*}
\mathcal{C}_{L,\mu+1}^{(1)}
\nonumber \\
&& \quad-\,\sqrt{(L-\mu)(L-\mu+1)}\,
a_{1/2}^{*}a_{-1/2}\mathcal{C}_{L,\mu-1}^{(1)}\bigg]
\label{4.26}
\end{eqnarray}
and
\begin{eqnarray}
\widetilde{\mathcal{M}}_{L+1,\mu}^{(1)}
&=& -(4\pi\epsilon_{0})c\,\frac{1}{(2L+1)(2L+3)}
R_{-L-1}^{(L+1,L)}\big(Q^{(0)},P^{(0)};P^{(0)},Q^{(0)}\big)
\nonumber \\
&& \times\,\bigg[\sqrt{(L+1)^{2}-\mu^{2}}\,
(|a_{1/2}|^{2}-|a_{-1/2}|^{2})\mathcal{C}_{L\mu}^{(1)}
+\sqrt{(L-\mu)(L-\mu+1)}\,a_{1/2}a_{-1/2}^{*}
\mathcal{C}_{L,\mu+1}^{(1)}
\nonumber \\
&& \quad -\,\sqrt{(L+\mu)(L+\mu+1)}
\,a_{1/2}^{*}a_{-1/2}\mathcal{C}_{L,\mu-1}^{(1)}\bigg],
\label{4.27}
\end{eqnarray}
respectively. On combining Eqs.\ (\ref{4.19}),
(\ref{4.25})--(\ref{4.27}) and (\ref{4.15}), we find that
\begin{equation}
\mathcal{M}_{\lambda\mu}^{(1)}
=\mathcal{M}_{\lambda\mu}^{(1)}\delta_{\lambda,L-1}
+\mathcal{M}_{\lambda\mu}^{(1)}\delta_{\lambda,L+1},
\label{4.28}
\end{equation}
where
\begin{eqnarray}
\hspace*{-5em}
\mathcal{M}_{L-1,\mu}^{(1)}
&=& -(4\pi\epsilon_{0})c\,\frac{2(L-1)}{L(4L^{2}-1)}
R_{L}^{(L-1,L)}\big(Q^{(0)},P^{(0)};P^{(0)},Q^{(0)}\big)
\bigg[-\sqrt{L^{2}-\mu^{2}}\,\nu_{0}\mathcal{C}_{L\mu}^{(1)}
\nonumber \\
&& +\,\sqrt{{\textstyle\frac{1}{2}}(L+\mu)(L+\mu+1)}\,
\nu_{-1}\mathcal{C}_{L,\mu+1}^{(1)}
+\sqrt{{\textstyle\frac{1}{2}}(L-\mu)(L-\mu+1)}\,
\nu_{1}\mathcal{C}_{L,\mu-1}^{(1)}\bigg]
\label{4.29}
\end{eqnarray}
and
\begin{eqnarray}
\mathcal{M}_{L+1,\mu}^{(1)}
&=& -(4\pi\epsilon_{0})c\,\frac{2}{(2L+1)(2L+3)}
R_{-L-1}^{(L+1,L)}\big(Q^{(0)},P^{(0)};P^{(0)},Q^{(0)}\big)
\bigg[\sqrt{(L+1)^{2}-\mu^{2}}\,\nu_{0}\mathcal{C}_{L\mu}^{(1)}
\nonumber \\
&& +\,\sqrt{{\textstyle\frac{1}{2}}(L-\mu)(L-\mu+1)}\,
\nu_{-1}\mathcal{C}_{L,\mu+1}^{(1)}
+\sqrt{{\textstyle\frac{1}{2}}(L+\mu)(L+\mu+1)}\,
\nu_{1}\mathcal{C}_{L,\mu-1}^{(1)}\bigg].
\label{4.30}
\end{eqnarray}
Thus, we see that, to the first-order of accuracy, the $2^{L}$-pole
electric field induces in the ground state of the atom \emph{two\/}
magnetic multipole moments, being the $2^{L-1}$-pole and the
$2^{L+1}$-pole ones, i.e.,
\begin{equation}
\boldsymbol{\mathsf{M}}_{\lambda}
\simeq\boldsymbol{\mathsf{M}}_{\lambda}^{(0)}\delta_{\lambda1}
+\boldsymbol{\mathsf{M}}_{\lambda}^{(1)}
(\delta_{\lambda,L-1}+\delta_{\lambda,L+1}).
\label{4.31}
\end{equation}
However, it is evident from Eq.\ (\ref{4.29}) that an exception
occurs in the case of the perturbing electric dipole ($L=1$) field,
when only the \emph{quadrupole\/} ($\lambda=2$) moment is induced
(cf.\ Ref.\ \cite{Szmy14}).

Consider now the irreducible spherical tensor product of rank
$\lambda$ of the vector $\boldsymbol{\nu}$, defined in Eq.\
(\ref{4.15}) and characterizing the unperturbed atomic state, and the
tensor $\boldsymbol{\mathsf{C}}_{L}^{(1)}$, characterizing the
perturbing multipole electric field. According to the general theory
of such products \cite[Sec.\ 3.1.7]{Vars75}, its components are given
by
\begin{equation}
\big\{\boldsymbol{\nu}
\otimes\boldsymbol{\mathsf{C}}_{L}^{(1)}\big\}_{\lambda\mu}
=\sum_{m=-1}^{1}\sum_{M=-L}^{L}
\langle1mLM|\lambda\mu\rangle\nu_{m}\mathcal{C}_{LM}^{(1)},
\label{4.32}
\end{equation}
where $\langle1mLM|\lambda\mu\rangle$ is a particular Clebsch--Gordan
coefficient. A look at a table of these coefficients (e.g., Ref.\
\cite[Table 8.2]{Vars75}) shows that the two induced magnetic moments
$\boldsymbol{\mathsf{M}}_{L\mp1}^{(1)}$, components of which are
displayed in Eqs.\ (\ref{4.29}) and (\ref{4.30}), may be compactly
written as
\begin{eqnarray}
\boldsymbol{\mathsf{M}}_{\lambda}^{(1)}
&=& -(4\pi\epsilon_{0})c\,\frac{2\sqrt{2}\,\lambda}
{(2\lambda+1)\sqrt{(2L+1)(\lambda+L+1)}}
R_{\kappa_{\lambda}}^{(\lambda,L)}
\big(Q^{(0)},P^{(0)};P^{(0)},Q^{(0)}\big)
\big\{\boldsymbol{\nu}
\otimes\boldsymbol{\mathsf{C}}_{L}^{(1)}\big\}_{\lambda}
\nonumber \\
&& \hspace*{25em} (\lambda=L\mp1),
\label{4.33}
\end{eqnarray}
with
\begin{equation}
\kappa_{\lambda}=-\frac{1}{2}(\lambda-L)(\lambda+L+1)
=\left\{
\begin{array}{ll}
L & \textrm{for $\lambda=L-1$} \\
-L-1 & \textrm{for $\lambda=L+1$}.
\end{array}
\right.
\label{4.34}
\end{equation}
\subsection{Atomic multipole E$L\to$M$(L\mp1)$
cross-susceptibilities} 
\label{IV.2}
We define the atomic electric-to-magnetic multipole
cross-susceptibilities $\alpha_{\mathrm{E}L\to\mathrm{M}(L\mp1)}$
through the relation
\begin{equation}
\boldsymbol{\mathsf{M}}_{\lambda}^{(1)}
=(4\pi\epsilon_{0})c\,\alpha_{\mathrm{E}L\to\mathrm{M}\lambda}
\frac{\big\{\boldsymbol{\nu}
\otimes\boldsymbol{\mathsf{C}}_{L}^{(1)}\big\}_{\lambda}}
{\langle10L0|\lambda0\rangle} 
\qquad (\lambda=L\mp1),
\label{4.35}
\end{equation}
where the Clebsch--Gordan coefficient standing in the denominator in
the fraction on the right-hand side is
\begin{equation}
\langle10L0|\lambda0\rangle
=(\lambda-L)\sqrt{\frac{\lambda+L+1}{2(2L+1)}}
\qquad (\lambda=L\mp1).
\label{4.36}
\end{equation}
Combining Eq.\ (\ref{4.35}) with the representation (\ref{4.33}) of
the tensor $\boldsymbol{\mathsf{M}}_{\lambda}^{(1)}$ links the
cross-susceptibility to the double radial integral appearing therein:
\begin{equation}
\alpha_{\mathrm{E}L\to\mathrm{M}\lambda} 
=-\frac{2\lambda(\lambda-L)}{(2\lambda+1)(2L+1)}
R_{\kappa_{\lambda}}^{(\lambda,L)}
\big(Q^{(0)},P^{(0)};P^{(0)},Q^{(0)}\big)
\qquad (\lambda=L\mp1).
\label{4.37}
\end{equation}
To tackle the latter, we exploit the Sturmian expansion (\ref{3.26}),
obtaining
\begin{eqnarray}
\alpha_{\mathrm{E}L\to\mathrm{M}\lambda} 
&=& -\frac{2\lambda(\lambda-L)}{(2\lambda+1)(2L+1)}
\sum_{n_{r}=-\infty}^{\infty}
\frac{1}{\mu_{n_{r}\kappa_{\lambda}}^{(0)}-1}
\int_{0}^{\infty}\mathrm{d}r\:r^{\lambda}
\left[Q^{(0)}(r)S_{n_{r}\kappa_{\lambda}}^{(0)}(r)
+P^{(0)}(r)T_{n_{r}\kappa_{\lambda}}^{(0)}(r)\right]
\nonumber \\
&& \times\int_{0}^{\infty}\mathrm{d}r'\:r^{\prime L}
\left[\mu_{n_{r}\kappa_{\lambda}}^{(0)}
P^{(0)}(r')S_{n_{r}\kappa_{\lambda}}^{(0)}(r')
+Q^{(0)}(r')T_{n_{r}\kappa_{\lambda}}^{(0)}(r')\right]
\qquad (\lambda=L\mp1).
\label{4.38}
\end{eqnarray}
The second of the two, now separated, radial integrals in Eq.\
(\ref{4.38}) is seen to be identical with the one we have evaluated
in Eq.\ (\ref{3.34}), while the first integral, with the aid of Eqs.\
(\ref{2.11}), (\ref{3.27}) and (\ref{3.31}), is found to be
\begin{eqnarray}
&& \hspace*{-5em} 
\int_{0}^{\infty}\mathrm{d}r\:r^{\lambda}
\left[Q^{(0)}(r)S_{n_{r}\kappa_{\lambda}}^{(0)}(r)
+P^{(0)}(r)T_{n_{r}\kappa_{\lambda}}^{(0)}(r)\right]
\nonumber \\
&=& -\alpha Z\left(\frac{a_{0}}{2Z}\right)^{\lambda+1}
\frac{\sqrt{2}\,(N_{n_{r}\kappa_{\lambda}}-\kappa_{\lambda})}
{\sqrt{a_{0}|n_{r}|!N_{n_{r}\kappa_{\lambda}}
(N_{n_{r}\kappa_{\lambda}}-\kappa_{\lambda})
\Gamma(2\gamma_{1}+1)\Gamma(|n_{r}|+2\gamma_{\kappa_{\lambda}}+1)}}
\nonumber \\
&& \times\,\frac{\Gamma(\gamma_{\kappa_{\lambda}}
+\gamma_{1}+\lambda+1)
\Gamma(|n_{r}|+\gamma_{\kappa_{\lambda}}-\gamma_{1}-\lambda)}
{\Gamma(\gamma_{\kappa_{\lambda}}-\gamma_{1}-\lambda)}.
\label{4.39}
\end{eqnarray}
Plugging Eqs.\ (\ref{4.39}) and (\ref{3.34}) into Eq.\ (\ref{4.38}),
then transforming the resulting series
$\sum_{n_{r}=-\infty}^{\infty}(\cdots)$ into a one of the sort
$\sum_{n_{r}=0}^{\infty}(\cdots)$, and identifying subsequently the
two ${}_{3}F_{2}(1)$ functions, yields the cross-susceptibility
$\alpha_{\mathrm{E}L\to\mathrm{M}\lambda}$ in the form
\begin{eqnarray}
\alpha_{\mathrm{E}L\to\mathrm{M}\lambda}
&=& \frac{\alpha a_{0}^{\lambda+L+1}}{Z^{\lambda+L+1}}
\frac{\lambda(\lambda-L)
\Gamma(\gamma_{\kappa_{\lambda}}+\gamma_{1}+\lambda+1)
\Gamma(\gamma_{\kappa_{\lambda}}+\gamma_{1}+L+1)}
{2^{\lambda+L}(2\lambda+1)
(2L+1)\Gamma(2\gamma_{1}+1)\Gamma(2\gamma_{\kappa_{\lambda}}+1)}
\nonumber \\
&& \times\,\bigg[\frac{\gamma_{1}(\lambda+1)}
{\gamma_{\kappa_{\lambda}}-\gamma_{1}+1}\,
{}_{3}F_{2}
\left(
\begin{array}{c}
\gamma_{\kappa_{\lambda}}-\gamma_{1}-\lambda,\:
\gamma_{\kappa_{\lambda}}-\gamma_{1}-L,\:
\gamma_{\kappa_{\lambda}}-\gamma_{1}+1 \\
\gamma_{\kappa_{\lambda}}-\gamma_{1}+2,\:
2\gamma_{\kappa_{\lambda}}+1 
\end{array}
;1
\right)
\nonumber \\
&& \quad -\,\frac{\gamma_{\kappa_{\lambda}}+\gamma_{1}}
{\kappa_{\lambda}+1}\,
{}_{3}F_{2}
\left(
\begin{array}{c}
\gamma_{\kappa_{\lambda}}-\gamma_{1}-\lambda,\:
\gamma_{\kappa_{\lambda}}-\gamma_{1}-L,\:
\gamma_{\kappa_{\lambda}}-\gamma_{1} \\
\gamma_{\kappa_{\lambda}}-\gamma_{1}+1,\:
2\gamma_{\kappa_{\lambda}}+1 
\end{array}
;1
\right)
\bigg]
\qquad (\lambda=L\mp1).
\nonumber \\
&&
\label{4.40}
\end{eqnarray}
Eliminating the second ${}_{3}F_{2}(1)$ in favor of the first one
with the help of Eq.\ (\ref{3.38}), we finally arrive at the
following general expression for the cross-susceptibility in
question:
\begin{eqnarray}
\hspace*{-2em}
\alpha_{\mathrm{E}L\to\mathrm{M}\lambda} 
&=& \frac{\alpha a_{0}^{\lambda+L+1}}{Z^{\lambda+L+1}}
\frac{\lambda(\lambda-L)\Gamma(2\gamma_{1}+\lambda+L+2)}
{2^{\lambda+L}(\kappa_{\lambda}+1)(2\lambda+1)(2L+1)
\Gamma(2\gamma_{1}+1)}
\nonumber \\
&& \times\,\bigg\{-1+\frac{(\lambda+1)
[\gamma_{1}(\kappa_{\lambda}+1)+L+1]
\Gamma(\gamma_{\kappa_{\lambda}}+\gamma_{1}+\lambda+1)
\Gamma(\gamma_{\kappa_{\lambda}}+\gamma_{1}+L+1)}
{(\gamma_{\kappa_{\lambda}}-\gamma_{1}+1)
\Gamma(2\gamma_{1}+\lambda+L+2)
\Gamma(2\gamma_{\kappa_{\lambda}}+1)}
\nonumber \\
&& \quad \times\,
{}_{3}F_{2}
\left(
\begin{array}{c}
\gamma_{\kappa_{\lambda}}-\gamma_{1}-\lambda,\:
\gamma_{\kappa_{\lambda}}-\gamma_{1}-L,\:
\gamma_{\kappa_{\lambda}}-\gamma_{1}+1 \\
\gamma_{\kappa_{\lambda}}-\gamma_{1}+2,\:
2\gamma_{\kappa_{\lambda}}+1 
\end{array}
;1
\right)
\bigg\}
\qquad (\lambda=L\mp1),
\nonumber \\
&&
\label{4.41}
\end{eqnarray}
where, we recall, $\kappa_{\lambda}$ has been defined in Eq.\
(\ref{4.34}). If in the above formula the explicit values of
$\lambda$ and $\kappa_{\lambda}$ are set, this gives explicitly
\begin{eqnarray}
\alpha_{\mathrm{E}L\to\mathrm{M}(L-1)}
&=& \frac{\alpha a_{0}^{2L}}{Z^{2L}}
\frac{(L-1)\Gamma(2\gamma_{1}+2L+1)}
{2^{2L-1}(L+1)(4L^{2}-1)\Gamma(2\gamma_{1}+1)}
\nonumber \\
&& \times\,\bigg[1-\frac{L(L+1)(\gamma_{1}+1)
\Gamma(\gamma_{L}+\gamma_{1}+L)\Gamma(\gamma_{L}+\gamma_{1}+L+1)}
{(\gamma_{L}-\gamma_{1}+1)\Gamma(2\gamma_{1}+2L+1)
\Gamma(2\gamma_{L}+1)}
\nonumber \\
&& \quad \times\,
{}_{3}F_{2}
\left(
\begin{array}{c}
\gamma_{L}-\gamma_{1}-L,\:
\gamma_{L}-\gamma_{1}-L+1,\:
\gamma_{L}-\gamma_{1}+1 \\
\gamma_{L}-\gamma_{1}+2,\:
2\gamma_{L}+1
\end{array}
;1
\right)
\bigg]
\label{4.42}
\end{eqnarray}
and
\begin{eqnarray}
\hspace*{-2em}
\alpha_{\mathrm{E}L\to\mathrm{M}(L+1)}
&=& \frac{\alpha a_{0}^{2L+2}}{Z^{2L+2}}
\frac{(L+1)\Gamma(2\gamma_{1}+2L+3)}
{2^{2L+1}L(2L+1)(2L+3)\Gamma(2\gamma_{1}+1)}
\nonumber \\
&& \times\,\bigg\{1+\frac{(L+2)(L\gamma_{1}-L-1)
\Gamma(\gamma_{L+1}+\gamma_{1}+L+1)
\Gamma(\gamma_{L+1}+\gamma_{1}+L+2)}
{(\gamma_{L+1}-\gamma_{1}+1)\Gamma(2\gamma_{1}+2L+3)
\Gamma(2\gamma_{L+1}+1)}
\nonumber \\
&& \quad \times\,
{}_{3}F_{2}
\left(
\begin{array}{c}
\gamma_{L+1}-\gamma_{1}-L-1,\:
\gamma_{L+1}-\gamma_{1}-L,\:
\gamma_{L+1}-\gamma_{1}+1 \\
\gamma_{L+1}-\gamma_{1}+2,\:
2\gamma_{L+1}+1
\end{array}
;1
\right)
\bigg\}.
\label{4.43}
\end{eqnarray}

In the particular case of the dipole ($L=1$) perturbing electric
field, the right-hand side of Eq.\ (\ref{4.42}) vanishes, while Eq.\
(\ref{4.43}) becomes
\begin{eqnarray}
\alpha_{\mathrm{E}1\to\mathrm{M}2}
&=& \frac{\alpha a_{0}^{4}}{Z^{4}}
\frac{\Gamma(2\gamma_{1}+5)}{60\Gamma(2\gamma_{1}+1)}
\bigg[1+\frac{3(\gamma_{1}-2)
\Gamma(\gamma_{2}+\gamma_{1}+2)
\Gamma(\gamma_{2}+\gamma_{1}+3)}
{(\gamma_{2}-\gamma_{1}+1)\Gamma(2\gamma_{1}+5)
\Gamma(2\gamma_{2}+1)}
\nonumber \\
&& \times\,
{}_{3}F_{2}
\left(
\begin{array}{c}
\gamma_{2}-\gamma_{1}-2,\:
\gamma_{2}-\gamma_{1}-1,\:
\gamma_{2}-\gamma_{1}+1 \\
\gamma_{2}-\gamma_{1}+2,\:
2\gamma_{2}+1 
\end{array}
;1
\right)
\bigg].
\label{4.44}
\end{eqnarray}
With the use of the relation (\ref{3.38}), the latter formula may be
transformed into the following one:
\begin{eqnarray}
\alpha_{\mathrm{E}1\to\mathrm{M}2}
&=& \frac{\alpha a_{0}^{4}}{Z^{4}}
\frac{\Gamma(2\gamma_{1}+5)}{240\Gamma(2\gamma_{1})}
\bigg[1-\frac{(\gamma_{1}-2)(\gamma_{2}+\gamma_{1})
\Gamma(\gamma_{2}+\gamma_{1}+2)
\Gamma(\gamma_{2}+\gamma_{1}+3)}
{\gamma_{1}\Gamma(2\gamma_{1}+5)\Gamma(2\gamma_{2}+1)}
\nonumber \\
&& \times\,
{}_{3}F_{2}
\left(
\begin{array}{c}
\gamma_{2}-\gamma_{1}-2,\:
\gamma_{2}-\gamma_{1}-1,\:
\gamma_{2}-\gamma_{1} \\
\gamma_{2}-\gamma_{1}+1,\:
2\gamma_{2}+1 
\end{array}
;1
\right)
\bigg],
\label{4.45}
\end{eqnarray}
which is identical with the one we have arrived at in Ref.\
\cite[Eq.\ (4.24)]{Szmy14}.

Numerical values of the cross-susceptibilities
$\alpha_{\mathrm{E}L\to\mathrm{M}(L\mp1)}$ with $1\leqslant
L\leqslant4$ for selected hydrogenic ions, computed from Eqs.\
(\ref{4.42}) and (\ref{4.43}), are presented in Tables \ref{T.4} and
\ref{T.5}.
\begin{center}
[Place for Tables \ref{T.4} and \ref{T.5}]
\end{center}

A derivation of quasi-relativistic approximations to the two
cross-susceptibilities $\alpha_{\mathrm{E}L\to\mathrm{M}(L\mp1)}$ is
very much analogous to the procedure we have adopted in Sec.\
\ref{III} for the polarizabilities $\alpha_{L}$. Exploiting the
relations (\ref{3.43}) and (\ref{3.44}), one deduces the estimates
\begin{equation}
{}_{3}F_{2}
\left(
\begin{array}{c}
\gamma_{L}-\gamma_{1}-L,\:
\gamma_{L}-\gamma_{1}-L+1,\:
\gamma_{L}-\gamma_{1}+1 \\
\gamma_{L}-\gamma_{1}+2,\:
2\gamma_{L}+1
\end{array}
;1
\right)
\simeq 1-(\alpha Z)^{2}\frac{L-1}{2(L+1)(2L+1)}
\label{4.46}
\end{equation}
and
\begin{equation}
{}_{3}F_{2}
\left(
\begin{array}{c}
\gamma_{L+1}-\gamma_{1}-L-1,\:
\gamma_{L+1}-\gamma_{1}-L,\:
\gamma_{L+1}-\gamma_{1}+1 \\
\gamma_{L+1}-\gamma_{1}+2,\:
2\gamma_{L+1}+1
\end{array}
;1
\right)
\simeq 1-(\alpha Z)^{2}\frac{L}{2(L+2)(2L+3)}.
\label{4.47}
\end{equation}
When the approximation (\ref{4.46}) is inserted into Eq.\
(\ref{4.42}), after some play with the recurrence relation
(\ref{3.50}) one finds that
\begin{equation}
\alpha_{\mathrm{E}L\to\mathrm{M}(L-1)}
\simeq\frac{\alpha a_{0}^{2L}}{Z^{2L}}
(\alpha Z)^{2}\frac{(L-1)^{2}(2L^{3}+5L^{2}+4L+2)(2L-2)!}
{2^{2L}L(L+1)(2L+1)}.
\label{4.48}
\end{equation}
Similarly, combining Eqs.\ (\ref{4.43}) and (\ref{4.47}) yields
\begin{eqnarray}
\alpha_{\mathrm{E}L\to\mathrm{M}(L+1)}
&\simeq& \frac{\alpha a_{0}^{2L+2}}{Z^{2L+2}}
\frac{(L+2)(2L+2)!}{2^{2L+2}L}
\nonumber \\
&& \times\,\bigg\{1-(\alpha Z)^{2}\bigg[\psi(2L+4)-\psi(2)
-\frac{L(2L^{4}+9L^{3}+17L^{2}+17L+8)}
{2(L+1)^{2}(L+2)(2L+1)(2L+3)}\bigg]\bigg\}.
\nonumber \\
&&
\label{4.49}
\end{eqnarray}
It is worthwhile to notice that irrespective of the value assumed by
$L$, the cross-susceptibility
$\alpha_{\mathrm{E}L\to\mathrm{M}(L-1)}$ (hence, also the induced
moment $\boldsymbol{\mathsf{M}}_{L-1}^{(1)}$) vanishes as $\alpha
Z\to0$, while in the same limit both
$\alpha_{\mathrm{E}L\to\mathrm{M}(L+1)}$ and
$\boldsymbol{\mathsf{M}}_{L+1}^{(1)}$ remain finite. 

Explicit forms of the expressions standing on the right-hand sides of
Eqs.\ (\ref{4.48}) and (\ref{4.49}), with $L$ restricted to the range
$1\leqslant L\leqslant4$, are collected in Table \ref{T.6}.
\begin{center}
[Place for Table \ref{T.6}]
\end{center}
%
%\newpage
%
\section{Magnetic toroidal multipole moments of the atom in the
multipole electric field and atomic E$L\to$T$L$ multipole
cross-sus\-ce\-pti\-bi\-li\-ties} 
\label{V} 
\setcounter{equation}{0}
\subsection{Decomposition of the atomic magnetic toroidal multipole
moments into the permanent and the first-order electric-field-induced
components}
\label{V.1}
The last member of the family of far-field atomic multipole moments
we wish to consider in this work are the magnetic toroidal moments.
We show in Appendix \ref{C} that spherical components of the
$2^{\lambda}$-pole magnetic toroidal moment
$\boldsymbol{\mathsf{T}}_{\lambda}$ due to a solenoidal current
density $\boldsymbol{j}(\boldsymbol{r})$ may be written in several
equivalent forms, out of which for the use in this section we choose
the following one:
\begin{equation}
\mathcal{T}_{\lambda\mu}=\frac{1}{\lambda+1}
\sqrt{\frac{4\pi}{2\lambda+1}}
\int_{\mathbb{R}^{3}}\mathrm{d}^{3}\boldsymbol{r}\:
r^{\lambda}Y_{\lambda\mu}(\boldsymbol{n}_{r})
\boldsymbol{r}\cdot\boldsymbol{j}(\boldsymbol{r}).
\label{5.1}
\end{equation}
Proceeding along the route sketched in Sec.\ \ref{IV}, after the
current is approximated as in Eq.\ (\ref{4.6}), we obtain
\begin{equation}
\mathcal{T}_{\lambda\mu}\simeq \mathcal{T}_{\lambda\mu}^{(0)}
+\mathcal{T}_{\lambda\mu}^{(1)},
\label{5.2}
\end{equation}
where
\begin{equation}
\mathcal{T}_{\lambda\mu}^{(0)}
=\frac{1}{\lambda+1}\sqrt{\frac{4\pi}{2\lambda+1}}
\int_{\mathbb{R}^{3}}\mathrm{d}^{3}\boldsymbol{r}\:
r^{\lambda}Y_{\lambda\mu}(\boldsymbol{n}_{r})
\boldsymbol{r}\cdot\boldsymbol{j}^{(0)}(\boldsymbol{r})
\label{5.3}
\end{equation}
and
\begin{equation}
\mathcal{T}_{\lambda\mu}^{(1)}
=\frac{1}{\lambda+1}\sqrt{\frac{4\pi}{2\lambda+1}}
\int_{\mathbb{R}^{3}}\mathrm{d}^{3}\boldsymbol{r}\:
r^{\lambda}Y_{\lambda\mu}(\boldsymbol{n}_{r})
\boldsymbol{r}\cdot\boldsymbol{j}^{(1)}(\boldsymbol{r}),
\label{5.4}
\end{equation}
with $\boldsymbol{j}^{(0)}(\boldsymbol{r})$ and
$\boldsymbol{j}^{(1)}(\boldsymbol{r})$ given, respectively, by Eqs.\
(\ref{4.7}) and (\ref{4.8}).

At first, we shall show that in the ground state of an isolated atom
all permanent toroidal multipole moments do vanish. To this end, we
insert the expression (\ref{4.7}) for
$\boldsymbol{j}^{(0)}(\boldsymbol{r})$ into the right-hand side of
Eq.\ (\ref{5.3}) and then make use of Eqs.\ (\ref{2.9}) and
(\ref{2.10}), together with the identity \cite[Eq.\ (3.1.3)]{Szmy07}
\begin{equation}
\boldsymbol{n}_{r}\cdot\boldsymbol{\sigma}
\Omega_{\kappa m_{\kappa}}(\boldsymbol{n}_{r})
=-\Omega_{-\kappa m_{\kappa}}(\boldsymbol{n}_{r}),
\label{5.5}
\end{equation}
obtaining
\begin{eqnarray}
\mathcal{T}_{\lambda\mu}^{(0)}
&=& \frac{\mathrm{i}ec}{\lambda+1}\sqrt{\frac{4\pi}{2\lambda+1}}
\int_{0}^{\infty}\mathrm{d}r\:r^{\lambda+1}P^{(0)}(r)Q^{(0)}(r)
\nonumber \\
&& \times\sum_{m=-1/2}^{1/2}\sum_{m'=-1/2}^{1/2}a_{m}^{*}a_{m'}
\left[\langle\Omega_{-1m}\big|Y_{\lambda\mu}\Omega_{-1m'}\rangle
-\langle\Omega_{1m}\big|Y_{\lambda\mu}\Omega_{1m'}\rangle\right].
\label{5.6}
\end{eqnarray}
It follows from the property displayed in Eq.\ (\ref{2.20}) [being,
in fact, the consequence of the identity (\ref{5.5})] that for all
$\lambda$, $\mu$, $m$ and $m'$ the expression in the square bracket
on the right-hand side of the above equation is zero. Hence, we
arrive at the aforementioned result
\begin{equation}
\mathcal{T}_{\lambda\mu}^{(0)}=0.
\label{5.7}
\end{equation}

Next, we turn our attention to the first-order induced moments
$\boldsymbol{\mathsf{T}}_{\lambda}^{(1)}$. From Eqs.\ (\ref{5.4}),
(\ref{4.8}) and (\ref{3.10}), we have
\begin{equation}
\mathcal{T}_{\lambda\mu}^{(1)}
=\widetilde{\mathcal{T}}_{\lambda\mu}^{(1)}
+(-)^{\mu}\widetilde{\mathcal{T}}_{\lambda,-\mu}^{(1)*},
\label{5.8}
\end{equation}
with
\begin{equation}
\widetilde{\mathcal{T}}_{\lambda\mu}^{(1)}=-\frac{ec}{\lambda+1}
\sqrt{\frac{4\pi}{2\lambda+1}}
\int_{\mathbb{R}^{3}}\mathrm{d}^{3}\boldsymbol{r}\:r^{\lambda+1}
Y_{\lambda\mu}(\boldsymbol{n}_{r})\Psi^{(0)\dag}(\boldsymbol{r})
\boldsymbol{n}_{r}\cdot\boldsymbol{\alpha}\Psi^{(1)}(\boldsymbol{r}).
\label{5.9}
\end{equation}
Skipping details that should be already obvious to the reader who has
gone through Secs.\ \ref{III} and \ref{IV}, we come to the inference
that Eq.\ (\ref{5.9}) may be converted into
\begin{eqnarray}
\widetilde{\mathcal{T}}_{\lambda\mu}^{(1)}
&=& \frac{\mathrm{i}(4\pi\epsilon_{0})c}{\lambda+1}
\frac{4\pi}{\sqrt{(2\lambda+1)(2L+1)}}
\sum_{\substack{\kappa=-\infty \\ (\kappa\neq0)}}^{\infty}
R_{\kappa}^{(\lambda+1,L)}\big(Q^{(0)},-P^{(0)};P^{(0)},Q^{(0)}\big)
\nonumber \\
&& \times\sum_{M=-L}^{L}\sum_{m_{\kappa}=-|\kappa|+1/2}^{|\kappa|-1/2}
\sum_{m=-1/2}^{1/2}\sum_{m'=-1/2}^{1/2}
a_{m}^{*}a_{m'}\mathcal{C}_{LM}^{(1)*}
\langle\Omega_{-1m}|Y_{\lambda\mu}\Omega_{\kappa m_{\kappa}}\rangle
\langle\Omega_{\kappa m_{\kappa}}|Y_{LM}\Omega_{-1m'}\rangle.
\nonumber \\
&&
\label{5.10}
\end{eqnarray}
Comparison of the multiple sum on the right-hand side of Eq.\
(\ref{5.10}) with the one that appears in Eq.\ (\ref{3.14}), shows
that the two are identical. This allows us to save labor, and after
exploiting results of Sec.\ \ref{III.1}, we quickly find that
\begin{equation}
\mathcal{T}_{\lambda\mu}^{(1)}=\mathcal{T}_{\lambda\mu}^{(1)}
\delta_{\lambda L},
\label{5.11}
\end{equation}
where
\begin{equation}
\mathcal{T}_{L\mu}^{(1)}=\mathcal{T}_{L\mu,L}^{(1)}
+\mathcal{T}_{L\mu,-L-1}^{(1)},
\label{5.12}
\end{equation}
with
\begin{eqnarray}
\mathcal{T}_{L\mu,\kappa}^{(1)} &=& (4\pi\epsilon_{0})c\,
\frac{2\mathrm{i}\sgn(\kappa)}{(L+1)(2L+1)^{2}}
R_{\kappa}^{(L+1,L)}\big(Q^{(0)},-P^{(0)};P^{(0)},Q^{(0)}\big)
\bigg[\mu(|a_{1/2}|^{2}-|a_{-1/2}|^{2})\mathcal{C}_{L\mu}^{(1)}
\nonumber \\
&& +\,\sqrt{(L-\mu)(L+\mu+1)}\,a_{1/2}a_{-1/2}^{*}
\mathcal{C}_{L,\mu+1}^{(1)}
+\sqrt{(L+\mu)(L-\mu+1)}\,a_{1/2}^{*}a_{-1/2}
\mathcal{C}_{L,\mu-1}^{(1)}\bigg]
\nonumber \\
&& \hspace*{25em} (\kappa=L,-L-1).
\label{5.13}
\end{eqnarray}
In Sec.\ \ref{IV.1}, we have succeeded to express the two induced
magnetic moments $\boldsymbol{\mathsf{M}}_{L\mp1}^{(1)}$ in terms of
certain irreducible tensor products of the vector (rank-1 tensor)
$\boldsymbol{\nu}$, introduced in Eq.\ (\ref{4.15}), and the
$2^{L}$-pole tensor $\boldsymbol{\mathsf{C}}_{L}^{(1)}$,
characterizing the perturbing electrostatic field. An analogous
simplification is possible in the case of the toroidal moments
considered here. Invoking Eq.\ (\ref{4.15}) and rewriting the
right-hand side of Eq.\ (\ref{5.13}) in terms of the spherical
components of the vector $\boldsymbol{\nu}$, after making use of a
table of the Clebsch--Gordan coefficients \cite[Table 8.2]{Vars75},
we arrive at
\begin{eqnarray}
\mathcal{T}_{L\mu,\kappa}^{(1)} &=& -(4\pi\epsilon_{0})c\,
\frac{2\mathrm{i}\sgn(\kappa)}{(2L+1)^{2}}\sqrt{\frac{L}{L+1}}\,
R_{\kappa}^{(L+1,L)}\big(Q^{(0)},-P^{(0)};P^{(0)},Q^{(0)}\big)
\big\{\boldsymbol{\nu}
\otimes\boldsymbol{\mathsf{C}}_{L}^{(1)}\big\}_{L\mu}
\nonumber \\
&& \hspace*{20em} (\kappa=L,-L-1).
\label{5.14}
\end{eqnarray}

In summary, in this section we have shown that in the ground state of
an isolated atom all magnetic toroidal multipole moments due to
electronic movement vanish, while, to the first order of
approximation, a perturbing $2^{L}$-pole static electric field
induces the toroidal moment of the same multipolar symmetry as that
of the perturbing field, i.e.,
\begin{equation}
\boldsymbol{\mathsf{T}}_{\lambda}
\simeq\boldsymbol{\mathsf{T}}_{\lambda}^{(1)}\delta_{\lambda L}.
\label{5.15}
\end{equation}
\subsection{Atomic multipole E$L\to$T$L$ cross-susceptibilities}
\label{V.2}
We define the atomic multipole electric-to-magnetic toroidal
cross-susceptibilities $\alpha_{\mathrm{E}L\to\mathrm{T}L}$ according
to
\begin{equation}
\boldsymbol{\mathsf{T}}_{L}^{(1)}=\mathrm{i}(4\pi\epsilon_{0})c\,
\alpha_{\mathrm{E}L\to\mathrm{T}L}
\sqrt{L(L+1)}\,\big\{\boldsymbol{\nu}
\otimes\boldsymbol{\mathsf{C}}_{L}^{(1)}\big\}_{L}.
\label{5.16}
\end{equation}
Comparison of Eq.\ (\ref{5.16}) with Eqs.\ (\ref{5.12}) and
(\ref{5.14}) gives $\alpha_{\mathrm{E}L\to\mathrm{T}L}$ in the form
of the sum
\begin{equation}
\alpha_{\mathrm{E}L\to\mathrm{T}L}
=\alpha_{\mathrm{E}L\to\mathrm{T}L,L}
+\alpha_{\mathrm{E}L\to\mathrm{T}L,-L-1},
\label{5.17}
\end{equation}
with the two addends given by
\begin{equation}
\alpha_{\mathrm{E}L\to\mathrm{T}L,\kappa}
=-\frac{2\sgn(\kappa)}{(L+1)(2L+1)^{2}}
R_{\kappa}^{(L+1,L)}\big(Q^{(0)},-P^{(0)};P^{(0)},Q^{(0)}\big)
\qquad (\kappa=L,-L-1)
\label{5.18}
\end{equation}
or, by virtue of the definition (\ref{3.15}) and the expansion
(\ref{3.26}), by
\begin{eqnarray}
\alpha_{\mathrm{E}L\to\mathrm{T}L,\kappa}
&=& -\frac{2\sgn(\kappa)}{(L+1)(2L+1)^{2}}
\nonumber \\
&& \times\sum_{n_{r}=-\infty}^{\infty}
\frac{1}{\mu_{n_{r}\kappa}^{(0)}-1}
\int_{0}^{\infty}\mathrm{d}r\:r^{L+1}
\left[Q^{(0)}(r)S_{n_{r}\kappa}^{(0)}(r)
-P^{(0)}(r)T_{n_{r}\kappa}^{(0)}(r)\right]
\nonumber \\
&& \quad \times\int_{0}^{\infty}\mathrm{d}r'\:r^{\prime L}
\left[\mu_{n_{r}\kappa}^{(0)}P^{(0)}(r')S_{n_{r}\kappa}^{(0)}(r')
+Q^{(0)}(r')T_{n_{r}\kappa}^{(0)}(r')\right]
\qquad (\kappa=L,-L-1).
\nonumber \\
&&
\label{5.19}
\end{eqnarray}
The integral over $r'$ is seen to be identical to the one evaluated
in Eq.\ (\ref{3.34}), while the one over $r$, after the use is made
of Eqs.\ (\ref{3.27}) and (\ref{3.31}), is found to be
\begin{eqnarray}
&& \hspace*{-5em} 
\int_{0}^{\infty}\mathrm{d}r\:r^{L+1}
\left[Q^{(0)}(r)S_{n_{r}\kappa}^{(0)}(r) 
-P^{(0)}(r)T_{n_{r}\kappa}^{(0)}(r)\right]
\nonumber \\
&=& \alpha Z\left(\frac{a_{0}}{2Z}\right)^{L+2}
\frac{\sqrt{2}\,|n_{r}|(|n_{r}|+2\gamma_{\kappa})}
{\sqrt{a_{0}|n_{r}|!N_{n_{r}\kappa}(N_{n_{r}\kappa}-\kappa)
\Gamma(2\gamma_{1}+1)\Gamma(|n_{r}|+2\gamma_{\kappa}+1)}}
\nonumber \\
&& \times\,\frac{\Gamma(\gamma_{\kappa}+\gamma_{1}+L+2)
\Gamma(|n_{r}|+\gamma_{\kappa}-\gamma_{1}-L-2)}
{\Gamma(\gamma_{\kappa}-\gamma_{1}-L-1)}.
\label{5.20}
\end{eqnarray}
Inserting Eqs.\ (\ref{3.34}) and (\ref{5.20}) into the right-hand
side of Eq.\ (\ref{5.19}) and summing the resulting series, with the
same procedure we have applied before in Secs.\ \ref{III.2} and
\ref{IV.2}, to a form involving a particular ${}_{3}F_{2}(1)$
function, yields the following general expression for
$\alpha_{\mathrm{E}L\to\mathrm{T}L,\kappa}$:
\begin{eqnarray}
\alpha_{\mathrm{E}L\to\mathrm{T}L,\kappa}
&=& \frac{\alpha a_{0}^{2L+2}}{Z^{2L+2}}
\frac{\sgn(\kappa)[\gamma_{1}(\kappa+1)+L+1]
\Gamma(\gamma_{\kappa}+\gamma_{1}+L+1)
\Gamma(\gamma_{\kappa}+\gamma_{1}+L+2)}
{2^{2L+1}(L+1)(2L+1)^{2}(\gamma_{\kappa}-\gamma_{1}+1)
\Gamma(2\gamma_{1}+1)\Gamma(2\gamma_{\kappa}+1)}
\nonumber \\
&& \times\,{}_{3}F_{2}
\left(
\begin{array}{c}
\gamma_{\kappa}-\gamma_{1}-L-1,\:
\gamma_{\kappa}-\gamma_{1}-L,\:
\gamma_{\kappa}-\gamma_{1}+1 \\
\gamma_{\kappa}-\gamma_{1}+2,\:
2\gamma_{\kappa}+1
\end{array}
;1
\right)
\qquad (\kappa=L,-L-1).
\nonumber \\
&&
\label{5.21}
\end{eqnarray}
Hence, we find that the cross-susceptibility
$\alpha_{\mathrm{E}L\to\mathrm{T}L}$ is the sum of
\begin{subequations}
\begin{eqnarray}
\alpha_{\mathrm{E}L\to\mathrm{T}L,L}
&=& \frac{\alpha a_{0}^{2L+2}}{Z^{2L+2}}
\frac{(\gamma_{1}+1)\Gamma(\gamma_{L}+\gamma_{1}+L+1)
\Gamma(\gamma_{L}+\gamma_{1}+L+2)}
{2^{2L+1}(2L+1)^{2}(\gamma_{L}-\gamma_{1}+1)
\Gamma(2\gamma_{1}+1)\Gamma(2\gamma_{L}+1)}
\nonumber \\
&& \times\,{}_{3}F_{2}
\left(
\begin{array}{c}
\gamma_{L}-\gamma_{1}-L-1,\:
\gamma_{L}-\gamma_{1}-L,\:
\gamma_{L}-\gamma_{1}+1 \\
\gamma_{L}-\gamma_{1}+2,\:
2\gamma_{L}+1
\end{array}
;1
\right)
\label{5.22a}
\end{eqnarray}
and
\begin{eqnarray}
\hspace*{-2em}
\alpha_{\mathrm{E}L\to\mathrm{T}L,-L-1}
&=& \frac{\alpha a_{0}^{2L+2}}{Z^{2L+2}}
\frac{(L\gamma_{1}-L-1)\Gamma(\gamma_{L+1}+\gamma_{1}+L+1)
\Gamma(\gamma_{L+1}+\gamma_{1}+L+2)}
{2^{2L+1}(L+1)(2L+1)^{2}(\gamma_{L+1}-\gamma_{1}+1)
\Gamma(2\gamma_{1}+1)\Gamma(2\gamma_{L+1}+1)}
\nonumber \\
&& \times\,{}_{3}F_{2}
\left(
\begin{array}{c}
\gamma_{L+1}-\gamma_{1}-L-1,\:
\gamma_{L+1}-\gamma_{1}-L,\:
\gamma_{L+1}-\gamma_{1}+1 \\
\gamma_{L+1}-\gamma_{1}+2,\:
2\gamma_{L+1}+1
\end{array}
;1
\right),
\label{5.22b}
\end{eqnarray}
\label{5.22}%
\end{subequations}
i.e., one has
\begin{eqnarray}
\alpha_{\mathrm{E}L\to\mathrm{T}L} 
&=& \frac{\alpha a_{0}^{2L+2}}{Z^{2L+2}}
\frac{1}{2^{2L+1}(2L+1)^{2}\Gamma(2\gamma_{1}+1)}
\bigg\{\frac{(\gamma_{1}+1)\Gamma(\gamma_{L}+\gamma_{1}+L+1)
\Gamma(\gamma_{L}+\gamma_{1}+L+2)}
{(\gamma_{L}-\gamma_{1}+1)\Gamma(2\gamma_{L}+1)}
\nonumber \\
&& \quad \times\,{}_{3}F_{2}
\left(
\begin{array}{c}
\gamma_{L}-\gamma_{1}-L-1,\:
\gamma_{L}-\gamma_{1}-L,\:
\gamma_{L}-\gamma_{1}+1 \\
\gamma_{L}-\gamma_{1}+2,\:
2\gamma_{L}+1
\end{array}
;1
\right)
\nonumber \\
&& +\,\frac{(L\gamma_{1}-L-1)\Gamma(\gamma_{L+1}+\gamma_{1}+L+1)
\Gamma(\gamma_{L+1}+\gamma_{1}+L+2)}
{(L+1)(\gamma_{L+1}-\gamma_{1}+1)\Gamma(2\gamma_{L+1}+1)}
\nonumber \\
&& \quad \times\,{}_{3}F_{2}
\left(
\begin{array}{c}
\gamma_{L+1}-\gamma_{1}-L-1,\:
\gamma_{L+1}-\gamma_{1}-L,\:
\gamma_{L+1}-\gamma_{1}+1 \\
\gamma_{L+1}-\gamma_{1}+2,\:
2\gamma_{L+1}+1
\end{array}
;1
\right)
\bigg\}.
\label{5.23}
\end{eqnarray}
Tabulation of numerical values of the cross-susceptibilities
$\alpha_{\mathrm{E}L\to\mathrm{T}L}$ with $1\leqslant L\leqslant4$
for selected hydrogenic ions is done in Table \ref{T.7}.
\begin{center}
[Place for Table \ref{T.7}]
\end{center}

In the particular case when the perturbing electric field is of
dipolar ($L=1$) character, the first ${}_{3}F_{2}(1)$ series on the
right-hand side of Eq.\ (\ref{5.23}) is a terminating one and
$\alpha_{\mathrm{E}1\to\mathrm{T}1}$ appears to have the relatively
simple form
\begin{eqnarray}
\alpha_{\mathrm{E}1\to\mathrm{T}1}
&=& \frac{\alpha a_{0}^{4}}{Z^{4}}
\bigg[\frac{(\gamma_{1}+1)^{3}(2\gamma_{1}+1)}{18}
+\frac{(\gamma_{1}-2)\Gamma(\gamma_{2}+\gamma_{1}+2)
\Gamma(\gamma_{2}+\gamma_{1}+3)}{144(\gamma_{2}-\gamma_{1}+1)
\Gamma(2\gamma_{1}+1)\Gamma(2\gamma_{2}+1)}
\nonumber \\
&& \times\,{}_{3}F_{2}
\left(
\begin{array}{c}
\gamma_{2}-\gamma_{1}-2,\:
\gamma_{2}-\gamma_{1}-1,\:
\gamma_{2}-\gamma_{1}+1 \\
\gamma_{2}-\gamma_{1}+2,\:
2\gamma_{2}+1
\end{array}
;1
\right)\bigg].
\label{5.24}
\end{eqnarray}
The dipolar case was studied by our group a decade ago in Ref.\
\cite{Miel06}, where the following representation of
$\alpha_{\mathrm{E}1\to\mathrm{T}1}$ was derived\footnote{~Equation
(\ref{5.25}) follows from Eq.\ (4.32) in Ref.\ \cite{Miel06}, where
the expression for $\tau=c\alpha_{\mathrm{E}1\to\mathrm{T}1}$ was
presented. The reader might be surprised that setting $L=1$ in Eqs.\
(\ref{5.22a}) and (\ref{5.22b}), and multiplying the results by the
speed of light, one does not reproduce Eqs.\ (4.24) and (4.31) in
Ref.\ \cite{Miel06}. The origin of this apparent paradox is that
splitting the components $\mathcal{T}_{L\mu}^{(1)}$ of the induced
toroidal moment into a sum of two $\kappa$-dependent addends is not
(and, in fact, need not to be) unique, and depends on a particular
integral representations of $\mathcal{T}_{\lambda\mu}$ chosen as a
starting point (in this connection, see the discussion in Appendix
\ref{C}). However, the sum of the two addends is, of course, always
the same.}:
\begin{eqnarray}
\alpha_{\mathrm{E}1\to\mathrm{T1}} &=& \frac{\alpha a_{0}^{4}}{Z^{4}}
\bigg[\frac{(\gamma_{1}+1)(2\gamma_{1}+1)
(8\gamma_{1}^{3}+54\gamma_{1}^{2}+67\gamma_{1}+18)}{864}
\nonumber \\
&& -\,\frac{(\gamma_{1}-2)(4\gamma_{1}+1)
\Gamma^{2}(\gamma_{2}+\gamma_{1}+2)}{576(\gamma_{2}-\gamma_{1})
\Gamma(2\gamma_{1}+1)\Gamma(2\gamma_{2}+1)}
%\nonumber \\
%&& \quad \times\,
{}_{3}F_{2}
\left(
\begin{array}{c}
\gamma_{2}-\gamma_{1}-1,\:
\gamma_{2}-\gamma_{1}-1,\:
\gamma_{2}-\gamma_{1} \\
\gamma_{2}-\gamma_{1}+1,\:
2\gamma_{2}+1
\end{array}
;1
\right)
\bigg].
\nonumber \\
&&
\label{5.25}
\end{eqnarray}
Equivalence of the expressions in Eqs.\ (\ref{5.24}) and (\ref{5.25})
may be proved with the aid of the identity
\begin{eqnarray}
{}_{3}F_{2}
\left(
\begin{array}{c}
a_{1},\:
a_{2},\:
a_{3} \\
a_{3}+1,\:
b
\end{array}
;1
\right)
&=& \left[1-\frac{(a_{1}-a_{3}-1)(a_{2}-a_{3}-1)}
{(a_{1}-1)(b-a_{1})}\right]\frac{\Gamma(b)\Gamma(b-a_{1}-a_{2}+1)}
{(b-a_{3}-1)\Gamma(b-a_{1})\Gamma(b-a_{2})}
\nonumber \\
&& -\,\frac{(a_{1}-a_{3}-1)(a_{1}-a_{3}-2)(a_{2}-a_{3}-1)}
{(a_{1}-1)(a_{3}+1)(b-a_{3}-1)}
\nonumber \\
&& \quad\times\,
{}_{3}F_{2}
\left(
\begin{array}{c}
a_{1}-1,\:
a_{2},\:
a_{3}+1 \\
a_{3}+2,\:
b
\end{array}
;1
\right)
\qquad
\Real(b-a_{1}-a_{2})>1.
\label{5.26}
\end{eqnarray}

To find a quasi-relativistic limit of the general expression for the
cross-susceptibilities under study, we approximate the hypergeometric
function appearing in Eq.\ (\ref{5.22a}) with the formula
\begin{eqnarray}
&& \hspace*{-5em}
{}_{3}F_{2}
\left(
\begin{array}{c}
\gamma_{L}-\gamma_{1}-L-1,\:
\gamma_{L}-\gamma_{1}-L,\:
\gamma_{L}-\gamma_{1}+1 \\
\gamma_{L}-\gamma_{1}+2,\:
2\gamma_{L}+1
\end{array}
;1
\right)
\nonumber \\
&\simeq& 
\frac{2L^{2}+5L+1}{(L+1)(2L+1)}
-(\alpha Z)^{2}\frac{12L^{5}+32L^{4}-15L^{3}-68L^{2}-17L+8}
{4L(L+1)^{2}(L+2)(2L+1)^{2}}
\label{5.27}
\end{eqnarray}
and the one in Eq.\ (\ref{5.22b}) using Eq.\ (\ref{4.47}). This yields
the following estimations of the two addends in Eq.\ (\ref{5.17}):
\begin{subequations}
\begin{eqnarray}
\alpha_{\mathrm{E}L\to\mathrm{T}L,L}
&\simeq& \frac{\alpha a_{0}^{2L+2}}{Z^{2L+2}}
\frac{(2L^{2}+5L+1)(2L)!}{2^{2L}L(2L+1)}
\nonumber \\
&& \times\,\left\{1-(\alpha Z)^{2}\left[\psi(2L+2)-\psi(2)
-\frac{2L^{6}-L^{5}-26L^{4}-31L^{3}+4L+4}
{4L^{2}(L+1)(L+2)(2L^{2}+5L+1)}\right]\right\},
\nonumber \\
&&
\label{5.28a}
\end{eqnarray}
\begin{eqnarray}
\alpha_{\mathrm{E}L\to\mathrm{T}L,-L-1}
&\simeq& -\frac{\alpha a_{0}^{2L+2}}{Z^{2L+2}}
\frac{(2L+3)(2L)!}{2^{2L+1}(L+1)(2L+1)}
\nonumber \\
&& \times\,\left\{1-(\alpha Z)^{2}
\left[\psi(2L+2)-\psi(2)
-\frac{2L^{5}+13L^{4}+27L^{3}+18L^{2}-3L-4}
{2(L+1)^{2}(L+2)(2L+3)}\right]\right\}.
\nonumber \\
&&
\label{5.28b}
\end{eqnarray}
\label{5.28}%
\end{subequations}
Adding the right-hand sides of Eqs.\ (\ref{5.28a}) and (\ref{5.28b}),
after some algebra we arrive at the sought quasi-relativistic
representation of $\alpha_{\mathrm{E}L\to\mathrm{T}L}$:
\begin{eqnarray}
\alpha_{\mathrm{E}L\to\mathrm{T}L}
&\simeq& \frac{\alpha a_{0}^{2L+2}}{Z^{2L+2}}
\frac{(L+2)(2L+1)!}{2^{2L+1}L(L+1)}
\bigg\{1-(\alpha Z)^{2}\bigg[\psi(2L+3)-\psi(2)
\nonumber \\
&& +\,\frac{6L^{7}+29L^{6}+49L^{5}+32L^{4}-L^{3}-16L^{2}-12L-4}
{2L^{2}(L+1)^{2}(L+2)^{2}(2L+1)^{2}}\bigg]\bigg\}.
\label{5.29}
\end{eqnarray}
Table \ref{T.8} collects the quasi-relativistic approximations of
$\alpha_{\mathrm{E}L\to\mathrm{T}L}$ obtained from Eq.\ (\ref{5.29})
for $1\leqslant L\leqslant4$.
\begin{center}
[Place for Table \ref{T.8}]
\end{center}
%
%\newpage
%
\section{Near-nucleus electric multipole moments of the atom in the
multipole electric field and electric multipole nuclear shielding
constants} 
\label{VI} 
\setcounter{equation}{0}
\subsection{Decomposition of the near-nucleus electric multipole
moments into the permanent and the first-order electric-field-induced
components} 
\label{VI.1}
With this section, we start the second part of the paper, in which we
shall be analyzing near-field (henceforth, in view of the present
context, the term \emph{near-nucleus\/} will be used instead)
multipole moments characterizing electrostatic and magnetostatic
potentials and fields generated by the atomic electron in the region
near a point where the atomic nucleus is located. The first of these
moments is the near-nucleus electric multipole moment
$\boldsymbol{\mathsf{R}}_{\lambda}$, the spherical components of
which are defined to be
\begin{equation}
\mathcal{R}_{\lambda\mu}
=\sqrt{\frac{4\pi}{2\lambda+1}}
\int_{\mathbb{R}^{3}}\mathrm{d}^{3}\boldsymbol{r}\:
r^{-\lambda-1}Y_{\lambda\mu}(\boldsymbol{n}_{r})\rho(\boldsymbol{r}),
\label{6.1}
\end{equation}
where $\rho(\boldsymbol{r})$ is the electronic charge density given
by Eq.\ (\ref{3.2}). Approximating the density as in Eq.\ (\ref{3.3})
yields
\begin{equation}
\mathcal{R}_{\lambda\mu}\simeq\mathcal{R}_{\lambda\mu}^{(0)}
+\mathcal{R}_{\lambda\mu}^{(1)},
\label{6.2}
\end{equation}
with the permanent and the first-order induced parts given by
\begin{equation}
\mathcal{R}_{\lambda\mu}^{(0)}
=\sqrt{\frac{4\pi}{2\lambda+1}}
\int_{\mathbb{R}^{3}}\mathrm{d}^{3}\boldsymbol{r}\:
r^{-\lambda-1}Y_{\lambda\mu}(\boldsymbol{n}_{r})
\rho^{(0)}(\boldsymbol{r})
\label{6.3}
\end{equation}
and
\begin{equation}
\mathcal{R}_{\lambda\mu}^{(1)}
=\sqrt{\frac{4\pi}{2\lambda+1}}
\int_{\mathbb{R}^{3}}\mathrm{d}^{3}\boldsymbol{r}\:
r^{-\lambda-1}Y_{\lambda\mu}(\boldsymbol{n}_{r})
\rho^{(1)}(\boldsymbol{r}),
\label{6.4}
\end{equation}
respectively. 

Evaluation of both $\mathcal{R}_{\lambda\mu}^{(0)}$ and
$\mathcal{R}_{\lambda\mu}^{(1)}$ is much simplified by the fact that
angular integrations involved are identical to these encountered in
Sec.\ \ref{III.1}, where the counterpart far-field moments
$\mathcal{Q}_{\lambda\mu}^{(0)}$ and $\mathcal{Q}_{\lambda\mu}^{(1)}$
have been analyzed. In result, one finds that
\begin{equation}
\boldsymbol{\mathsf{R}}_{\lambda}
\simeq\boldsymbol{\mathsf{R}}_{\lambda}^{(0)}\delta_{\lambda0}
+\boldsymbol{\mathsf{R}}_{\lambda}^{(1)}\delta_{\lambda L},
\label{6.5}
\end{equation}
with the non-vanishing components given by
\begin{equation}
\mathcal{R}_{00}^{(0)}=-e\int_{0}^{\infty}\mathrm{d}r\:r^{-1}
\left\{[P^{(0)}(r)]^{2}+[Q^{(0)}(r)]^{2}\right\}
\label{6.6}
\end{equation}
and
\begin{equation}
\mathcal{R}_{L\mu}^{(1)}=\mathcal{R}_{L\mu,L}^{(1)}
+\mathcal{R}_{L\mu,-L-1}^{(1)},
\label{6.7}
\end{equation}
where
\begin{equation}
\mathcal{R}_{L\mu,\kappa}^{(1)}=(4\pi\epsilon_{0})
\frac{2|\kappa|}{(2L+1)^{2}}\,
R_{\kappa}^{(-L-1,L)}\big(P^{(0)},Q^{(0)};P^{(0)},Q^{(0)}\big)
\mathcal{C}_{L\mu}^{(1)}
\qquad (\kappa=L,-L-1).
\label{6.8}
\end{equation}
In the last equation,
$R_{\kappa}^{(-L-1,L)}\big(P^{(0)},Q^{(0)};P^{(0)},Q^{(0)}\big)$ is a
particular form of the double radial integral defined in Eq.\
(\ref{3.15}). Invoking the explicit forms (\ref{2.11}) of the
electronic radial functions, one immediately finds that
\begin{equation}
\mathcal{R}_{00}^{(0)}=-\frac{Ze}{a_{0}\gamma_{1}}.
\label{6.9}
\end{equation}
\subsection{Atomic electric multipole nuclear shielding constants}
\label{VI.2}
In analogy with the far-field case, the present analysis leads in a
natural way to a definition of a near-nucleus electric $2^{L}$-pole
polarizability of the atom, $\sigma_{\mathrm{E}L\to\mathrm{E}L}$,
through the relation
\begin{equation}
\boldsymbol{\mathsf{R}}_{L}^{(1)}=(4\pi\epsilon_{0})
\sigma_{\mathrm{E}L\to\mathrm{E}L}\boldsymbol{\mathsf{C}}_{L}^{(1)}.
\label{6.10}
\end{equation}
In the literature, the near-nucleus polarizability
$\sigma_{\mathrm{E}L\to\mathrm{E}L}$ is usually named the electric
multipole nuclear shielding constant, and in what follows, we shall
adopt that nomenclature. Combining Eq.\ (\ref{6.10}) with Eqs.\
(\ref{6.7}) and (\ref{6.8}), we deduce that
\begin{equation}
\sigma_{\mathrm{E}L\to\mathrm{E}L}
=\sigma_{\mathrm{E}L\to\mathrm{E}L,L}
+\sigma_{\mathrm{E}L\to\mathrm{E}L,-L-1},
\label{6.11}
\end{equation}
with
\begin{equation}
\sigma_{\mathrm{E}L\to\mathrm{E}L,\kappa}
=\frac{2|\kappa|}{(2L+1)^{2}}
R_{\kappa}^{(-L-1,L)}\big(P^{(0)},Q^{(0)};P^{(0)},Q^{(0)}\big)
\qquad (\kappa=L,-L-1),
\label{6.12}
\end{equation}
where, by virtue of Eqs.\ (\ref{3.15}) and (\ref{3.26}), the double
radial integral may be written as
\begin{eqnarray}
&& \hspace*{-6em}
R_{\kappa}^{(-L-1,L)}\big(P^{(0)},Q^{(0)};P^{(0)},Q^{(0)}\big)
\nonumber \\
&=& \sum_{n_{r}=-\infty}^{\infty}\frac{1}{\mu_{n_{r}\kappa}^{(0)}-1}
\int_{0}^{\infty}\mathrm{d}r\:r^{-L-1}
\left[P^{(0)}(r)S_{n_{r}\kappa}^{(0)}(r)
+Q^{(0)}(r)T_{n_{r}\kappa}^{(0)}(r)\right]
\nonumber \\
&& \times\int_{0}^{\infty}\mathrm{d}r'\:r^{\prime L}
\left[\mu_{n_{r}\kappa}^{(0)}P^{(0)}(r')S_{n_{r}\kappa}^{(0)}(r')
+Q^{(0)}(r')T_{n_{r}\kappa}^{(0)}(r')\right]
\qquad (\kappa=L,-L-1).
\label{6.13}
\end{eqnarray}
As the integral over $r'$ is the one from Eq.\ (\ref{3.34}), while
that over $r$ is found to be
\begin{eqnarray}
&& \hspace*{-5em} 
\int_{0}^{\infty}\mathrm{d}r\:r^{-L-1}
\left[P^{(0)}(r)S_{n_{r}\kappa}^{(0)}(r)
+Q^{(0)}(r)T_{n_{r}\kappa}^{(0)}(r)\right]
\nonumber \\
&=& -\,\left(\frac{2Z}{a_{0}}\right)^{L}
\frac{\sqrt{2}(N_{n_{r}\kappa}-\kappa)
[\gamma_{1}(N_{n_{r}\kappa}+\kappa)
-(|n_{r}|+\gamma_{\kappa}-\gamma_{1}+L)]}
{\sqrt{a_{0}|n_{r}|!N_{n_{r}\kappa}(N_{n_{r}\kappa}-\kappa)
\Gamma(2\gamma_{1}+1)\Gamma(|n_{r}|+2\gamma_{\kappa}+1)}}
\nonumber \\
&& \times\,\frac{\Gamma(\gamma_{\kappa}+\gamma_{1}-L)
\Gamma(|n_{r}|+\gamma_{\kappa}-\gamma_{1}+L)}
{\Gamma(\gamma_{\kappa}-\gamma_{1}+L+1)}
%\nonumber \\
%&& \hspace*{10em}
\qquad
\bigg(\textrm{$Z<\alpha^{-1}\frac{\sqrt{4L^{2}-1}}{2L}$ for 
$\kappa=L$}\bigg),
\nonumber \\
&&
\label{6.14}
\end{eqnarray}
by means of the by now familiar procedure of summation of the series
over $n_{r}$ to a closed form involving a single ${}_{3}F_{2}(1)$
function, we arrive at the following expression for
$\sigma_{\mathrm{E}L\to\mathrm{E}L,\kappa}$:
\begin{eqnarray}
\sigma_{\mathrm{E}L\to\mathrm{E}L,\kappa}
&=& \frac{2|\kappa|}{Z(\kappa+1)(2L+1)^{2}}
\nonumber \\
&& \times\,
\bigg\{-1+\frac{[\gamma_{1}(\kappa+1)-L][\gamma_{1}(\kappa+1)+L+1]
\Gamma(\gamma_{\kappa}+\gamma_{1}-L)
\Gamma(\gamma_{\kappa}+\gamma_{1}+L+1)}{(\gamma_{\kappa}-\gamma_{1}+1)
\Gamma(2\gamma_{1}+1)\Gamma(2\gamma_{\kappa}+1)}
\nonumber \\
&& \quad \times\,
{}_{3}F_{2}
\left(
\begin{array}{c}
\gamma_{\kappa}-\gamma_{1}+L+1,\:
\gamma_{\kappa}-\gamma_{1}-L,\:
\gamma_{\kappa}-\gamma_{1}+1 \\
\gamma_{\kappa}-\gamma_{1}+2,\:
2\gamma_{\kappa}+1
\end{array}
;1
\right)
\bigg\}
\qquad (\kappa=L,-L-1).
\nonumber \\
&&
\label{6.15}
\end{eqnarray}
At the first sight, the expression on the right-hand side of Eq.\
(\ref{6.15}) looks equally complicated as its counterparts displayed
in Eqs.\ (\ref{3.39}), (\ref{4.41}) and (\ref{5.21}). It appears,
however, that because of particular forms of the parameters in the
${}_{3}F_{2}(1)$ function being involved, it may be transformed to a
much simpler representation. To this end, we exploit the identity
\cite[Eq.\ (7.4.4.1)]{Prud03}
\begin{eqnarray}
\hspace*{-3em}
{}_{3}F_{2}
\left(
\begin{array}{c}
a_{1},\:
a_{2},\:
a_{3} \\
b_{1},\:
b_{2}
\end{array}
;1
\right)
&=& \frac{\Gamma(b_{2})\Gamma(s)}
{\Gamma(b_{2}-a_{2})\Gamma(s+a_{2})}\,
{}_{3}F_{2}
\left(
\begin{array}{c}
b_{1}-a_{1},\:
b_{1}-a_{3},\:
a_{2} \\
b_{1},\:
s+a_{2}
\end{array}
;1
\right)
\nonumber \\
&& [\textrm{$s=b_{1}+b_{2}-a_{1}-a_{2}-a_{3}$; $\Real s>0$; 
$\Real(b_{2}-a_{2})>0$}],
\label{6.16}
\end{eqnarray}
and this casts Eq.\ (\ref{6.15}) into
\begin{eqnarray}
\hspace*{-3em}
\sigma_{\mathrm{E}L\to\mathrm{E}L,\kappa}
&=& \frac{2|\kappa|}{Z(\kappa+1)(2L+1)^{2}}
\bigg\{-1+\frac{[\gamma_{1}(\kappa+1)-L][\gamma_{1}(\kappa+1)+L+1]}
{(\gamma_{\kappa}-\gamma_{1}+1)(\gamma_{\kappa}+\gamma_{1}-L)}
\nonumber \\
&& \times\,
{}_{3}F_{2}
\left(
\begin{array}{c}
-L+1,\:
1,\:
\gamma_{\kappa}-\gamma_{1}-L \\
\gamma_{\kappa}-\gamma_{1}+2,\:
\gamma_{\kappa}+\gamma_{1}-L+1
\end{array}
;1
\right)
\bigg\}
\qquad (\kappa=L,-L-1).
\label{6.17}
\end{eqnarray}
Hence, the two addends on the right-hand side of Eq.\ (\ref{6.11})
may be explicitly written as
\begin{subequations}
\begin{eqnarray}
\sigma_{\mathrm{E}L\to\mathrm{E}L,L}
&=& \frac{2L}{Z(L+1)(2L+1)^{2}}
\bigg\{-1+\frac{(L+1)(\gamma_{1}+1)[\gamma_{1}(L+1)-L]}
{(\gamma_{L}-\gamma_{1}+1)(\gamma_{L}+\gamma_{1}-L)}
\nonumber \\
&& \times\,
{}_{3}F_{2}
\left(
\begin{array}{c}
-L+1,\:
1,\:
\gamma_{L}-\gamma_{1}-L \\
\gamma_{L}-\gamma_{1}+2,\:
\gamma_{L}+\gamma_{1}-L+1
\end{array}
;1
\right)
\bigg\}
\label{6.18a}
\end{eqnarray}
and
\begin{eqnarray}
\sigma_{\mathrm{E}L\to\mathrm{E}L,-L-1}
&=& \frac{2(L+1)}{ZL(2L+1)^{2}}
\bigg\{1-\frac{L(\gamma_{1}+1)(L\gamma_{1}-L-1)}
{(\gamma_{L+1}-\gamma_{1}+1)(\gamma_{L+1}+\gamma_{1}-L)}
\nonumber \\
&& \times\,
{}_{3}F_{2}
\left(
\begin{array}{c}
-L+1,\:
1,\:
\gamma_{L+1}-\gamma_{1}-L \\
\gamma_{L+1}-\gamma_{1}+2,\:
\gamma_{L+1}+\gamma_{1}-L+1
\end{array}
;1
\right)
\bigg\},
\label{6.18b}
\end{eqnarray}
\label{6.18}%
\end{subequations}
and consequently the sought formula for the $2^{L}$-pole electric
shielding constants is
\begin{eqnarray}
\sigma_{\mathrm{E}L\to\mathrm{E}L} 
&=& \frac{2}{ZL(L+1)(2L+1)}
\bigg\{1+\frac{L^{2}(L+1)(\gamma_{1}+1)[\gamma_{1}(L+1)-L]}
{(2L+1)(\gamma_{L}-\gamma_{1}+1)(\gamma_{L}+\gamma_{1}-L)}
\nonumber \\
&& \quad \times\,
{}_{3}F_{2}
\left(
\begin{array}{c}
-L+1,\:
1,\:
\gamma_{L}-\gamma_{1}-L \\
\gamma_{L}-\gamma_{1}+2,\:
\gamma_{L}+\gamma_{1}-L+1
\end{array}
;1
\right)
\nonumber \\
&& -\,\frac{L(L+1)^{2}(\gamma_{1}+1)(L\gamma_{1}-L-1)}
{(2L+1)(\gamma_{L+1}-\gamma_{1}+1)(\gamma_{L+1}+\gamma_{1}-L)}
\nonumber \\
&& \quad \times\,
{}_{3}F_{2}
\left(
\begin{array}{c}
-L+1,\:
1,\:
\gamma_{L+1}-\gamma_{1}-L \\
\gamma_{L+1}-\gamma_{1}+2,\:
\gamma_{L+1}+\gamma_{1}-L+1
\end{array}
;1
\right)
\bigg\}.
\label{6.19}
\end{eqnarray}

At the first sight, it might seem that the constraint on $Z$, under
which the above formula is valid, is the one in Eq.\ (\ref{6.14}).
This is indeed the case if $L\geqslant2$, but for $L=1$ the situation
is different. A closer look at the left-hand side of Eq.\
(\ref{6.14}) and at Eq.\ (\ref{3.27}) shows that in the dipole case
the convergence condition in the former equation is rooted in the
presence of the constant terms in the Laguerre polynomials in the
radial Sturmians $S_{n_{r}1}^{(0)}(r)$ and $T_{n_{r}1}^{(0)}(r)$.
However, it is easy to show that the series
\begin{eqnarray}
&& \hspace*{-6em}
\tilde{R}_{1}^{(-2,1)}\big(P^{(0)},Q^{(0)};P^{(0)},Q^{(0)}\big)
\nonumber \\
&=& \sum_{n_{r}=-\infty}^{\infty}\frac{1}{\mu_{n_{r}1}^{(0)}-1}
\int_{0}^{\infty}\mathrm{d}r\:r^{-2}
\left[P^{(0)}(r)\tilde{S}_{n_{r}1}^{(0)}(r)
+Q^{(0)}(r)\tilde{T}_{n_{r}1}^{(0)}(r)\right]
\nonumber \\
&& \times\int_{0}^{\infty}\mathrm{d}r'\:r^{\prime}
\left[\mu_{n_{r}1}^{(0)}P^{(0)}(r')S_{n_{r}1}^{(0)}(r')
+Q^{(0)}(r')T_{n_{r}1}^{(0)}(r')\right],
\label{6.20}
\end{eqnarray}
where
\begin{subequations}
\begin{eqnarray}
\hspace*{-5em}
\tilde{S}_{n_{r}1}^{(0)}(r)
&=& \sqrt{\frac{(1+\gamma_{1})(|n_{r}|+2\gamma_{1})|n_{r}|!}
{2ZN_{n_{r}1}(N_{n_{r}1}-1)\Gamma(|n_{r}|+2\gamma_{1})}}
\nonumber \\
&& \times\,\left(\frac{2Zr}{a_{0}}\right)^{\gamma_{1}}
\textrm{e}^{-Zr/a_{0}}\left[L_{|n_{r}|-1}^{(2\gamma_{1})}(0)
+\frac{1-N_{n_{r}1}}{|n_{r}|+2\gamma_{1}}
L_{|n_{r}|}^{(2\gamma_{1})}(0)\right]
\label{6.21a}
\end{eqnarray}
and
\begin{eqnarray}
\hspace*{-5em}
\tilde{T}_{n_{r}1}^{(0)}(r)
&=& \sqrt{\frac{(1-\gamma_{1})(|n_{r}|+2\gamma_{1})|n_{r}|!}
{2ZN_{n_{r}1}(N_{n_{r}1}-1)\Gamma(|n_{r}|+2\gamma_{1})}}
\nonumber \\
&& \times\,\left(\frac{2Zr}{a_{0}}\right)^{\gamma_{1}}
\textrm{e}^{-Zr/a_{0}}\left[L_{|n_{r}|-1}^{(2\gamma_{1})}(0)
-\frac{1-N_{n_{r}1}}{|n_{r}|+2\gamma_{1}}
L_{|n_{r}|}^{(2\gamma_{1})}(0)\right],
\label{6.21b}
\end{eqnarray}
\label{6.21}%
\end{subequations}
gives a null contribution to the integral
$R_{1}^{(-2,1)}\big(P^{(0)},Q^{(0)};P^{(0)},Q^{(0)}\big)$. In
consequence, in the dipole case the limitation on $Z$ is weaker than
for higher multipoles, being simply the natural one $Z<\alpha^{-1}$.
Hence, in summary, the constraint on the validity of the formula for
$\sigma_{\mathrm{E}L\to\mathrm{E}L}$ displayed in Eq.\ (\ref{6.19})
is
\begin{equation}
Z<
\left\{
\begin{array}{ll}
\alpha^{-1} & \textrm{for $L=1$} \\*[1ex]
\displaystyle
\alpha^{-1}\frac{\sqrt{4L^{2}-1}}{2L} 
& \textrm{for $L\geqslant2$}.
\end{array}
\right.
\label{6.22}
\end{equation}

As it holds that $L\geqslant1$, both ${}_{3}F_{2}(1)$ functions that
appear in Eq.\ (\ref{6.19}) are seen to be terminating ones. This
implies that the susceptibilities of the sort considered here may be
expressed in terms of elementary functions. Explicit formulas for
$\sigma_{\mathrm{E}L\to\mathrm{E}L}$ with $L$ constrained by
$1\leqslant L\leqslant4$ are displayed in Table \ref{T.9}.
\begin{center}
[Place for Table \ref{T.9}]
\end{center}

Since only elementary functions are involved, numerical values of
$\sigma_{\mathrm{E}L\to\mathrm{E}L}$, if desired, may be computed to
any required accuracy, even without having access to any specialized
software. For this reason, we have decided not to provide tabulation
of such data here.

To complete the task, we shall derive the quasi-relativistic
representations for the shielding constants
$\sigma_{\mathrm{E}L\to\mathrm{E}L}$. Somewhat surprisingly, this
appears to be a bit more involved than in the case of the three
far-field susceptibilities analyzed in Secs.\ \ref{III} to \ref{V}.
Referring to Eqs.\ (\ref{3.43}) and (\ref{3.44}), the
quasi-relativistic approximations for the two ${}_{3}F_{2}(1)$
functions in Eqs.\ (\ref{6.18a}) and (\ref{6.18b}) are found to be
\begin{eqnarray}
&& \hspace*{-3em}
{}_{3}F_{2}
\left(
\begin{array}{c}
-L+1,\:
1,\:
\gamma_{L}-\gamma_{1}-L \\
\gamma_{L}-\gamma_{1}+2,\:
\gamma_{L}+\gamma_{1}-L+1
\end{array}
;1
\right)
\nonumber \\
&\simeq& \frac{3L+1}{2(L+1)}-(\alpha Z)^{2}\frac{L-1}{4L(L+1)}
\bigg[\frac{L^{2}-5}{2(L+1)}+\frac{(L-1)(L-2)}{3(L+2)}\,
{}_{3}F_{2}
\left(
\begin{array}{c}
-L+3,\:
1,\:
1 \\
L+3,\:
4
\end{array}
;1
\right)
\bigg]
\nonumber \\
&&
\label{6.23}
\end{eqnarray}
and
\begin{eqnarray}
&& \hspace*{-10em}
{}_{3}F_{2}
\left(
\begin{array}{c}
-L+1,\:
1,\:
\gamma_{L+1}-\gamma_{1}-L \\
\gamma_{L+1}-\gamma_{1}+2,\:
\gamma_{L+1}+\gamma_{1}-L+1
\end{array}
;1
\right)
\nonumber \\
&\simeq& 1-(\alpha Z)^{2}\frac{L(L-1)}{6(L+1)(L+2)}\,
{}_{3}F_{2}
\left(
\begin{array}{c}
-L+2,\:
1,\:
1 \\
L+3,\:
4
\end{array}
;1
\right),
\label{6.24}
\end{eqnarray}
respectively. It is seen that in both cases the coefficients at
$(\alpha Z)^{2}$ involve the hypergeometric functions of the sort
considered in Appendix \ref{E}. From Eq.\ (\ref{E.10}), we obtain
\begin{eqnarray}
{}_{3}F_{2}
\left(
\begin{array}{c}
-L+2,\:
1,\:
1 \\
L+3,\:
4
\end{array}
;1
\right)
&=& -\frac{3(L+2)(5L+4)}{2L(L+1)}
+\frac{6(L+2)(2L+1)}{L(L-1)}[\psi(2L+1)-\psi(L+2)]
\nonumber \\
&&
\label{6.25}
\end{eqnarray}
and
\begin{eqnarray}
{}_{3}F_{2}
\left(
\begin{array}{c}
-L+3,\:
1,\:
1 \\
L+3,\:
4
\end{array}
;1
\right)
&=& -\frac{3(L+2)(5L+1)}{2L(L-1)}
+\frac{6(L+2)(2L+1)}{(L-1)(L-2)}[\psi(2L)-\psi(L+2)]
\nonumber \\
&&
\label{6.26}
\end{eqnarray}
(singularities at $L=1$ and $L=2$ in the above two equations are
apparent only and are removable through the application of the
L'Hospital's rule), which allows us to rewrite Eqs.\ (\ref{6.23}) and
(\ref{6.24}) as
\begin{eqnarray}
&& \hspace*{-5em}
{}_{3}F_{2}
\left(
\begin{array}{c}
-L+1,\:
1,\:
\gamma_{L}-\gamma_{1}-L \\
\gamma_{L}-\gamma_{1}+2,\:
\gamma_{L}+\gamma_{1}-L+1
\end{array}
;1
\right)
\nonumber \\
&\simeq& \frac{3L+1}{2(L+1)}-(\alpha Z)^{2}\frac{(L-1)(2L+1)}{2L(L+1)}
\left[\psi(2L+1)-\psi(L+1)-\frac{L+1}{2L+1}\right]
\label{6.27}
\end{eqnarray}
and
\begin{eqnarray}
&& \hspace*{-5em}
{}_{3}F_{2}
\left(
\begin{array}{c}
-L+1,\:
1,\:
\gamma_{L+1}-\gamma_{1}-L \\
\gamma_{L+1}-\gamma_{1}+2,\:
\gamma_{L+1}+\gamma_{1}-L+1
\end{array}
;1
\right)
\nonumber \\
&\simeq& 1-(\alpha Z)^{2}\frac{2L+1}{L+1}
\bigg[\psi(2L+1)-\psi(L+1)-\frac{L(5L+7)}{4(L+1)(2L+1)}\bigg],
\label{6.28}
\end{eqnarray}
respectively. Inserting the above two estimates into Eqs.\
(\ref{6.18a}) and (\ref{6.18b}) and passing to the quasi-relativistic
limit with the factors multiplying the two ${}_{3}F_{2}(1)$'s, one
arrives at
\begin{subequations}
\begin{eqnarray}
\sigma_{\mathrm{E}L\to\mathrm{E}L,L}
&\simeq& \frac{2}{Z(L+1)(2L+1)}
\nonumber \\
&& \times\,\bigg\{1-(\alpha Z)^{2}\frac{L-1}{L}
\bigg[\psi(2L+1)-\psi(L+1)
+\frac{6L^{4}+L^{3}+L^{2}-2L-2}{4L(L-1)(2L+1)}\bigg]\bigg\}
\nonumber \\
&&
\label{6.29a}
\end{eqnarray}
and
\begin{equation}
\sigma_{\mathrm{E}L\to\mathrm{E}L,-L-1}
\simeq\frac{2}{ZL(2L+1)}
\bigg\{1-(\alpha Z)^{2}\frac{L}{L+1}
\bigg[\psi(2L+1)-\psi(L+1)-\frac{2L^{2}+7L+1}
{4(2L+1)}\bigg]\bigg\}.
\label{6.29b}
\end{equation}
\label{6.29}%
\end{subequations}
Hence, we infer the following formula for the quasi-relativistic
approximations of the electric multipole shielding constants:
\begin{equation}
\sigma_{\mathrm{E}L\to\mathrm{E}L}
\simeq\frac{2}{ZL(L+1)}
\bigg\{1-(\alpha Z)^{2}\frac{2L-1}{2L+1}
\bigg[\psi(2L+1)-\psi(L+1)+\frac{L^{3}-2L^{2}+L-1}
{2L(2L-1)}\bigg]\bigg\}.
\label{6.30}
\end{equation}
Estimates of $\sigma_{\mathrm{E}L\to\mathrm{E}L}$ resulting from Eq.\
(\ref{6.30}) are presented in Table \ref{T.10} for $1\leqslant
L\leqslant4$.
\begin{center}
[Place for Table \ref{T.10}]
\end{center}

We have verified numerically that the quasi-relativistic formula in
Eq.\ (\ref{6.30}) is equivalent to a much more complicated one given
in Ref.\ \cite[Eq.\ (37)]{Kane77}, provided one corrects the latter
and replaces $(2l-2)!$ by $(2l-n)!$. In addition, we remark that the
quasi-relativistic approximations to
$\sigma_{\mathrm{E}L\to\mathrm{E}L}$ supplied in Refs.\ \cite[Eq.\
(3)]{Zapr74} and \cite[Eq.\ (4.41)]{Zapr85} are correct for $L=1$ and
$L=3$, but for $L=2$ the factors $k_{2}$ and $K_{2}$ displayed
therein should take the value $2/5$ instead of $59/150$.
%
%\newpage
%
\section{Near-nucleus magnetic multipole moments of the atom in the
multipole electric field and near-nucleus E$L\to$M$(L\mp1)$ multipole
cross-susceptibilities} 
\label{VII} 
\setcounter{equation}{0}
\subsection{Decomposition of the near-nucleus magnetic multipole
moments into the permanent and the first-order electric-field-induced
components} 
\label{VII.1}
Next, we shall consider the near-nucleus magnetic multipole moments
of the atom in the $2^{L}$-pole electric field. In agreement with
Eq.\ (\ref{B.38}), the spherical components of the $2^{\lambda}$-pole
moment of that sort are given by
\begin{equation}
\mathcal{N}_{\lambda\mu}
=\mathrm{i}\,\sqrt{\frac{4\pi(\lambda+1)}{\lambda(2\lambda+1)}}
\int_{\mathbb{R}^{3}}\mathrm{d}^{3}\boldsymbol{r}\:
r^{-\lambda-1}\boldsymbol{Y}_{\lambda\mu}^{\lambda}(\boldsymbol{n}_{r})
\cdot\boldsymbol{j}(\boldsymbol{r}),
\label{7.1}
\end{equation}
where $\boldsymbol{j}(\boldsymbol{r})$ is the electronic current in
the atom, defined as in Eq.\ (\ref{4.5}). The same argument that has
been applied to transform Eq.\ (\ref{4.1}) into Eq.\ (\ref{4.4}),
allows us to rewrite the above definition as
\begin{equation}
\mathcal{N}_{\lambda\mu}
=-\frac{\mathrm{i}}{\lambda}\sqrt{\frac{4\pi}{2\lambda+1}}
\int_{\mathbb{R}^{3}}\mathrm{d}^{3}\boldsymbol{r}\:
r^{-\lambda-1}Y_{\lambda\mu}(\boldsymbol{n}_{r})
\boldsymbol{\Lambda}\cdot\boldsymbol{j}(\boldsymbol{r}).
\label{7.2}
\end{equation}
Approximating the current as in Eq.\ (\ref{4.6}), yields
\begin{equation}
\mathcal{N}_{\lambda\mu}\simeq\mathcal{N}_{\lambda\mu}^{(0)}
+\mathcal{N}_{\lambda\mu}^{(1)},
\label{7.3}
\end{equation}
with
\begin{equation}
\mathcal{N}_{\lambda\mu}^{(0)}
=-\frac{\mathrm{i}}{\lambda}\sqrt{\frac{4\pi}{2\lambda+1}}
\int_{\mathbb{R}^{3}}\mathrm{d}^{3}\boldsymbol{r}\:
r^{-\lambda-1}Y_{\lambda\mu}(\boldsymbol{n}_{r})
\boldsymbol{\Lambda}\cdot\boldsymbol{j}^{(0)}(\boldsymbol{r})
\label{7.4}
\end{equation}
and
\begin{equation}
\mathcal{N}_{\lambda\mu}^{(1)}
=-\frac{\mathrm{i}}{\lambda}\sqrt{\frac{4\pi}{2\lambda+1}}
\int_{\mathbb{R}^{3}}\mathrm{d}^{3}\boldsymbol{r}\:
r^{-\lambda-1}Y_{\lambda\mu}(\boldsymbol{n}_{r})
\boldsymbol{\Lambda}\cdot\boldsymbol{j}^{(1)}(\boldsymbol{r}).
\label{7.5}
\end{equation}
Evidently, angular integrations which arise when the integrals in
Eqs.\ (\ref{7.4}) and (\ref{7.5}) are evaluated in the spherical
coordinates are identical to these in Eqs.\ (\ref{4.10}) and
(\ref{4.11}), respectively. This immediately allows us to write
\begin{equation}
\boldsymbol{\mathsf{N}}_{\lambda}
\simeq\boldsymbol{\mathsf{N}}_{\lambda}^{(0)}\delta_{\lambda1}
+\boldsymbol{\mathsf{N}}_{\lambda}^{(0)}
(\delta_{\lambda,L-1}+\delta_{\lambda,L+1}), 
\label{7.6}
\end{equation}
where
\begin{equation}
\boldsymbol{\mathsf{N}}_{1}^{(0)}=-\frac{4}{3}ec\boldsymbol{\nu}
\int_{0}^{\infty}\mathrm{d}r\:r^{-2}P^{(0)}(r)Q^{(0)}(r)
\label{7.7}
\end{equation}
and
\begin{eqnarray}
\boldsymbol{\mathsf{N}}_{\lambda}^{(1)}
&=& (4\pi\epsilon_{0})c\,\frac{2\sqrt{2}\,(\lambda+1)}
{(2\lambda+1)\sqrt{(2L+1)(\lambda+L+1)}}
R_{\kappa_{\lambda}}^{(-\lambda-1,L)}
\big(Q^{(0)},P^{(0)};P^{(0)},Q^{(0)}\big)
\big\{\boldsymbol{\nu}
\otimes\boldsymbol{\mathsf{C}}_{L}^{(1)}\big\}_{\lambda}
\nonumber \\
&& \hspace*{25em} (\textrm{$\lambda=L\mp1$, $\lambda\neq0$}),
\label{7.8}
\end{eqnarray}
with $\kappa_{\lambda}$ defined as in Eq.\ (\ref{4.34}). We see that
an isolated atom possesses only the permanent dipole moment of the
sort considered, while the first-order moments of that kind induced
by the $2^{L}$-pole electric field are of ranks $L-1$ and $L+1$,
except for the dipole ($L=1$) case when only the quadrupole moment
arises [cf.\ Eq.\ (\ref{B.42})].

Straightforward evaluation of the radial integral in Eq.\ (\ref{7.7})
gives the following closed-form representation of the permanent
dipole moment $\boldsymbol{\mathsf{N}}_{1}^{(0)}$:
\begin{equation}
\boldsymbol{\mathsf{N}}_{1}^{(0)}
=\frac{8}{3\gamma_{1}(2\gamma_{1}-1)}
\frac{\mu_{\mathrm{B}}Z^{3}}{a_{0}^{3}}\boldsymbol{\nu}
\qquad \left(Z<\alpha^{-1}\frac{\sqrt{3}}{2}\right),
\label{7.9}
\end{equation}
where the constraint on $Z$ results from the convergence condition
for the integral at its lower limit.
\subsection{Atomic near-nucleus multipole E$L\to$M$(L\mp1)$
cross-susceptibilities}
\label{VII.2}
The near-nucleus multipole electric-to-magnetic
cross-susceptibilities are defined through the relation
\begin{equation}
\boldsymbol{\mathsf{N}}_{\lambda}^{(1)}
=(4\pi\epsilon_{0})c\,\sigma_{\mathrm{E}L\to\mathrm{M}\lambda}
\frac{\big\{\boldsymbol{\nu}
\otimes\boldsymbol{\mathsf{C}}_{L}^{(1)}\big\}_{\lambda}}
{\langle10L0|\lambda0\rangle} 
\qquad (\textrm{$\lambda=L\mp1$; $\lambda\neq0$}),
\label{7.10}
\end{equation}
where $\boldsymbol{\mathsf{N}}_{\lambda}^{(1)}$ has been given in
Eq.\ (\ref{7.8}). Recalling the expression (\ref{4.36}) for the
Clebsch--Gordan coefficient $\langle10L0|\lambda0\rangle$ with
$\lambda=L\mp1$, yields the susceptibility
$\sigma_{\mathrm{E}L\to\mathrm{M}\lambda}$ in the form
\begin{equation}
\sigma_{\mathrm{E}L\to\mathrm{M}\lambda}
=\frac{2(\lambda+1)(\lambda-L)}{(2\lambda+1)(2L+1)}
R_{\kappa_{\lambda}}^{(-\lambda-1,L)}
\big(Q^{(0)},P^{(0)};P^{(0)},Q^{(0)}\big)
\qquad (\textrm{$\lambda=L\mp1$; $\lambda\neq0$}).
\label{7.11}
\end{equation}
In view of Eqs.\ (\ref{3.15}) and (\ref{3.26}), we may write
\begin{eqnarray}
&& \hspace*{-5em}
R_{\kappa_{\lambda}}^{(-\lambda-1,L)}
\big(Q^{(0)},P^{(0)};P^{(0)},Q^{(0)}\big)
\nonumber \\
&=& \sum_{n_{r}=-\infty}^{\infty}
\frac{1}{\mu_{n_{r}\kappa_{\lambda}}^{(0)}-1}
\int_{0}^{\infty}\mathrm{d}r\:r^{-\lambda-1}
\left[Q^{(0)}(r)S_{n_{r}\kappa_{\lambda}}^{(0)}(r)
+P^{(0)}(r)T_{n_{r}\kappa_{\lambda}}^{(0)}(r)\right]
\nonumber \\
&& \times\int_{0}^{\infty}\mathrm{d}r'\:r^{\prime L}
\left[\mu_{n_{r}\kappa_{\lambda}}^{(0)}
P^{(0)}(r')S_{n_{r}\kappa_{\lambda}}^{(0)}(r')
+Q^{(0)}(r')T_{n_{r}\kappa_{\lambda}}^{(0)}(r')\right]
\nonumber \\
&& \hspace*{20em} (\textrm{$\lambda=L\mp1$; $\lambda\neq0$}).
\label{7.12}
\end{eqnarray}
The second of the two integrals on the right-hand side of Eq.\
(\ref{7.12}) is the same which has appeared in the last four
sections, and its value has been given in Eq.\ (\ref{3.34}). As
regards the first one, its value may be deduced from Eq.\
(\ref{4.39}), after the replacement $\lambda\to-\lambda-1$ is made in
the latter, and this gives
\begin{eqnarray}
&& \hspace*{-5em} 
\int_{0}^{\infty}\mathrm{d}r\:r^{-\lambda-1}
\left[Q^{(0)}(r)S_{n_{r}\kappa_{\lambda}}^{(0)}(r)
+P^{(0)}(r)T_{n_{r}\kappa_{\lambda}}^{(0)}(r)\right]
\nonumber \\
&=& -\,\alpha Z\left(\frac{2Z}{a_{0}}\right)^{\lambda}
\frac{\sqrt{2}(N_{n_{r}\kappa_{\lambda}}-\kappa_{\lambda})}
{\sqrt{a_{0}|n_{r}|!N_{n_{r}\kappa_{\lambda}}
(N_{n_{r}\kappa_{\lambda}}-\kappa_{\lambda})
\Gamma(2\gamma_{1}+1)\Gamma(|n_{r}|+2\gamma_{\kappa_{\lambda}}+1)}}
\nonumber \\
&& \times\,\frac{\Gamma(\gamma_{\kappa_{\lambda}}+\gamma_{1}-\lambda)
\Gamma(|n_{r}|+\gamma_{\kappa_{\lambda}}-\gamma_{1}+\lambda+1)}
{\Gamma(\gamma_{\kappa_{\lambda}}-\gamma_{1}+\lambda+1)}
\nonumber \\
&& \hspace*{12em}
\left(\textrm{$Z<\alpha^{-1}\frac{\sqrt{(2L+1)(2L+3)}}{2(L+1)}$
 for $\lambda=L+1$}\right),
\label{7.13}
\end{eqnarray}
where the constraint, henceforth tacitly assumed to hold, guarantees
the integral in question converges at its lower limit. Plugging Eqs.\
(\ref{3.34}) and (\ref{7.13}) into Eq.\ (\ref{7.12}), and then the
latter into Eq.\ (\ref{7.11}), after transformations which are
already routine at this stage, we obtain
\begin{eqnarray}
\sigma_{\mathrm{E}L\to\mathrm{M}\lambda}
&=& \frac{\alpha a_{0}^{L-\lambda}}{Z^{L-\lambda}}
\frac{(\lambda+1)(\lambda-L)\Gamma(2\gamma_{1}-\lambda+L+1)}
{2^{L-\lambda-1}(\kappa_{\lambda}+1)(2\lambda+1)(2L+1)
\Gamma(2\gamma_{1}+1)}
\nonumber \\
&& \times\,\bigg\{1+\frac{\lambda[\gamma_{1}(\kappa_{\lambda}+1)+L+1]
\Gamma(\gamma_{\kappa_{\lambda}}+\gamma_{1}-\lambda)
\Gamma(\gamma_{\kappa_{\lambda}}+\gamma_{1}+L+1)}
{(\gamma_{\kappa_{\lambda}}-\gamma_{1}+1)
\Gamma(2\gamma_{1}-\lambda+L+1)
\Gamma(2\gamma_{\kappa_{\lambda}}+1)}
\nonumber \\
&& \quad \times\,
{}_{3}F_{2}
\left(
\begin{array}{c}
\gamma_{\kappa_{\lambda}}-\gamma_{1}+\lambda+1,\:
\gamma_{\kappa_{\lambda}}-\gamma_{1}-L,\:
\gamma_{\kappa_{\lambda}}-\gamma_{1}+1 \\
\gamma_{\kappa_{\lambda}}-\gamma_{1}+2,\:
2\gamma_{\kappa_{\lambda}}+1 
\end{array}
;1
\right)
\bigg\}
\nonumber \\
&& \hspace*{20em} (\textrm{$\lambda=L\mp1$; $\lambda\neq0$}).
\label{7.14}
\end{eqnarray}
Applying the identity (\ref{6.16}) to the hypergeometric function of
the right-hand side of Eq.\ (\ref{7.14}), casts the latter formula to
the simpler form
\begin{eqnarray}
\sigma_{\mathrm{E}L\to\mathrm{M}\lambda}
&=& \frac{\alpha a_{0}^{L-\lambda}}{Z^{L-\lambda}}
\frac{(\lambda+1)(\lambda-L)\Gamma(2\gamma_{1}-\lambda+L+1)}
{2^{L-\lambda-1}(\kappa_{\lambda}+1)(2\lambda+1)(2L+1)
\Gamma(2\gamma_{1}+1)}
\nonumber \\
&& \times\,
\bigg\{1+\frac{\lambda[\gamma_{1}(\kappa_{\lambda}+1)+L+1]}
{(\gamma_{\kappa_{\lambda}}-\gamma_{1}+1)
(\gamma_{\kappa_{\lambda}}+\gamma_{1}-\lambda)}\,
{}_{3}F_{2}
\left(
\begin{array}{c}
-\lambda+1,\:
1,\:
\gamma_{\kappa_{\lambda}}-\gamma_{1}-L \\
\gamma_{\kappa_{\lambda}}-\gamma_{1}+2,\:
\gamma_{\kappa_{\lambda}}+\gamma_{1}-\lambda+1 
\end{array}
;1
\right)
\bigg\}
\nonumber \\
&& \hspace*{20em} (\textrm{$\lambda=L\mp1$; $\lambda\neq0$}),
\label{7.15}
\end{eqnarray}
where the ${}_{3}F_{2}(1)$ function is a truncating one. Hence, we
find that the explicit representations of the two
cross-susceptibilities $\sigma_{\mathrm{E}L\to\mathrm{M}(L\mp1)}$ are
\begin{eqnarray}
\sigma_{\mathrm{E}L\to\mathrm{M}(L-1)}
&=& -\frac{\alpha a_{0}}{Z}
\frac{L(2\gamma_{1}+1)}{(L+1)(4L^{2}-1)}
\bigg\{1+\frac{(L^{2}-1)(\gamma_{1}+1)}
{(\gamma_{L}-\gamma_{1}+1)(\gamma_{L}+\gamma_{1}-L+1)}
\nonumber \\
&& \times\,
{}_{3}F_{2}
\left(
\begin{array}{c}
-L+2,\:
1,\:
\gamma_{L}-\gamma_{1}-L \\
\gamma_{L}-\gamma_{1}+2,\:
\gamma_{L}+\gamma_{1}-L+2 
\end{array}
;1
\right)
\bigg\}
\qquad (L\neq1)
\label{7.16}
\end{eqnarray}
and
\begin{eqnarray}
\sigma_{\mathrm{E}L\to\mathrm{M}(L+1)}
&=& -\frac{\alpha Z}{a_{0}}\frac{2(L+2)}{\gamma_{1}L(2L+1)(2L+3)}
\bigg\{1-\frac{(L+1)(L\gamma_{1}-L-1)}
{(\gamma_{L+1}-\gamma_{1}+1)(\gamma_{L+1}+\gamma_{1}-L-1)}
\nonumber \\
&& \times\,
{}_{3}F_{2}
\left(
\begin{array}{c}
-L,\:
1,\:
\gamma_{L+1}-\gamma_{1}-L \\
\gamma_{L+1}-\gamma_{1}+2,\:
\gamma_{L+1}+\gamma_{1}-L 
\end{array}
;1
\right)
\bigg\}
\nonumber \\
&& \hspace*{13em}
\left(Z<\alpha^{-1}\frac{\sqrt{(2L+1)(2L+3)}}{2(L+1)}\right).
\label{7.17}
\end{eqnarray}
Elementary expressions for $\sigma_{\mathrm{E}L\to\mathrm{M}(L-1)}$
with $2\leqslant L\leqslant4$ and
$\sigma_{\mathrm{E}L\to\mathrm{M}(L+1)}$ with $1\leqslant
L\leqslant4$, inferred from Eqs.\ (\ref{7.16}) and (\ref{7.17}), are
displayed in Table \ref{T.11}.
\begin{center}
[Place for Table \ref{T.11}]
\end{center}

To establish the quasi-relativistic approximations to the formulas in
Eqs.\ (\ref{7.16}) and (\ref{7.17}), we consider the estimates
\begin{eqnarray}
&& \hspace*{-3em}
{}_{3}F_{2}
\left(
\begin{array}{c}
-L+2,\:
1,\:
\gamma_{L}-\gamma_{1}-L \\
\gamma_{L}-\gamma_{1}+2,\:
\gamma_{L}+\gamma_{1}-L+2 
\end{array}
;1
\right)
\nonumber \\
&\simeq& \frac{4L+1}{3(L+1)}
-(\alpha Z)^{2}\frac{L-2}{6L(L+1)}
\left[\frac{2L^{2}+L-7}{3(L+1)}+\frac{(L-1)(L-3)}{4(L+2)}\,
{}_{3}F_{2}
\left(
\begin{array}{c}
-L+4,\:
1,\:
1 \\
L+3,\:
5
\end{array}
;1
\right)
\right]
\nonumber \\
&&
\label{7.18}
\end{eqnarray}
and
\begin{equation}
{}_{3}F_{2}
\left(
\begin{array}{c}
-L,\:
1,\:
\gamma_{L+1}-\gamma_{1}-L \\
\gamma_{L+1}-\gamma_{1}+2,\:
\gamma_{L+1}+\gamma_{1}-L 
\end{array}
;1
\right)
\simeq 1-(\alpha Z)^{2}\frac{L^{2}}{4(L+1)(L+2)}\,
{}_{3}F_{2}
\left(
\begin{array}{c}
-L+1,\:
1,\:
1 \\
L+3,\:
3 
\end{array}
;1
\right),
\label{7.19}
\end{equation}
which, with the aid of the identities
\begin{eqnarray}
&& \hspace*{-5em}
{}_{3}F_{2}
\left(
\begin{array}{c}
-L+4,\:
1,\:
1 \\
L+3,\:
5
\end{array}
;1
\right)
\nonumber \\
&=& \frac{8(L+2)(4L^{2}-1)}{(L-1)(L-2)(L-3)}
[\psi(2L+1)-\psi(L+2)]-\frac{2(L+2)(32L-19)}{3(L-2)(L-3)}
\label{7.20}
\end{eqnarray}
[cf.\ Eq.\ (\ref{E.11})] and
\begin{equation}
{}_{3}F_{2}
\left(
\begin{array}{c}
-L+1,\:
1,\:
1 \\
L+3,\:
3
\end{array}
;1
\right)
=\frac{4(L+2)}{L}[\psi(2L+2)-\psi(L+2)]-\frac{2(L+2)}{L+1}
\label{7.21}
\end{equation}
[cf.\ Eq.\ (\ref{E.3})], may be rewritten as
\begin{eqnarray}
&& \hspace*{-5em}
{}_{3}F_{2}
\left(
\begin{array}{c}
-L+2,\:
1,\:
\gamma_{L}-\gamma_{1}-L \\
\gamma_{L}-\gamma_{1}+2,\:
\gamma_{L}+\gamma_{1}-L+2 
\end{array}
;1
\right)
\nonumber \\
&\simeq& \frac{4L+1}{3(L+1)}
-(\alpha Z)^{2}\frac{4L^{2}-1}{3L(L+1)}\bigg[\psi(2L+1)-\psi(L+1)
-\frac{7(L+1)(4L-3)}{12(4L^{2}-1)}\bigg]
\label{7.22}
\end{eqnarray}
and
\begin{equation}
{}_{3}F_{2}
\left(
\begin{array}{c}
-L,\:
1,\:
\gamma_{L+1}-\gamma_{1}-L \\
\gamma_{L+1}-\gamma_{1}+2,\:
\gamma_{L+1}+\gamma_{1}-L 
\end{array}
;1
\right)
\simeq 1-(\alpha Z)^{2}\frac{L}{L+1}
\bigg[\psi(2L+2)-\psi(L+2)-\frac{L}{2(L+1)}\bigg],
\label{7.23}
\end{equation}
respectively. Applying the estimates in Eqs.\ (\ref{3.43}),
(\ref{7.22}) and (\ref{7.23}) to the right-hand sides of Eqs.\
(\ref{7.16}) and (\ref{7.17}), after some algebra we arrive at the
sought quasi-relativistic approximations
\begin{equation}
\sigma_{\mathrm{E}L\to\mathrm{M}(L-1)}
\simeq-\frac{\alpha a_{0}}{Z}\frac{1}{L+1}
\bigg\{1-(\alpha Z)^{2}
\frac{L-1}{L}\bigg[\psi(2L)-\psi(L)
-\frac{L(4L^{2}-3L-5)}{4(L-1)(4L^{2}-1)}\bigg]\bigg\}
\qquad (L\neq1)
\label{7.24}
\end{equation}
and
\begin{eqnarray}
\hspace*{-3em}
\sigma_{\mathrm{E}L\to\mathrm{M}(L+1)}
&\simeq& -\frac{\alpha Z}{a_{0}}
\frac{4(L+2)}{L(2L+1)(2L+3)}
\nonumber \\
&& \times\,\bigg\{1-(\alpha Z)^{2}\frac{L}{2(L+1)}
\bigg[\psi(2L+2)-\psi(L+1)-\frac{(L+1)(L+4)}{2L}\bigg]\bigg\}.
\label{7.25}
\end{eqnarray}
Particular cases of the later two formulas are displayed in Table
\ref{T.12}.
\begin{center}
[Place for Table \ref{T.12}]
\end{center}
%
%\newpage
%
\section{Near-nucleus magnetic toroidal multipole moments of the atom
in the multipole electric field and near-nucleus E$L\to$T$L$
multipole cross-susceptibilities} 
\label{VIII}
\setcounter{equation}{0}
\subsection{Decomposition of the near-nucleus magnetic toroidal
multipole moments into the permanent and the first-order
electric-field-induced components} 
\label{VIII.1}
The last family of the atomic multipole moments we wish to look at in
the present work are the near-nucleus magnetic toroidal multipole
moments $\boldsymbol{\mathsf{U}}_{\lambda}$. According to Eq.\
(\ref{D.10}), their spherical components may be defined as
\begin{equation}
\mathcal{U}_{\lambda\mu}=-\frac{1}{\lambda}
\sqrt{\frac{4\pi}{2\lambda+1}}
\int_{\mathbb{R}^{3}}\mathrm{d}^{3}\boldsymbol{r}\:r^{-\lambda-1}
Y_{\lambda\mu}(\boldsymbol{n}_{r})
\boldsymbol{r}\cdot\boldsymbol{j}(\boldsymbol{r}).
\label{8.1}
\end{equation}
In the weak-perturbing-field regime, which we consider in this paper,
after exploiting Eq.\ (\ref{4.6}), we have
\begin{equation}
\mathcal{U}_{\lambda\mu}\simeq\mathcal{U}_{\lambda\mu}^{(0)}
+\mathcal{U}_{\lambda\mu}^{(1)},
\label{8.2}
\end{equation}
with
\begin{equation}
\mathcal{U}_{\lambda\mu}^{(0)}=-\frac{1}{\lambda}
\sqrt{\frac{4\pi}{2\lambda+1}}
\int_{\mathbb{R}^{3}}\mathrm{d}^{3}\boldsymbol{r}\:r^{-\lambda-1}
Y_{\lambda\mu}(\boldsymbol{n}_{r})
\boldsymbol{r}\cdot\boldsymbol{j}^{(0)}(\boldsymbol{r})
\label{8.3}
\end{equation}
and
\begin{equation}
\mathcal{U}_{\lambda\mu}^{(1)}=-\frac{1}{\lambda}
\sqrt{\frac{4\pi}{2\lambda+1}}
\int_{\mathbb{R}^{3}}\mathrm{d}^{3}\boldsymbol{r}\:r^{-\lambda-1}
Y_{\lambda\mu}(\boldsymbol{n}_{r})
\boldsymbol{r}\cdot\boldsymbol{j}^{(1)}(\boldsymbol{r})
\label{8.4}
\end{equation}
being the permanent and the first-order induced parts, respectively,
of the moment under study in the atomic ground state. Exactly in the
same manner as in Sec.\ \ref{V}, one may show that the isolated atom
in the ground state does not possess any non vanishing moments of the
sort considered,
\begin{equation}
\boldsymbol{\mathsf{U}}_{\lambda}^{(0)}=\boldsymbol{\mathsf{0}},
\label{8.5}
\end{equation}
and that the only induced moment is the one with the multipolar
symmetry identical to that of the perturbing electric field, i.e.,
\begin{equation}
\boldsymbol{\mathsf{U}}_{\lambda}
\simeq\boldsymbol{\mathsf{U}}_{\lambda}^{(1)}\delta_{\lambda L},
\label{8.6}
\end{equation}
with
\begin{eqnarray}
\boldsymbol{\mathsf{U}}_{L}^{(1)}
&=& (4\pi\epsilon_{0})c\,\frac{2\mathrm{i}}{(2L+1)^{2}}
\sqrt{\frac{L+1}{L}}
\nonumber \\
&& \times\,
\left[R_{L}^{(-L,L)}\big(Q^{(0)},-P^{(0)};P^{(0)},Q^{(0)}\big)
-R_{-L-1}^{(-L,L)}\big(Q^{(0)},-P^{(0)};P^{(0)},Q^{(0)}\big)\right]
\big\{\boldsymbol{\nu}
\otimes\boldsymbol{\mathsf{C}}_{L}^{(1)}\big\}_{L}
\nonumber \\
&&
\label{8.7}
\end{eqnarray}
[for the definition of the double radial integrals
$R_{\kappa}^{(-L,L)}\big(Q^{(0)},-P^{(0)};P^{(0)},Q^{(0)}\big)$, see
Eq.\ (\ref{3.15})].
\subsection{Atomic near-nucleus multipole E$L\to$T$L$
cross-susceptibilities} 
\label{VIII.2}
In complete analogy to the far-field case discussed in Sec.\
\ref{V.2}, we define the near-nucleus multipole
$\mathrm{E}L\to\mathrm{T}L$ cross-susceptibilities through the
relation
\begin{equation}
\boldsymbol{\mathsf{U}}_{L}^{(1)}=\mathrm{i}(4\pi\epsilon_{0})c
\sigma_{\mathrm{E}L\to\mathrm{T}L}
\sqrt{L(L+1)}\,\big\{\boldsymbol{\nu}
\otimes\boldsymbol{\mathsf{C}}_{L}^{(1)}\big\}_{L}.
\label{8.8}
\end{equation}
Comparison of Eqs.\ (\ref{8.7}) and (\ref{8.8}) yields
$\sigma_{\mathrm{E}L\to\mathrm{T}L}$ as the sum
\begin{equation}
\sigma_{\mathrm{E}L\to\mathrm{T}L}
=\sigma_{\mathrm{E}L\to\mathrm{T}L,L}
+\sigma_{\mathrm{E}L\to\mathrm{T}L,-L-1},
\label{8.9}
\end{equation}
where
\begin{equation}
\sigma_{\mathrm{E}L\to\mathrm{T}L,\kappa}
=\frac{2\sgn(\kappa)}{L(2L+1)^{2}}
R_{\kappa}^{(-L,L)}\big(Q^{(0)},-P^{(0)};P^{(0)},Q^{(0)}\big)
\qquad (\kappa=L,-L-1).
\label{8.10}
\end{equation}
After the Sturmian expansion (\ref{3.26}) is used, the double radial
integral appearing in Eq.\ (\ref{8.10}) takes the form of the series
\begin{eqnarray}
&& \hspace*{-5em}
R_{\kappa}^{(-L,L)}\big(Q^{(0)},-P^{(0)};P^{(0)},Q^{(0)}\big)
\nonumber \\
&=& \sum_{n_{r}=-\infty}^{\infty}
\frac{1}{\mu_{n_{r}\kappa}^{(0)}-1}
\int_{0}^{\infty}\mathrm{d}r\:r^{-L}
\left[Q^{(0)}(r)S_{n_{r}\kappa}^{(0)}(r)
-P^{(0)}(r)T_{n_{r}\kappa}^{(0)}(r)\right]
\nonumber \\
&& \times\int_{0}^{\infty}\mathrm{d}r'\:r^{\prime L}
\left[\mu_{n_{r}\kappa}^{(0)}
P^{(0)}(r')S_{n_{r}\kappa}^{(0)}(r')
+Q^{(0)}(r')T_{n_{r}\kappa}^{(0)}(r')\right].
\label{8.11}
\end{eqnarray}
With no difficulty, from Eqs.\ (\ref{2.11}), (\ref{3.27}) and
(\ref{3.31}) one finds that
\begin{eqnarray}
&& \hspace*{-5em} 
\int_{0}^{\infty}\mathrm{d}r\:r^{-L}
\left[Q^{(0)}(r)S_{n_{r}\kappa}^{(0)}(r) 
-P^{(0)}(r)T_{n_{r}\kappa}^{(0)}(r)\right]
\nonumber \\
&=& \alpha Z\left(\frac{2Z}{a_{0}}\right)^{L-1}
\frac{\sqrt{2}\,|n_{r}|(|n_{r}|+2\gamma_{\kappa})}
{\sqrt{a_{0}|n_{r}|!N_{n_{r}\kappa}(N_{n_{r}\kappa}-\kappa)
\Gamma(2\gamma_{1}+1)\Gamma(|n_{r}|+2\gamma_{\kappa}+1)}}
\nonumber \\
&& \times\,\frac{\Gamma(\gamma_{\kappa}+\gamma_{1}-L+1)
\Gamma(|n_{r}|+\gamma_{\kappa}-\gamma_{1}+L-1)}
{\Gamma(\gamma_{\kappa}-\gamma_{1}+L)}.
\label{8.12}
\end{eqnarray}
Inserting Eqs.\ (\ref{3.28}), (\ref{3.34}) and (\ref{8.12}) into Eq.\
(\ref{8.11}) and proceeding then along the same path as in the
preceding sections to transform the series
$\sum_{n_{r}=-\infty}^{\infty}(\cdots)$ into the one of the form 
$\sum_{n_{r}=0}^{\infty}(\cdots)$, we arrive at
\begin{eqnarray}
\sigma_{\mathrm{E}L\to\mathrm{T}L,\kappa}
&=& -\frac{\alpha a_{0}}{Z}\frac{\sgn(\kappa)
[\gamma_{1}(\kappa+1)+L+1]\Gamma(\gamma_{\kappa}+\gamma_{1}-L+1)
\Gamma(\gamma_{\kappa}+\gamma_{1}+L+1)}
{L(2L+1)^{2}(\gamma_{\kappa}-\gamma_{1}+1)
\Gamma(2\gamma_{1}+1)\Gamma(2\gamma_{\kappa}+1)}
\nonumber \\
&& \times\,{}_{3}F_{2}
\left(
\begin{array}{c}
\gamma_{\kappa}-\gamma_{1}+L,\:
\gamma_{\kappa}-\gamma_{1}-L,\:
\gamma_{\kappa}-\gamma_{1}+1 \\
\gamma_{\kappa}-\gamma_{1}+2,\:
2\gamma_{\kappa}+1
\end{array}
;1
\right)
\qquad (\kappa=L,-L-1).
\nonumber \\
&&
\label{8.13}
\end{eqnarray}
In the final step, we apply the hypergeometric identity (\ref{6.16})
to convert the ${}_{3}F_{2}(1)$ series in Eq.\ (\ref{8.13}) into a
more suitable one, which yields
\begin{eqnarray}
\sigma_{\mathrm{E}L\to\mathrm{T}L,\kappa} 
&=& -\frac{\alpha a_{0}}{Z}
\frac{\sgn(\kappa)(2\gamma_{1}+1)[\gamma_{1}(\kappa+1)+L+1]}
{L(2L+1)^{2}(\gamma_{\kappa}-\gamma_{1}+1)
(\gamma_{\kappa}+\gamma_{1}-L+1)}
\nonumber \\
&& \times\,{}_{3}F_{2}
\left(
\begin{array}{c}
-L+2,\:
1,\:
\gamma_{\kappa}-\gamma_{1}-L \\
\gamma_{\kappa}-\gamma_{1}+2,\:
\gamma_{\kappa}+\gamma_{1}-L+2
\end{array}
;1
\right)
\qquad (\kappa=L,-L-1).
\label{8.14}
\end{eqnarray}
Hence, the cross-susceptibility $\sigma_{\mathrm{E}L\to\mathrm{T}L}$
is the sum of
\begin{subequations}
\begin{eqnarray}
\sigma_{\mathrm{E}L\to\mathrm{T}L,L} 
&=& -\frac{\alpha a_{0}}{Z}\frac{(L+1)(\gamma_{1}+1)(2\gamma_{1}+1)}
{L(2L+1)^{2}(\gamma_{L}-\gamma_{1}+1)
(\gamma_{L}+\gamma_{1}-L+1)}
\nonumber \\
&& \times\,{}_{3}F_{2}
\left(
\begin{array}{c}
-L+2,\:
1,\:
\gamma_{L}-\gamma_{1}-L \\
\gamma_{L}-\gamma_{1}+2,\:
\gamma_{L}+\gamma_{1}-L+2
\end{array}
;1
\right)
\label{8.15a}
\end{eqnarray}
and
\begin{eqnarray}
\sigma_{\mathrm{E}L\to\mathrm{T}L,-L-1} 
&=& -\frac{\alpha a_{0}}{Z}
\frac{(2\gamma_{1}+1)(L\gamma_{1}-L-1)}
{L(2L+1)^{2}(\gamma_{L+1}-\gamma_{1}+1)
(\gamma_{L+1}+\gamma_{1}-L+1)}
\nonumber \\
&& \times\,{}_{3}F_{2}
\left(
\begin{array}{c}
-L+2,\:
1,\:
\gamma_{L+1}-\gamma_{1}-L \\
\gamma_{L+1}-\gamma_{1}+2,\:
\gamma_{L+1}+\gamma_{1}-L+2
\end{array}
;1
\right),
\label{8.15b}
\end{eqnarray}
\label{8.15}%
\end{subequations}
and is explicitly given by
\begin{eqnarray}
\sigma_{\mathrm{E}L\to\mathrm{T}L} 
&=& -\frac{\alpha a_{0}}{Z}\frac{(L+1)(2\gamma_{1}+1)}{L(2L+1)^{2}}
\bigg\{\frac{\gamma_{1}+1}{(\gamma_{L}-\gamma_{1}+1)
(\gamma_{L}+\gamma_{1}-L+1)}
\nonumber \\
&& \quad \times\,
{}_{3}F_{2}
\left(
\begin{array}{c}
-L+2,\:
1,\:
\gamma_{L}-\gamma_{1}-L \\
\gamma_{L}-\gamma_{1}+2,\:
\gamma_{L}+\gamma_{1}-L+2
\end{array}
;1
\right) 
\nonumber \\
&& +\,\frac{L\gamma_{1}-L-1}{(L+1)(\gamma_{L+1}-\gamma_{1}+1)
(\gamma_{L+1}+\gamma_{1}-L+1)}
\nonumber \\
&& \quad \times\,
{}_{3}F_{2}
\left(
\begin{array}{c}
-L+2,\:
1,\:
\gamma_{L+1}-\gamma_{1}-L \\
\gamma_{L+1}-\gamma_{1}+2,\:
\gamma_{L+1}+\gamma_{1}-L+2
\end{array}
;1
\right)
\bigg\}.
\label{8.16}
\end{eqnarray}
For $L\geqslant2$, both hypergeometric series on the right-hand side
of Eq.\ (\ref{8.16}) truncate, and consequently the corresponding
cross-susceptibilities $\sigma_{\mathrm{E}L\to\mathrm{T}L}$ may be
written in terms of elementary functions. The dipole ($L=1$) case is
different since then the second ${}_{3}F_{2}(1)$ function remains
transcendental. Explicit analytical expressions for the
cross-susceptibilities $\sigma_{\mathrm{E}L\to\mathrm{T}L}$ with
$1\leqslant L\leqslant4$ are presented in Table \ref{T.13}. In turn,
in Table \ref{T.14} we provide numerical data for the dipole
cross-susceptibility $\sigma_{\mathrm{E}1\to\mathrm{T}1}$ for
selected values of the nuclear charge number $Z$.
\begin{center}
[Place for Tables \ref{T.13} and \ref{T.14}]
\end{center}

We move to the derivation of the quasi-relativistic limit of the
expression in Eq.\ (\ref{8.16}). The ${}_{3}F_{2}(1)$ function in
Eq.\ (\ref{8.15a}) is the one we have already come across in Sec.\
\ref{VII.2}, and the approximation to it is given in Eq.\
(\ref{7.22}). In turn, for the hypergeometric function in Eq.\
(\ref{8.15b}) we have
\begin{eqnarray}
&& \hspace*{-5em}
{}_{3}F_{2}
\left(
\begin{array}{c}
-L+2,\:
1,\:
\gamma_{L+1}-\gamma_{1}-L \\
\gamma_{L+1}-\gamma_{1}+2,\:
\gamma_{L+1}+\gamma_{1}-L+2
\end{array}
;1
\right)
\nonumber \\
&\simeq& 1-(\alpha Z)^{2}\frac{L(L-2)}{8(L+1)(L+2)}\,
{}_{3}F_{2}
\left(
\begin{array}{c}
-L+3,\:
1,\:
1 \\
L+3,\:
5
\end{array}
;1
\right).
\label{8.17}
\end{eqnarray}
From Eq.\ (\ref{E.11}) we have
\begin{equation}
{}_{3}F_{2}
\left(
\begin{array}{c}
-L+3,\:
1,\:
1 \\
L+3,\:
5
\end{array}
;1
\right)
=\frac{16(L+2)(2L+1)}{(L-1)(L-2)}[\psi(2L)-\psi(L+2)]
-\frac{4(L+2)(16L^{2}+17L+3)}{3L(L^{2}-1)}
\label{8.18}
\end{equation}
(we do not exclude the cases $L=1$ and $L=2$ as singularities at
these two values of $L$ in the expression on the right-hand side are
only apparent and are removable via the passage to the limit
procedure) and consequently the ${}_{3}F_{2}(1)$ function in Eq.\
(\ref{8.15b}) may be approximated as
\begin{eqnarray}
&& \hspace*{-5em}
{}_{3}F_{2}
\left(
\begin{array}{c}
-L+2,\:
1,\:
\gamma_{L+1}-\gamma_{1}-L \\
\gamma_{L+1}-\gamma_{1}+2,\:
\gamma_{L+1}+\gamma_{1}-L+2
\end{array}
;1
\right)
\nonumber \\
&\simeq& 1-(\alpha Z)^{2}\frac{2L(2L+1)}{L^{2}-1}
\left[\psi(2L+1)-\psi(L+1)
-\frac{16L^{2}+21L-1}{12(L+1)(2L+1)}\right].
\label{8.19}
\end{eqnarray}
Using Eqs.\ (\ref{7.22}) and (\ref{8.19}), after some algebraic
simplifications we obtain
\begin{subequations}
\begin{eqnarray}
\sigma_{\mathrm{E}L\to\mathrm{T}L,L} 
&\simeq& -\frac{\alpha a_{0}}{Z}\frac{4L+1}{L^{2}(2L+1)^2}
\nonumber \\
&& \times\,\left\{1-(\alpha Z)^{2}
\frac{4L^{2}-1}{L(4L+1)}\left[\psi(2L-1)-\psi(L)
-\frac{4L^{2}-3L-4}{4(4L^{2}-1)}\right]\right\}
\label{8.20a}
\end{eqnarray}
and
\begin{eqnarray}
\hspace*{-3em}
\sigma_{\mathrm{E}L\to\mathrm{T}L,-L-1}
&\simeq& \frac{\alpha a_{0}}{Z}\frac{1}{L(L+1)(2L+1)^{2}}
\nonumber \\
&& \times\,\left\{1-(\alpha Z)^{2}\frac{2L(2L+1)}{L^{2}-1}
\left[\psi(2L+1)-\psi(L+1)-\frac{L(L+5)}{4(2L+1)}\right]\right\}.
\label{8.20b}
\end{eqnarray}
\label{8.20}%
\end{subequations}
Hence, the sought approximate expression for the cross-susceptibility
$\sigma_{\mathrm{E}L\to\mathrm{T}L}$ is
\begin{eqnarray}
\sigma_{\mathrm{E}L\to\mathrm{T}L}
&\simeq& -\frac{\alpha a_{0}}{Z}
\frac{1}{L^{2}(L+1)}\bigg\{1-(\alpha
Z)^{2}\frac{2L^{4}-L^{3}-3L^{2}-L+1}{L(2L+1)(L^{2}-1)}
\nonumber \\
&& \times\,\bigg[\psi(2L)-\psi(L)-\frac{L(L^{3}-L^{2}-3L-5)}
{4(2L^{4}-L^{3}-3L^{2}-L+1)}\bigg]\bigg\}.
\label{8.21}
\end{eqnarray}
In the particular case of $L=1$, after the L'Hospital's rule is
applied (this is admissible since $L$ may be formally treated as a
continuous parameter, cf.\ Appendix \ref{E}) and the well-known
identities
\begin{equation}
\psi'(1)=\frac{\pi^{2}}{6},
\qquad
\psi'(2)=\frac{\pi^{2}}{6}-1
\label{8.22}
\end{equation}
are exploited, Eq.\ (\ref{8.21}) becomes
\begin{equation}
\sigma_{\mathrm{E}1\to\mathrm{T}1}\simeq-\frac{\alpha a_{0}}{Z}
\frac{1}{2}\left[1-\left(\frac{3}{4}-\frac{\pi^{2}}{18}\right)
(\alpha Z)^{2}\right].
\label{8.23}
\end{equation}
The explicit forms of the quasi-relativistic approximations to
$\sigma_{\mathrm{E}L\to\mathrm{T}L}$ for $1\leqslant L\leqslant4$,
resulting from Eqs.\ (\ref{8.23}) and (\ref{8.21}), are displayed in
Table \ref{T.15}.
\begin{center}
[Place for Table \ref{T.15}]
\end{center}
%
%\newpage
%
\section{Summary and future prospectives}
\label{IX}
\setcounter{equation}{0}
In this paper, we have considered various far- and near-field
electric and magnetic multipole moments induced in the ground state
of the Dirac one-electron atom by an external, weak, static electric
$2^{L}$-pole field. Strengths of all these induced moments have been
characterized by congruent atomic multipole susceptibilities, using
formulas brought together in Table \ref{T.16}. Table \ref{T.17} shows
how the susceptibilities in question enter the near- and far-zone
asymptotic representations of the lowest-order electric and magnetic
fields, and their potentials, generated by the atom in response to
the perturbation. For the reader's convenience, all exact closed-form
analytical expressions for the susceptibilities, derived by us in
Secs.\ \ref{III}--\ref{VIII} with the aid of the Sturmian expansion
of the Dirac--Coulomb Green function, are collected in Tables
\ref{T.18} and \ref{T.19}.
\begin{center}
[Place for Tables \ref{T.16}, \ref{T.17}, \ref{T.18} and \ref{T.19}].
\end{center}

Because of space limitations, in Tables \ref{T.2}, \ref{T.4},
\ref{T.5} and \ref{T.7} embedded in Secs.\ \ref{III}--\ref{V}, we
have provided numerical values of the far-field susceptibilities
$\alpha_{L}$ ($\equiv\alpha_{\mathrm{E}L\to\mathrm{E}L}$),
$\alpha_{\mathrm{E}L\to\mathrm{M}(L\mp1)}$ and
$\alpha_{\mathrm{E}L\to\mathrm{T}L}$, all with $1\leqslant
L\leqslant4$, only for selected values of the nuclear charge number
$Z$. A complete tabulation of values of these susceptibilities for
all integer values of $Z$ from the range $1\leqslant Z\leqslant137$
will be presented elsewhere.

There are two directions in which we would like to extend the current
research. First, we plan to analyze various electric and magnetic
moments induced in the ground state of the atom by an external, weak,
static $2^{L}$-pole \emph{magnetic\/} field; a result of such an
analysis would be a set of static \emph{magnetic\/} multipole
susceptibilities for the atomic ground state. Second, we intend to
carry out an analogous study, both for electric and magnetic
perturbing fields, for the atom in energetically excited states
belonging to the manifold characterized by the principal quantum
number $n=2$. Our preliminary insight into the latter problem shows
that such calculations, although significantly more complex than
those presented here, should nevertheless be feasible.
\section*{Acknowledgments}
We thank Dr.\ Patrycja Stefa{\'n}ska for valuable discussions.
%
%\newpage
%
\appendix
\section{Multipole polarizabilities vs.\ the second-order correction
to the atomic ground-state energy}
\label{A}
\setcounter{equation}{0}
The purpose of this appendix is to give a relationship between the
atomic multipole polarizability $\alpha_{L}$ and the second-order
correction to atomic energy due to a perturbing static electric
$2^{L}$-pole field defined in Sec.\ \ref{II}. It is assumed that
before the field was switched on, the atom had been in its ground
state.

If we go one step beyond the first-order perturbation theory used in
Sec.\ \ref{II}, the perturbed wave function of the atom may be
approximated by
\begin{equation}
\Psi(\boldsymbol{r})\simeq\Psi^{(0)}(\boldsymbol{r})
+\Psi^{(1)}(\boldsymbol{r})+\Psi^{(2)}(\boldsymbol{r}).
\label{A.1}
\end{equation}
The zeroth-order wave function and the first order correction to it
have been given in Eqs.\ (\ref{2.9}) and (\ref{2.28}), respectively;
we recall that the first-order perturbation theory has left the
coefficients $a_{\pm1/2}$ in Eq.\ (\ref{2.9}) undetermined.
Similarly, for the atomic energy in the field we may write
\begin{equation}
E\simeq E^{(0)}+E^{(1)}+E^{(2)}
\label{A.2}
\end{equation}
with $E^{(0)}$ and $E^{(1)}$ given by Eqs.\ (\ref{2.6}) and
(\ref{2.27}), respectively. Proceeding in the standard manner, from
Eqs.\ (\ref{2.4}), (\ref{A.1}) and (\ref{A.2}) we deduce that the
corrections $\Psi^{(2)}(\boldsymbol{r})$ and $E^{(2)}$ solve
\begin{equation}
\left[-\mathrm{i}c\hbar\boldsymbol{\alpha}\cdot\boldsymbol{\nabla}
+\beta m_{\mathrm{e}}c^{2}-\frac{Ze^{2}}{(4\pi\epsilon_{0})r}
-E^{(0)}\right]\Psi^{(2)}(\boldsymbol{r})
=-[V_{L}^{(1)}(\boldsymbol{r})-E^{(1)}]\Psi^{(1)}(\boldsymbol{r})
+E^{(2)}\Psi^{(0)}(\boldsymbol{r}),
\label{A.3}
\end{equation}
subject to the orthogonality constraints
\begin{equation}
\int_{\mathbb{R}^{3}}\mathrm{d}^{3}\boldsymbol{r}\:
\Psi_{m}^{(0)\dag}(\boldsymbol{r})\Psi^{(2)}(\boldsymbol{r})=0
\qquad (m=\pm{\textstyle\frac{1}{2}}).
\label{A.4}
\end{equation}
Projecting Eq.\ (\ref{A.3}) from the left onto the unperturbed basis
functions $\Psi_{\pm1/2}^{(0)}(\boldsymbol{r})$ and then making use
of Eqs.\ (\ref{2.9}), (\ref{2.12}), (\ref{2.16}), (\ref{2.27}) and
(\ref{2.28}), we arrive at the homogeneous algebraic system
\begin{equation}
\sum_{m'=-1/2}^{1/2}
\big[V_{L,mm'}^{(1,1)}-E^{(2)}\delta_{mm'}\big]a_{m'}=0
\qquad (m=\pm{\textstyle\frac{1}{2}}),
\label{A.5}
\end{equation}
in which
\begin{equation}
V_{L,mm'}^{(1,1)}=-\int_{\mathbb{R}^{3}}\mathrm{d}^{3}\boldsymbol{r}
\int_{\mathbb{R}^{3}}\mathrm{d}^{3}\boldsymbol{r}'\:
\Psi_{m}^{(0)\dag}(\boldsymbol{r})V_{L}^{(1)}(\boldsymbol{r})
\bar{G}\mbox{}^{(0)}(\boldsymbol{r},\boldsymbol{r}')
V_{L}^{(1)}(\boldsymbol{r}')\Psi_{m'}^{(0)}(\boldsymbol{r}').
\label{A.6}
\end{equation}
To simplify the expression on the right-hand side of Eq.\
(\ref{A.6}), we make use of Eqs.\ (\ref{2.3}), (\ref{2.10}),
(\ref{3.13}) and (\ref{2.20}), and also of the definitions
(\ref{3.16}) and (\ref{3.15}). This yields
\begin{eqnarray}
V_{L,mm'}^{(1,1)} &=& -(4\pi\epsilon_{0})\frac{4\pi}{2L+1}
\sum_{\substack{\kappa=-\infty \\ (\kappa\neq0)}}^{\infty}
R_{\kappa}^{(L,L)}\big(P^{(0)},Q^{(0)};P^{(0)},Q^{(0)}\big)
\sum_{M=-L}^{L}\sum_{M'=-L}^{L}
\mathcal{C}_{LM}^{(1)*}\mathcal{C}_{LM'}^{(1)*}
\nonumber \\
&& \times 
\sum_{m_{\kappa}=-|\kappa|+1/2}^{|\kappa|-1/2}
\langle\Omega_{-1m}\big|Y_{LM}\Omega_{\kappa m_{\kappa}}\rangle
\langle\Omega_{\kappa m_{\kappa}}\big|Y_{LM'}\Omega_{-1m'}\rangle.
\label{A.7}
\end{eqnarray}
Evaluation of the angular integrals with the help of Eq.\
(\ref{2.22}) casts Eq.\ (\ref{A.7}) into
\begin{eqnarray}
V_{L,mm'}^{(1,1)} &=& -(4\pi\epsilon_{0})
\sum_{\substack{\kappa=-\infty \\ (\kappa\neq0)}}^{\infty}
\frac{\delta_{\kappa L}+\delta_{\kappa,-L-1}}{2L+1}
R_{\kappa}^{(L,L)}\big(P^{(0)},Q^{(0)};P^{(0)},Q^{(0)}\big)
\nonumber \\
&& \times\bigg[\delta_{m,1/2}\delta_{m',1/2}
\sum_{M=-L}^{L}(-)^{M}\frac{\kappa+M}{2\kappa+1}\,
\mathcal{C}_{LM}^{(1)*}\mathcal{C}_{L,-M}^{(1)*}
\nonumber \\
&& \quad -\,\sgn(\kappa)\delta_{m,1/2}\delta_{m',-1/2}
\sum_{M=-L}^{L}(-)^{M}
\frac{\sqrt{(\kappa+M)(\kappa-M+1)}}{|2\kappa+1|}\,
\mathcal{C}_{LM}^{(1)*}\mathcal{C}_{L,-M+1}^{(1)*}
\nonumber \\
&& \quad -\,\sgn(\kappa)\delta_{m,-1/2}\delta_{m',1/2}
\sum_{M=-L}^{L}(-)^{M}
\frac{\sqrt{(\kappa-M)(\kappa+M+1)}}{|2\kappa+1|}\,
\mathcal{C}_{LM}^{(1)*}\mathcal{C}_{L,-M-1}^{(1)*}
\nonumber \\
&& \quad +\,\delta_{m,-1/2}\delta_{m',-1/2}
\sum_{M=-L}^{L}(-)^{M}\frac{\kappa-M}{2\kappa+1}\,
\mathcal{C}_{LM}^{(1)*}\mathcal{C}_{L,-M}^{(1)*}\bigg].
\label{A.8}
\end{eqnarray}
The four sums over $M$ may be simplified after evident symmetry
properties of their summands are taken into account. One finds that
\begin{equation}
\sum_{M=-L}^{L}(-)^{M}\frac{\kappa\pm M}{2\kappa+1}\,
\mathcal{C}_{LM}^{(1)*}\mathcal{C}_{L,-M}^{(1)*}
=\frac{|\kappa|}{2L+1}\sum_{M=-L}^{L}\mathcal{C}_{LM}^{(1)*}
\mathcal{C}_{LM}^{(1)}
\qquad (\kappa=L,-L-1)
\label{A.9}
\end{equation}
and
\begin{equation}
\sum_{M=-L}^{L}(-)^{M}
\frac{\sqrt{(\kappa\pm M)(\kappa\mp M+1)}}{|2\kappa+1|}\,
\mathcal{C}_{LM}^{(1)*}\mathcal{C}_{L,-M\pm1}^{(1)*}=0
\qquad (\kappa=L,-L-1).
\label{A.10}
\end{equation}
Hence, we deduce that
\begin{eqnarray}
V_{L,mm'}^{(1,1)} &=& -\delta_{mm'}(4\pi\epsilon_{0})
\bigg[\frac{L}{(2L+1)^{2}}
R_{L}^{(L,L)}\big(P^{(0)},Q^{(0)};P^{(0)},Q^{(0)}\big)
\nonumber \\
&& +\,\frac{L+1}{(2L+1)^{2}}
R_{-L-1}^{(L,L)}\big(P^{(0)},Q^{(0)};P^{(0)},Q^{(0)}\big)\bigg]
\sum_{M=-L}^{L}\mathcal{C}_{LM}^{(1)*}\mathcal{C}_{LM}^{(1)}.
\label{A.11}
\end{eqnarray}
The matrix formed by the elements $V_{L,mm'}^{(1,1)}$ is thus seen to
be a multiple of the unit $2\times2$ matrix. Recalling Eqs.\
(\ref{3.24}) and (\ref{3.25}), we are led to the conclusion that
application of the second-order perturbation theory does not remove
degeneracy of the atomic ground state and that the second-order
correction to energy of that state may be written as
\begin{equation}
E^{(2)}=-\frac{1}{2}(4\pi\epsilon_{0})\alpha_{L}
\boldsymbol{\mathsf{C}}_{L}^{(1)}
\cdot\boldsymbol{\mathsf{C}}_{L}^{(1)}.
\label{A.12}
\end{equation}
This result is very well known in the dipole ($L=1$) case.
%
%\newpage
%
\section{Far-field and near-field expansions of the magnetic vector
potential and the magnetic induction} 
\label{B}
\setcounter{equation}{0}
On our way to a deeper understanding how various sorts of multipole
moment tensors arise in the far- and near-field asymptotic expansions
of the magnetic vector potential, we have much benefited from
studying the methodological paper of Agre \cite{Agre11}, which we
wholeheartedly recommend to all readers interested in the subject.
\subsection{General considerations}
\label{B.I}
For a given stationary solenoidal current distribution
$\boldsymbol{j}(\boldsymbol{r})$, the magnetostatic vector potential
may be found from the formula
\begin{equation}
\boldsymbol{A}(\boldsymbol{r})=\frac{\mu_{0}}{4\pi}
\int_{\mathbb{R}^{3}}\mathrm{d}^{3}\boldsymbol{r}'\:
\frac{\boldsymbol{j}(\boldsymbol{r}')}
{|\boldsymbol{r}-\boldsymbol{r}'|}.
\label{B.1}
\end{equation}
Exploiting the multipole expansion
\begin{equation}
\frac{1}{|\boldsymbol{r}-\boldsymbol{r}'|}
=\sum_{L=0}^{\infty}\sum_{M_{L}=-L}^{L}
\frac{4\pi}{2L+1}\frac{r_{<}^{L}}{r_{>}^{L+1}}
Y_{LM_{L}}^{*}(\boldsymbol{n}_{r})
Y_{LM_{L}}(\boldsymbol{n}_{r}^{\prime}),
\label{B.2}
\end{equation}
one finds that in the far-field ($r\to\infty$) and near-field
($r\to0$) regions the vector potential
$\boldsymbol{A}(\boldsymbol{r})$ behaves as
\begin{equation}
\boldsymbol{A}(\boldsymbol{r})\stackrel{r\to\infty}{\longrightarrow}
\sum_{L=0}^{\infty}\boldsymbol{A}^{LL}(\boldsymbol{r})
\label{B.3}
\end{equation}
and
\begin{equation}
\boldsymbol{A}(\boldsymbol{r})\stackrel{r\to0}{\longrightarrow}
\sum_{L=0}^{\infty}\boldsymbol{A}^{L,-L-1}(\boldsymbol{r}),
\label{B.4}
\end{equation}
respectively, where
\begin{equation}
\boldsymbol{A}^{L\lambda}(\boldsymbol{r})
=\frac{\mu_{0}}{4\pi}\frac{4\pi}{2L+1}\,r^{-\lambda-1}
\sum_{M_{L}=-L}^{L}Y_{LM_{L}}^{*}(\boldsymbol{n}_{r})
\int_{\mathbb{R}^{3}}\mathrm{d}^{3}\boldsymbol{r}'\:
r^{\prime\lambda}Y_{LM_{L}}(\boldsymbol{n}_{r}^{\prime})
\boldsymbol{j}(\boldsymbol{r}')
\qquad (\lambda=L,-L-1).
\label{B.5}
\end{equation}
In the next step, we make use of the closure identity
\begin{equation}
\sum_{m=-1}^{1}\boldsymbol{e}_{m}^{*}\boldsymbol{e}_{m}
=\boldsymbol{\mathsf{I}}
\label{B.6}
\end{equation}
for the unit vectors of the cyclic basis, which gives
\begin{eqnarray}
\boldsymbol{A}^{L\lambda}(\boldsymbol{r})
&=& \frac{\mu_{0}}{4\pi}\frac{4\pi}{2L+1}\,r^{-\lambda-1}
\sum_{M_{L}=-L}^{L}\sum_{m=-1}^{1}
Y_{LM_{L}}^{*}(\boldsymbol{n}_{r})\boldsymbol{e}_{m}^{*}
\int_{\mathbb{R}^{3}}\mathrm{d}^{3}\boldsymbol{r}'\:
r^{\prime\lambda}Y_{LM_{L}}(\boldsymbol{n}_{r}^{\prime})
\boldsymbol{e}_{m}\cdot\boldsymbol{j}(\boldsymbol{r}')
\nonumber \\
&& \hspace*{20em} (\lambda=L,-L-1).
\label{B.7}
\end{eqnarray}
The product $Y_{LM_{L}}(\boldsymbol{n}_{r})\boldsymbol{e}_{m}$,
appearing in Eq.\ (\ref{B.7}), may be expanded in the basis of
vector spherical harmonics as
\begin{equation}
Y_{LM_{L}}(\boldsymbol{n}_{r})\boldsymbol{e}_{m}
=\sum_{J=|L-1|}^{L+1}\sum_{M_{J}=-J}^{J}
\langle LM_{L}1m\big|JM_{J}\rangle
\boldsymbol{Y}_{JM_{J}}^{L}(\boldsymbol{n}_{r}).
\label{B.8}
\end{equation}
Inserting this relation twice into Eq.\ (\ref{B.7}) and exploiting
the orthonormality property
\begin{equation}
\sum_{M_{a}=-L_{a}}^{L_{a}}\sum_{M_{b}=-L_{b}}^{L_{b}}
\langle L_{a}M_{a}L_{b}M_{b}\big|JM_{J}\rangle
\langle L_{a}M_{a}L_{b}M_{b}\big|J'M_{J'}\rangle
=\delta_{JJ'}\delta_{M_{J}M_{J'}}
\label{B.9}
\end{equation}
of the Clebsch--Gordan coefficients yields
\begin{eqnarray}
\boldsymbol{A}^{L\lambda}(\boldsymbol{r})
&=& \frac{\mu_{0}}{4\pi}\frac{4\pi}{2L+1}\,r^{-\lambda-1}
\sum_{J=|L-1|}^{L+1}\sum_{M_{J}=-J}^{J}
\boldsymbol{Y}_{JM_{J}}^{L*}(\boldsymbol{n}_{r})
\int_{\mathbb{R}^{3}}\mathrm{d}^{3}\boldsymbol{r}'\:r^{\prime\lambda}
\boldsymbol{Y}_{JM_{J}}^{L}(\boldsymbol{n}_{r}^{\prime})
\cdot\boldsymbol{j}(\boldsymbol{r}')
\nonumber \\
&& \hspace*{20em} (\lambda=L,-L-1).
\label{B.10}
\end{eqnarray}
If we define
\begin{equation}
\mathcal{Z}_{JM_{J}}^{L\lambda}=\sqrt{\frac{4\pi}{2L+1}}
\int_{\mathbb{R}^{3}}\mathrm{d}^{3}\boldsymbol{r}\:r^{\lambda}
\boldsymbol{Y}_{JM_{J}}^{L}(\boldsymbol{n}_{r})
\cdot\boldsymbol{j}(\boldsymbol{r})
\qquad (|L-1|\leqslant J\leqslant L+1),
\label{B.11}
\end{equation}
Eq.\ (\ref{B.10}) becomes
\begin{equation}
\boldsymbol{A}^{L\lambda}(\boldsymbol{r})
=\frac{\mu_{0}}{4\pi}\sqrt{\frac{4\pi}{2L+1}}\,
r^{-\lambda-1}\sum_{J=|L-1|}^{L+1}\sum_{M_{J}=-J}^{J}
\mathcal{Z}_{JM_{J}}^{L\lambda}
\boldsymbol{Y}_{JM_{J}}^{L*}(\boldsymbol{n}_{r}).
\label{B.12}
\end{equation}

We shall show that the coefficients $\mathcal{Z}_{JM_{J}}^{L\lambda}$
are components of a rank-$J$ irreducible spherical tensor
$\boldsymbol{\mathsf{Z}}_{J}^{L\lambda}$. Using the relation
\begin{equation}
\boldsymbol{Y}_{JM_{J}}^{L}(\boldsymbol{n}_{r})
=\sum_{M_{L}=-L}^{L}\sum_{m=-1}^{1}
\langle LM_{L}1m\big|JM_{J}\rangle
Y_{LM_{L}}(\boldsymbol{n}_{r})\boldsymbol{e}_{m},
\label{B.13}
\end{equation}
reciprocal to the one in Eq.\ (\ref{B.8}), and the fact that the
$m$-th cyclic component of the vector
$\boldsymbol{j}(\boldsymbol{r})$ is given by
\begin{equation}
j_{m}(\boldsymbol{r})
=\boldsymbol{e}_{m}\cdot\boldsymbol{j}(\boldsymbol{r}),
\label{B.14}
\end{equation}
we obtain
\begin{eqnarray}
\boldsymbol{Y}_{JM_{J}}^{L}(\boldsymbol{n}_{r})
\cdot\boldsymbol{j}(\boldsymbol{r})
&=& \sum_{M_{L}=-L}^{L}\sum_{m=-1}^{1}
\langle LM_{L}1m\big|JM_{J}\rangle 
Y_{LM_{L}}(\boldsymbol{n}_{r})j_{m}(\boldsymbol{r})
\nonumber \\
&=& \big\{\boldsymbol{\mathsf{Y}}_{L}(\boldsymbol{n}_{r})
\otimes\boldsymbol{j}(\boldsymbol{r})\big\}_{JM_{J}},
\label{B.15}
\end{eqnarray}
and further
\begin{equation}
\mathcal{Z}_{JM_{J}}^{L\lambda}=\sqrt{\frac{4\pi}{2L+1}}
\int_{\mathbb{R}^{3}}\mathrm{d}^{3}\boldsymbol{r}\:r^{\lambda}
\big\{\boldsymbol{\mathsf{Y}}_{L}(\boldsymbol{n}_{r})
\otimes\boldsymbol{j}(\boldsymbol{r})\big\}_{JM_{J}},
\label{B.16}
\end{equation}
which proves the statement.

Concluding this section, we observe that since in the definition
(\ref{B.11}) $J$ varies in the range $|L-1|\leqslant J\leqslant L+1$,
for $L=0$ there are only two tensors in the
$\boldsymbol{\mathsf{Z}}_{J}^{L\lambda}$ family:
$\boldsymbol{\mathsf{Z}}_{1}^{00}$ and
$\boldsymbol{\mathsf{Z}}_{1}^{0,-1}$, while for $L\geqslant1$ their
number formally increases to six:
$\boldsymbol{\mathsf{Z}}_{L-1}^{LL}$,
$\boldsymbol{\mathsf{Z}}_{L}^{LL}$,
$\boldsymbol{\mathsf{Z}}_{L+1}^{LL}$ and
$\boldsymbol{\mathsf{Z}}_{L-1}^{L,-L-1}$,
$\boldsymbol{\mathsf{Z}}_{L}^{L,-L-1}$,
$\boldsymbol{\mathsf{Z}}_{L+1}^{L,-L-1}$. However, in Secs.\
\ref{B.II} and \ref{B.III} we shall show that the tensors
$\boldsymbol{\mathsf{Z}}_{L+1}^{LL}$ and
$\boldsymbol{\mathsf{Z}}_{L-1}^{L,-L-1}$ vanish identically.
\subsection{The far-field case ($\lambda=L$). The tensors
$\boldsymbol{\mathsf{M}}_{L}$ and $\boldsymbol{\mathsf{T}}_{L-1}$}
\label{B.II}
In the far-field zone, the asymptotics of the magnetic vector
potential is that displayed in Eq.\ (\ref{B.3}), with
$\boldsymbol{A}^{LL}(\boldsymbol{r})$ given by Eq.\ (\ref{B.12})
specialized to the case $\lambda=L$. Components of the pertinent
far-field tensors $\boldsymbol{\mathsf{Z}}_{JM_{J}}^{LL}$ may be
found from Eq.\ (\ref{B.11}), in which one sets $\lambda=L$.

Consider the tensor $\boldsymbol{\mathsf{Z}}_{L+1,M_{L+1}}^{LL}$. In
accordance with what has been said above, its components are given by
\begin{equation}
\mathcal{Z}_{L+1,M_{L+1}}^{LL}=\sqrt{\frac{4\pi}{2L+1}}
\int_{\mathbb{R}^{3}}\mathrm{d}^{3}\boldsymbol{r}\:
r^{L}\boldsymbol{Y}_{L+1,M_{L+1}}^{L}(\boldsymbol{n}_{r})
\cdot\boldsymbol{j}(\boldsymbol{r}).
\label{B.17}
\end{equation}
However, from the differential relation \cite[Eq.\ (5.8.9)]{Vars75}
\begin{eqnarray}
\boldsymbol{\nabla}[r^{\lambda+1}Y_{JM_{J}}(\boldsymbol{n}_{r})]
&=& (J-\lambda-1)\sqrt{\frac{J+1}{2J+1}}\,r^{\lambda}
\boldsymbol{Y}_{JM_{J}}^{J+1}(\boldsymbol{n}_{r})
+(J+\lambda+2)\sqrt{\frac{J}{2J+1}}\,r^{\lambda}
\boldsymbol{Y}_{JM_{J}}^{J-1}(\boldsymbol{n}_{r})
\nonumber \\
&&
\label{B.18}
\end{eqnarray}
it follows that
\begin{equation}
r^{L}\boldsymbol{Y}_{L+1,M_{L+1}}^{L}(\boldsymbol{n}_{r})
=\frac{\boldsymbol{\nabla}[r^{L+1}
Y_{L+1,M_{L+1}}(\boldsymbol{n}_{r})]}{\sqrt{(L+1)(2L+3)}}.
\label{B.19}
\end{equation}
Consequently, the integrand in Eq.\ (\ref{B.17}) may be rewritten as
\begin{equation}
r^{L}\boldsymbol{Y}_{L+1,M_{L+1}}^{L}(\boldsymbol{n}_{r})
\cdot\boldsymbol{j}(\boldsymbol{r})
=\frac{\boldsymbol{\nabla}\cdot[r^{L+1}
Y_{L+1,M_{L+1}}(\boldsymbol{n}_{r})
\boldsymbol{j}(\boldsymbol{r})]}{\sqrt{(L+1)(2L+3)}}
-\frac{r^{L+1}Y_{L+1,M_{L+1}}(\boldsymbol{n}_{r})
\boldsymbol{\nabla}\cdot\boldsymbol{j}(\boldsymbol{r})}
{\sqrt{(L+1)(2L+3)}}.
\label{B.20}
\end{equation}
Since, by assumption, the current is solenoidal, the second term on
the right-hand side of Eq.\ (\ref{B.20}) vanishes. Hence, replacing
the integrand in Eq.\ (\ref{B.17}) by the first term on the
right-hand side of Eq.\ (\ref{B.20}), and then using the Gauss
divergence theorem, casts the former equation into
\begin{equation}
\mathcal{Z}_{L+1,M_{L+1}}^{LL}=\sqrt{\frac{4\pi}{(L+1)(2L+1)(2L+3)}}
\lim_{r\to\infty}\oint_{4\pi}\mathrm{d}^{2}\boldsymbol{n}_{r}\:
r^{L+3}Y_{L+1,M_{L+1}}(\boldsymbol{n}_{r})
\boldsymbol{n}_{r}\cdot\boldsymbol{j}(\boldsymbol{r}).
\label{B.21}
\end{equation}
Hence, provided that the current obeys the asymptotic constraint
\begin{equation}
\lim_{r\to\infty}r^{L+3}
\boldsymbol{n}_{r}\cdot\boldsymbol{j}(\boldsymbol{r})=0,
\label{B.22}
\end{equation}
one finds that
\begin{equation}
\mathcal{Z}_{L+1,M_{L+1}}^{LL}=0,
\label{B.23}
\end{equation}
i.e., the tensor $\boldsymbol{\mathsf{Z}}_{L+1}^{LL}$ vanishes
identically. In result, Eqs.\ (\ref{B.3}) and (\ref{B.12}) may be
combined into
\begin{equation}
\boldsymbol{A}(\boldsymbol{r})\stackrel{r\to\infty}{\longrightarrow}
\frac{\mu_{0}}{4\pi}\sum_{L=1}^{\infty}
\sqrt{\frac{4\pi}{2L+1}}\,r^{-L-1}
\sum_{J=|L-1|}^{L}\sum_{M_{J}=-J}^{J}\mathcal{Z}_{JM_{J}}^{LL}
\boldsymbol{Y}_{JM_{J}}^{L*}(\boldsymbol{n}_{r})
\label{B.24}
\end{equation}
(the sum starts from $L=1$, as we have proved above that the tensor
$\boldsymbol{\mathsf{Z}}_{1}^{00}$ vanishes).

In the literature, instead of the tensors
$\boldsymbol{\mathsf{Z}}_{L}^{LL}$ and
$\boldsymbol{\mathsf{Z}}_{L-1}^{LL}$, one encounters the tensors
$\boldsymbol{\mathsf{M}}_{L}$ and $\boldsymbol{\mathsf{T}}_{L-1}$,
components of which, given by
\begin{equation}
\mathcal{M}_{LM_{L}}=-\mathrm{i}\,\sqrt{\frac{4\pi L}{(L+1)(2L+1)}}
\int_{\mathbb{R}^{3}}\mathrm{d}^{3}\boldsymbol{r}\:
r^{L}\boldsymbol{Y}_{LM_{L}}^{L}(\boldsymbol{n}_{r})
\cdot\boldsymbol{j}(\boldsymbol{r})
\label{B.25}
\end{equation}
and
\begin{equation}
\mathcal{T}_{L-1,M_{L-1}}=-\frac{1}{2L+1}\sqrt{\frac{4\pi}{L}}
\int_{\mathbb{R}^{3}}\mathrm{d}^{3}\boldsymbol{r}\:
r^{L}\boldsymbol{Y}_{L-1,M_{L-1}}^{L}(\boldsymbol{n}_{r})
\cdot\boldsymbol{j}(\boldsymbol{r}),
\label{B.26}
\end{equation}
are related to those of $\boldsymbol{\mathsf{Z}}_{L}^{LL}$ and
$\boldsymbol{\mathsf{Z}}_{L-1}^{LL}$ through
\begin{equation}
\mathcal{M}_{LM_{L}}=-\mathrm{i}\,\sqrt{\frac{L}{L+1}}\,
\mathcal{Z}_{LM_{L}}^{LL}
\label{B.27}
\end{equation}
and
\begin{equation}
\mathcal{T}_{L-1,M_{L-1}}=-\frac{1}{\sqrt{L(2L+1)}}\,
\mathcal{Z}_{L-1,M_{L-1}}^{LL}.
\label{B.28}
\end{equation}
The tensor $\boldsymbol{\mathsf{M}}_{L}$ is the plain $2^{L}$-pole
magnetic moment, and the tensor $\boldsymbol{\mathsf{T}}_{L-1}$ is
named the $2^{L-1}$-pole magnetic toroidal moment \cite{Dubo90}
(various integral representations of components of the $2^{L}$-pole
moment $\boldsymbol{\mathsf{T}}_{L}$ are derived in Appendix
\ref{C}). With the use of components of these two more common
tensors, the expansion (\ref{B.24}) is replaced by
\begin{eqnarray}
\boldsymbol{A}(\boldsymbol{r})
&\stackrel{r\to\infty}{\longrightarrow}&
\frac{\mu_{0}}{4\pi}\sum_{L=1}^{\infty}
\sqrt{\frac{4\pi}{2L+1}}\,r^{-L-1}
\Bigg[\mathrm{i}\,\sqrt{\frac{L+1}{L}}\sum_{M_{L}=-L}^{L}
\mathcal{M}_{LM_{L}}
\boldsymbol{Y}_{LM_{L}}^{L*}(\boldsymbol{n}_{r})
\nonumber \\
&& \quad -\,(1-\delta_{L1})\sqrt{L(2L+1)}
\sum_{M_{L-1}=-L+1}^{L-1}\mathcal{T}_{L-1,M_{L-1}}
\boldsymbol{Y}_{L-1,M_{L-1}}^{L*}(\boldsymbol{n}_{r})\Bigg].
\label{B.29}
\end{eqnarray}
The factor $1-\delta_{L1}$ has been inserted into the second term in
the square bracket since, as it will be shown at the end of Appendix
\ref{C}, it holds that $\mathcal{T}_{00}=0$.

It remains to derive the expression for the asymptotic representation
of the magnetic induction with the aid of the well-known formula
\begin{equation}
\boldsymbol{B}(\boldsymbol{r})
=\boldsymbol{\nabla}\times\boldsymbol{A}(\boldsymbol{r}).
\label{B.30}
\end{equation}
If we exploit the differential relation (\ref{B.18}) with
$\lambda=-L-1$ and $J=L-1$, this allows us to write the product
$r^{-L-1}\boldsymbol{Y}_{L-1,M_{L-1}}^{L}(\boldsymbol{n}_{r})$ as a
gradient of a scalar field:
\begin{equation}
r^{-L-1}\boldsymbol{Y}_{L-1,M_{L-1}}^{L}(\boldsymbol{n}_{r})
=\frac{\boldsymbol{\nabla}[r^{-L}Y_{L-1,M_{L-1}}(\boldsymbol{n}_{r})]}
{\sqrt{L(2L-1)}}.
\label{B.31}
\end{equation}
This immediately implies that
\begin{equation}
\boldsymbol{\nabla}\times[r^{-L-1}
\boldsymbol{Y}_{L-1,M_{L-1}}^{L*}(\boldsymbol{n}_{r})]=\boldsymbol{0}.
\label{B.32}
\end{equation}
On the other hand, from the identity \cite[Eq.\ (7.3.55)]{Vars75}
\begin{eqnarray}
\boldsymbol{\nabla}
\times[f(r)\boldsymbol{Y}_{LM_{L}}^{L}(\boldsymbol{n}_{r})]
&=& \mathrm{i}\,\sqrt{\frac{L}{2L+1}}
\left(\frac{\partial}{\partial r}-\frac{L}{r}\right)f(r)
\boldsymbol{Y}_{LM_{L}}^{L+1}(\boldsymbol{n}_{r})
\nonumber \\
&& +\,\mathrm{i}\,\sqrt{\frac{L+1}{2L+1}}
\left(\frac{\partial}{\partial r}+\frac{L+1}{r}\right)f(r)
\boldsymbol{Y}_{LM_{L}}^{L-1}(\boldsymbol{n}_{r})
\label{B.33}
\end{eqnarray}
one infers that
\begin{equation}
\boldsymbol{\nabla}\times[r^{-L-1}
\boldsymbol{Y}_{LM_{L}}^{L*}(\boldsymbol{n}_{r})]
=\mathrm{i}\,\sqrt{L(2L+1)}\,r^{-L-2}
\boldsymbol{Y}_{LM_{L}}^{L+1*}(\boldsymbol{n}_{r}).
\label{B.34}
\end{equation}
On combining Eq.\ (\ref{B.30}) with Eqs.\ (\ref{B.29}), (\ref{B.32})
and (\ref{B.34}), one arrives at the sought asymptotic representation
\begin{equation}
\boldsymbol{B}(\boldsymbol{r})\stackrel{r\to\infty}{\longrightarrow}
-\frac{\mu_{0}}{4\pi}\sum_{L=1}^{\infty}\sqrt{4\pi(L+1)}\,r^{-L-2}
\sum_{M_{L}=-L}^{L}\mathcal{M}_{LM_{L}}
\boldsymbol{Y}_{LM_{L}}^{L+1*}(\boldsymbol{n}_{r})
\label{B.35}
\end{equation}
of the magnetic induction. It is seen that components of the magnetic
toroidal moments $\boldsymbol{\mathsf{T}}_{L}$ do not appear in Eq.\
(\ref{B.35}).
\subsection{The near-field case ($\lambda=-L-1$). The tensors
$\boldsymbol{\mathsf{N}}_{L}$ and $\boldsymbol{\mathsf{U}}_{L+1}$}
\label{B.III}
In the near-field region, the asymptotic representation of the vector
potential may be derived from Eqs.\ (\ref{B.4}) and (\ref{B.12}), the
latter with $\lambda=-L-1$. Components of the tensors
$\boldsymbol{\mathsf{Z}}_{J}^{L,-L-1}$ are given by Eq.\
(\ref{B.11}), with $\lambda$ specialized as above. Then, in complete
analogy with what has been presented above, the use of the identity
(\ref{B.31}) leads to the inference that
\begin{equation}
\boldsymbol{\mathsf{Z}}_{L-1}^{L,-L-1}=\boldsymbol{\mathsf{0}},
\label{B.36}
\end{equation}
provided that the current is constrained to obey
\begin{equation}
\lim_{r\to0}r^{-L+2}
\boldsymbol{n}_{r}\cdot\boldsymbol{j}(\boldsymbol{r})=0.
\label{B.37}
\end{equation}
If, pursuing further the analogy with the material of Sec.\
\ref{B.II}, we introduce the tensors $\boldsymbol{\mathsf{N}}_{L}$
and $\boldsymbol{\mathsf{U}}_{L+1}$ with components
\begin{equation}
\mathcal{N}_{LM_{L}}=\mathrm{i}\,\sqrt{\frac{4\pi(L+1)}{L(2L+1)}}
\int_{\mathbb{R}^{3}}\mathrm{d}^{3}\boldsymbol{r}\:
r^{-L-1}\boldsymbol{Y}_{LM_{L}}^{L}(\boldsymbol{n}_{r})
\cdot\boldsymbol{j}(\boldsymbol{r})
\label{B.38}
\end{equation}
and
\begin{equation}
\mathcal{U}_{L+1,M_{L+1}}=-\frac{1}{2L+1}\sqrt{\frac{4\pi}{L+1}}
\int_{\mathbb{R}^{3}}\mathrm{d}^{3}\boldsymbol{r}\:
r^{-L-1}\boldsymbol{Y}_{L+1,M_{L+1}}^{L}(\boldsymbol{n}_{r})
\cdot\boldsymbol{j}(\boldsymbol{r}),
\label{B.39}
\end{equation}
respectively, related to those of
$\boldsymbol{\mathsf{Z}}_{L}^{L,-L-1}$ and
$\boldsymbol{\mathsf{Z}}_{L+1}^{L,-L-1}$ through
\begin{equation}
\mathcal{N}_{LM_{L}}=\mathrm{i}\,\sqrt{\frac{L+1}{L}}\,
\mathcal{Z}_{LM_{L}}^{L,-L-1}
\label{B.40}
\end{equation}
and
\begin{equation}
\mathcal{U}_{L+1,M_{L+1}}=-\frac{1}{\sqrt{(L+1)(2L+1)}}\,
\mathcal{Z}_{L+1,M_{L+1}}^{L,-L-1},
\label{B.41}
\end{equation}
after some algebra we arrive at the following near-field limit for
the vector potential $\boldsymbol{A}(\boldsymbol{r})$:
\begin{eqnarray}
\boldsymbol{A}(\boldsymbol{r})
&\stackrel{r\to0}{\longrightarrow}&
-\frac{\mu_{0}}{4\pi}\sum_{L=0}^{\infty}\sqrt{\frac{4\pi}{2L+1}}\,
r^{L}\Bigg[(1-\delta_{L0})\mathrm{i}\,\sqrt{\frac{L}{L+1}}
\sum_{M_{L}=-L}^{L}\mathcal{N}_{LM_{L}}
\boldsymbol{Y}_{LM_{L}}^{L*}(\boldsymbol{n}_{r})
\nonumber \\
&& +\,\sqrt{(L+1)(2L+1)}
\sum_{M_{L+1}=-L-1}^{L+1}\mathcal{U}_{L+1,M_{L+1}}
\boldsymbol{Y}_{L+1,M_{L+1}}^{L*}(\boldsymbol{n}_{r})\Bigg].
\label{B.42}
\end{eqnarray}
The factor $1-\delta_{L0}$ has been inserted into the first term in
the square bracket since Eq.\ (\ref{B.12}) implies that the only
non-zero contribution to $\boldsymbol{A}^{0,-1}(\boldsymbol{r})$
comes from the term involving components of the tensor
$\boldsymbol{\mathsf{Z}}_{1}^{0,-1}$ [or equivalently, by virtue of
Eq.\ (\ref{B.41}), the tensor $\boldsymbol{\mathsf{U}}_{1}$].

The near-field multipole expansion of the magnetic induction is
obtained from Eqs.\ (\ref{B.30}) and (\ref{B.42}), with the aid of
the curl identities
\begin{equation}
\boldsymbol{\nabla}\times[r^{L}\boldsymbol{Y}_{L+1,M_{L+1}}^{L*}
(\boldsymbol{n}_{r})]=\boldsymbol{0}
\label{B.43}
\end{equation}
[cf.\ Eq.\ (\ref{B.19})] and
\begin{equation}
\boldsymbol{\nabla}\times[r^{L}
\boldsymbol{Y}_{LM_{L}}^{L*}(\boldsymbol{n}_{r})]
=-\mathrm{i}\,\sqrt{(L+1)(2L+1)}\,r^{L-1}
\boldsymbol{Y}_{LM_{L}}^{L-1*}(\boldsymbol{n}_{r})
\label{B.44}
\end{equation}
[cf.\ Eq.\ (\ref{B.33})]. The required result is
\begin{equation}
\boldsymbol{B}(\boldsymbol{r})\stackrel{r\to0}{\longrightarrow}
-\frac{\mu_{0}}{4\pi}\sum_{L=1}^{\infty}\sqrt{4\pi L}\,
r^{L-1}\sum_{M_{L}=-L}^{L}\mathcal{N}_{LM_{L}}
\boldsymbol{Y}_{LM_{L}}^{L-1*}(\boldsymbol{n}_{r}).
\label{B.45}
\end{equation}
%
%\newpage
%
\section{Alternative integral representations of the far-field
magnetic toroidal multipole moments $\boldsymbol{\mathsf{T}}_{L}$} 
\label{C}
\setcounter{equation}{0}
It follows from the material presented in Appendix \ref{B} that for a
given sourceless stationary current distribution
$\boldsymbol{j}(\boldsymbol{r})$, spherical components of the
magnetic toroidal $2^{L}$-pole moment $\boldsymbol{\mathsf{T}}_{L}$
may be defined as\footnote{~The components of the magnetic toroidal
multipole moments defined in Refs.\ \cite{Agre11,Dubo90} are complex
conjugates of ours. Moreover, in Refs.\ \cite{Agre11,Dubo90} the
Gauss system of units was used, while in this paper we conform to the
International System of Units; consequently, we omit the factor $1/c$
in the definition of $\mathcal{T}_{LM}$.}
\begin{equation}
\mathcal{T}_{LM}
=-\frac{1}{2L+3}\sqrt{\frac{4\pi}{L+1}}
\int_{\mathbb{R}^{3}}\mathrm{d}^{3}\boldsymbol{r}\:r^{L+1}
\boldsymbol{Y}_{LM}^{L+1}(\boldsymbol{n}_{r})
\cdot\boldsymbol{j}(\boldsymbol{r}),
\label{C.1}
\end{equation}
where $\boldsymbol{Y}_{JM_{J}}^{L}(\boldsymbol{n}_{r})$ denotes the
vector spherical harmonic (\ref{B.13}), or equivalently as
\begin{equation}
\mathcal{T}_{LM}
=-\frac{1}{2L+3}\sqrt{\frac{4\pi}{L+1}}
\int_{\mathbb{R}^{3}}\mathrm{d}^{3}\boldsymbol{r}\:r^{L+1}
\big\{\boldsymbol{\mathsf{Y}}_{L+1}(\boldsymbol{n}_{r})
\otimes\boldsymbol{j}(\boldsymbol{r})\big\}_{LM}
\label{C.2}
\end{equation}
(observe that the irreducible tensor product appearing in the above
equation is commutative).

Consider now the identity \cite[Eq.\ (5.8.9)]{Vars75}
\begin{eqnarray}
\boldsymbol{\nabla}[f(r)Y_{LM}(\boldsymbol{n}_{r})]
&=& -\,\sqrt{\frac{L+1}{2L+1}}
\left(\frac{\partial}{\partial r}-\frac{L}{r}\right)
f(r)\boldsymbol{Y}_{LM}^{L+1}(\boldsymbol{n}_{r})
\nonumber \\
&& +\,\sqrt{\frac{L}{2L+1}}
\left(\frac{\partial}{\partial r}+\frac{L+1}{r}\right)
f(r)\boldsymbol{Y}_{LM}^{L-1}(\boldsymbol{n}_{r}).
\label{C.3}
\end{eqnarray}
In the particular case $f(r)=r^{L+2}$, Eq.\ (\ref{C.3}) yields
\begin{equation}
\boldsymbol{\nabla}[r^{L+2}Y_{LM}(\boldsymbol{n}_{r})]
=-2\sqrt{\frac{L+1}{2L+1}}\,r^{L+1}
\boldsymbol{Y}_{LM}^{L+1}(\boldsymbol{n}_{r})
+(2L+3)\sqrt{\frac{L}{2L+1}}\,r^{L+1}
\boldsymbol{Y}_{LM}^{L-1}(\boldsymbol{n}_{r}).
\label{C.4}
\end{equation}
Hence, it follows that Eq.\ (\ref{C.1}) may be rewritten as
\begin{eqnarray}
&& \hspace*{-3em}
\mathcal{T}_{LM}
=\frac{\sqrt{\pi(2L+1)}}{(L+1)(2L+3)}
\int_{\mathbb{R}^{3}}\mathrm{d}^{3}\boldsymbol{r}\:
\boldsymbol{j}(\boldsymbol{r})\cdot\boldsymbol{\nabla}
[r^{L+2}Y_{LM}(\boldsymbol{n}_{r})]
-\frac{\sqrt{\pi L}}{L+1}
\int_{\mathbb{R}^{3}}\mathrm{d}^{3}\boldsymbol{r}\:r^{L+1}
\boldsymbol{Y}_{LM}^{L-1}(\boldsymbol{n}_{r})
\cdot\boldsymbol{j}(\boldsymbol{r}).
\nonumber \\
&&
\label{C.5}
\end{eqnarray}
Let us transform the first integrand as follows:
\begin{equation}
\boldsymbol{j}(\boldsymbol{r})\cdot\boldsymbol{\nabla}
[r^{L+2}Y_{LM}(\boldsymbol{n}_{r})]
=\boldsymbol{\nabla}\cdot[r^{L+2}
Y_{LM}(\boldsymbol{n}_{r})\boldsymbol{j}(\boldsymbol{r})]
-r^{L+2}Y_{LM}(\boldsymbol{n}_{r})
\boldsymbol{\nabla}\cdot\boldsymbol{j}(\boldsymbol{r}).
\label{C.6}
\end{equation}
As we have assumed that the current is sourceless, one has
$\boldsymbol{\nabla}\cdot\boldsymbol{j}(\boldsymbol{r})=0$ and the
second term on the right-hand side of the above equation is zero.
Furthermore, if the current density obeys the constraint
\begin{equation}
\lim_{r\to\infty}r^{L+4}
\boldsymbol{n}_{r}\cdot\boldsymbol{j}(\boldsymbol{r})=0
\label{C.7}
\end{equation}
(which is certainly the case for atomic currents which vanish
exponentially at infinity), application of the Gauss' integral
theorem to the first term on the right-hand side of Eq.\ (\ref{C.6})
leads to the inference that the first integral on the right-hand side
of Eq.\ (\ref{C.5}) vanishes. In that way, we have proved that
$\mathcal{T}_{LM}$, defined in Eq.\ (\ref{C.1}) in terms of the
vector harmonic $\boldsymbol{Y}_{LM}^{L+1}(\boldsymbol{n}_{r})$, may
be equivalently expressed as
\begin{equation}
\mathcal{T}_{LM}=-\frac{\sqrt{\pi L}}{L+1}
\int_{\mathbb{R}^{3}}\mathrm{d}^{3}\boldsymbol{r}\:r^{L+1}
\boldsymbol{Y}_{LM}^{L-1}(\boldsymbol{n}_{r})
\cdot\boldsymbol{j}(\boldsymbol{r}).
\label{C.8}
\end{equation}
Multiplying Eq.\ (\ref{C.8}) by $\eta\in\mathbb{C}$ and Eq.\
(\ref{C.1}) by $1-\eta$, and then adding, we find
\begin{equation}
\mathcal{T}_{LM}=-\sqrt{\frac{\pi}{L+1}}
\int_{\mathbb{R}^{3}}\mathrm{d}^{3}\boldsymbol{r}\:r^{L+1}
\left[\eta\sqrt{\frac{L}{L+1}}\,
\boldsymbol{Y}_{LM}^{L-1}(\boldsymbol{n}_{r})
+(1-\eta)\frac{2}{2L+3}
\boldsymbol{Y}_{LM}^{L+1}(\boldsymbol{n}_{r})\right]
\cdot\boldsymbol{j}(\boldsymbol{r}).
\label{C.9}
\end{equation}
Playing with the value of $\eta$, the above general formula may be
used to obtain particular expressions for $\mathcal{T}_{LM}$, some of
which have already appeared in the literature. For instance, with
\mbox{$\eta=(L+1)/(2L+1)$} Eq.\ (\ref{C.9}) becomes
\begin{equation}
\mathcal{T}_{LM}=-\frac{\sqrt{\pi L}}{2L+1}
\int_{\mathbb{R}^{3}}\mathrm{d}^{3}\boldsymbol{r}\:r^{L+1}
\left[\boldsymbol{Y}_{LM}^{L-1}(\boldsymbol{n}_{r})
+\frac{2}{2L+3}\sqrt{\frac{L}{L+1}}\,
\boldsymbol{Y}_{LM}^{L+1}(\boldsymbol{n}_{r})\right]
\cdot\boldsymbol{j}(\boldsymbol{r}),
\label{C.10}
\end{equation}
which coincides with Eq.\ (4.10) in Ref.\ \cite{Dubo90}. Furthermore,
if $\eta=-2/(2L+1)$, then Eq.\ (\ref{C.9}) reduces to
\begin{equation}
\mathcal{T}_{LM}
=\frac{1}{L+1}\sqrt{\frac{4\pi}{2L+1}}
\int_{\mathbb{R}^{3}}\mathrm{d}^{3}\boldsymbol{r}\:r^{L+1}
\left[\sqrt{\frac{L}{2L+1}}\,
\boldsymbol{Y}_{LM}^{L-1}(\boldsymbol{n}_{r})
-\sqrt{\frac{L+1}{2L+1}}\,
\boldsymbol{Y}_{LM}^{L+1}(\boldsymbol{n}_{r})\right]
\cdot\boldsymbol{j}(\boldsymbol{r}).
\label{C.11}
\end{equation}
Since it is known (cf.\ Ref.\ \cite[Eq.\ (7.3.70)]{Vars75}) that
\begin{equation}
\sqrt{\frac{L}{2L+1}}\,
\boldsymbol{Y}_{LM}^{L-1}(\boldsymbol{n}_{r})
-\sqrt{\frac{L+1}{2L+1}}\,
\boldsymbol{Y}_{LM}^{L+1}(\boldsymbol{n}_{r})
=\boldsymbol{n}_{r}Y_{LM}(\boldsymbol{n}_{r}),
\label{C.12}
\end{equation}
Eq.\ (\ref{C.11}) may be rewritten in the following compact form:
\begin{equation}
\mathcal{T}_{LM}
=\frac{1}{L+1}\sqrt{\frac{4\pi}{2L+1}}
\int_{\mathbb{R}^{3}}\mathrm{d}^{3}\boldsymbol{r}\:
r^{L}Y_{LM}(\boldsymbol{n}_{r})
\boldsymbol{r}\cdot\boldsymbol{j}(\boldsymbol{r}),
\label{C.13}
\end{equation}
given before in Refs.\ \cite[Eq.\ (B.4)]{Dubo90} and \cite[Eq.\
(2.2)]{Miel06}. The representation (\ref{C.13}) of $\mathcal{T}_{LM}$
has been found to be most suitable for the purposes of this work, and
the considerations presented in Sec.\ \ref{V} have been based upon
it. As the last example, in Eq.\ (\ref{C.9}) we put
$\eta=2(L+1)/[(L+2)(2L+1)]$. This casts the latter equation into
\begin{equation}
\mathcal{T}_{LM}=-\frac{1}{L+2}\sqrt{\frac{4\pi L}{(L+1)(2L+1)}}
\int_{\mathbb{R}^{3}}\mathrm{d}^{3}\boldsymbol{r}\:r^{L+1}
\left[\sqrt{\frac{L+1}{2L+1}}\,
\boldsymbol{Y}_{LM}^{L-1}(\boldsymbol{n}_{r})
+\sqrt{\frac{L}{2L+1}}\,
\boldsymbol{Y}_{LM}^{L+1}(\boldsymbol{n}_{r})\right]
\cdot\boldsymbol{j}(\boldsymbol{r}).
\label{C.14}
\end{equation}
The integrand in Eq.\ (\ref{C.14}) may be simplified with the aid of
the formula (cf.\ Ref.\ \cite[Eq.\ (7.3.73)]{Vars75})
\begin{equation}
\sqrt{\frac{L+1}{2L+1}}\,
\boldsymbol{Y}_{LM}^{L-1}(\boldsymbol{n}_{r})
+\sqrt{\frac{L}{2L+1}}\,
\boldsymbol{Y}_{LM}^{L+1}(\boldsymbol{n}_{r})
=-\mathrm{i}\boldsymbol{n}_{r}
\times\boldsymbol{Y}_{LM}^{L}(\boldsymbol{n}_{r}).
\label{C.15}
\end{equation}
This yields
\begin{equation}
\mathcal{T}_{LM}=-\frac{\mathrm{i}}{L+2}
\sqrt{\frac{4\pi L}{(L+1)(2L+1)}}
\int_{\mathbb{R}^{3}}\mathrm{d}^{3}\boldsymbol{r}\:r^{L}
\boldsymbol{Y}_{LM}^{L}(\boldsymbol{n}_{r})
\cdot[\boldsymbol{r}\times\boldsymbol{j}(\boldsymbol{r})].
\label{C.16}
\end{equation}
Since it holds that \cite[Eqs.\ (7.3.9) and (7.3.6)]{Vars75}
\begin{equation}
\boldsymbol{Y}_{LM}^{L}(\boldsymbol{n}_{r})
=\frac{\boldsymbol{\Lambda}Y_{LM}(\boldsymbol{n}_{r})}{\sqrt{L(L+1)}},
\label{C.17}
\end{equation}
the third particular expression for $\mathcal{T}_{LM}$ we
wish to present here is
\begin{equation}
\mathcal{T}_{LM}=\frac{\mathrm{i}}{(L+1)(L+2)}
\sqrt{\frac{4\pi}{2L+1}}
\int_{\mathbb{R}^{3}}\mathrm{d}^{3}\boldsymbol{r}\:
r^{L}Y_{LM}(\boldsymbol{n}_{r})
\boldsymbol{\Lambda}\cdot[\boldsymbol{r}
\times\boldsymbol{j}(\boldsymbol{r})].
\label{C.18}
\end{equation}

Concluding this appendix, we observe that the monopole toroidal
moment vanishes identically. This is immediately seen if in Eq.\
(\ref{C.13}) one sets $L=M=0$, obtaining
\begin{equation}
\mathcal{T}_{00}=\int_{\mathbb{R}^{3}}\mathrm{d}^{3}\boldsymbol{r}\:
\boldsymbol{r}\cdot\boldsymbol{j}(\boldsymbol{r}),
\label{C.19}
\end{equation}
and then one replaces the integrand by the right-hand side of the
obvious identity
\begin{equation}
\boldsymbol{r}\cdot\boldsymbol{j}(\boldsymbol{r})=\frac{1}{2}
\boldsymbol{\nabla}\cdot[r^{2}\boldsymbol{j}(\boldsymbol{r})]
-\frac{1}{2}r^{2}
\boldsymbol{\nabla}\cdot\boldsymbol{j}(\boldsymbol{r}).
\label{C.20}
\end{equation}
%
%\newpage
%
\section{Alternative integral representations of the near-field
magnetic toroidal multipole moments $\boldsymbol{\mathsf{U}}_{L}$} 
\label{D}
\setcounter{equation}{0}
In Appendix \ref{B}, we have come across a set of the near-nucleus
magnetic toroidal multipole moments $\boldsymbol{\mathsf{U}}_{L}$,
with $L\geqslant1$, the spherical components of which are given by
\begin{equation}
\mathcal{U}_{LM}=-\frac{1}{2L-1}\sqrt{\frac{4\pi}{L}}
\int_{\mathbb{R}^{3}}\mathrm{d}^{3}\boldsymbol{r}\:
r^{-L}\boldsymbol{Y}_{LM}^{L-1}
(\boldsymbol{n}_{r})\cdot\boldsymbol{j}(\boldsymbol{r}),
\label{D.1}
\end{equation}
or equivalently by
\begin{equation}
\mathcal{U}_{LM}=-\frac{1}{2L-1}\sqrt{\frac{4\pi}{L}}
\int_{\mathbb{R}^{3}}\mathrm{d}^{3}\boldsymbol{r}\:
r^{-L}\big\{\boldsymbol{\mathsf{Y}}_{L-1}
(\boldsymbol{n}_{r})\otimes\boldsymbol{j}(\boldsymbol{r})
\big\}_{LM}.
\label{D.2}
\end{equation}
In this appendix, we aim to show that there is a one-parameter family
of representations of $\mathcal{U}_{LM}$ which, under some
constraints imposed on the current density
$\boldsymbol{j}(\boldsymbol{r})$, are equivalent to the one given in
Eq.\ (\ref{D.1}). We shall be brief, as in many details the reasoning
is similar to that presented in Appendix \ref{C}, where the
counterpart set of the far-field moments has been considered.

Substitution of $f(r)=r^{-L+1}$ into Eq.\ (\ref{C.3})
transforms the latter into
\begin{equation}
\boldsymbol{\nabla}[r^{-L+1}Y_{LM}(\boldsymbol{n}_{r})]
=(2L-1)\sqrt{\frac{L+1}{2L+1}}\,r^{-L}
\boldsymbol{Y}_{LM}^{L+1}(\boldsymbol{n}_{r})
+2\sqrt{\frac{L}{2L+1}}\,r^{-L}
\boldsymbol{Y}_{LM}^{L-1}(\boldsymbol{n}_{r}),
\label{D.3}
\end{equation}
hence, it follows that Eq.\ (\ref{D.1}) may be rewritten as
\begin{eqnarray}
\mathcal{U}_{LM} &=& -\frac{\sqrt{\pi(2L+1)}}{L(2L-1)}
\int_{\mathbb{R}^{3}}\mathrm{d}^{3}\boldsymbol{r}\:
\boldsymbol{j}(\boldsymbol{r})\cdot\boldsymbol{\nabla}
[r^{-L+1}Y_{LM}(\boldsymbol{n}_{r})]
\nonumber \\
&& +\,\frac{\sqrt{\pi(L+1)}}{L}
\int_{\mathbb{R}^{3}}\mathrm{d}^{3}\boldsymbol{r}\:r^{-L}
\boldsymbol{Y}_{LM}^{L+1}(\boldsymbol{n}_{r})
\cdot\boldsymbol{j}(\boldsymbol{r}).
\label{D.4}
\end{eqnarray}
Now, it is evident that
\begin{equation}
\boldsymbol{j}(\boldsymbol{r})\cdot\boldsymbol{\nabla}
[r^{-L+1}Y_{LM}(\boldsymbol{n}_{r})]
=\boldsymbol{\nabla}\cdot[r^{-L+1}
Y_{LM}(\boldsymbol{n}_{r})\boldsymbol{j}(\boldsymbol{r})]
-r^{-L+1}Y_{LM}(\boldsymbol{n}_{r})
\boldsymbol{\nabla}\cdot\boldsymbol{j}(\boldsymbol{r}).
\label{D.5}
\end{equation}
Consequently, if the current is solenoidal and such that
\begin{equation}
\lim_{r\to0}r^{-L+3}
\boldsymbol{n}_{r}\cdot\boldsymbol{j}(\boldsymbol{r})=0,
\qquad
\lim_{r\to\infty}r^{-L+3}
\boldsymbol{n}_{r}\cdot\boldsymbol{j}(\boldsymbol{r})=0,
\label{D.6}
\end{equation}
the first integral on the right-hand side of Eq.\ (\ref{D.4})
vanishes, yielding
\begin{equation}
\mathcal{U}_{LM}=\frac{\sqrt{\pi(L+1)}}{L}
\int_{\mathbb{R}^{3}}\mathrm{d}^{3}\boldsymbol{r}\:r^{-L}
\boldsymbol{Y}_{LM}^{L+1}(\boldsymbol{n}_{r})
\cdot\boldsymbol{j}(\boldsymbol{r}).
\label{D.7}
\end{equation}
Multiplying Eq.\ (\ref{D.1}) by a parameter $\eta\in\mathbb{C}$ and
Eq.\ (\ref{D.7}) by $1-\eta$, and then adding, we obtain the sought
one-parameter family of equivalent expressions
\begin{equation}
\mathcal{U}_{LM}=-\sqrt{\frac{\pi}{L}}
\int_{\mathbb{R}^{3}}\mathrm{d}^{3}\boldsymbol{r}\:r^{-L}
\left[\eta\frac{2}{2L-1}
\boldsymbol{Y}_{LM}^{L-1}(\boldsymbol{n}_{r})
-(1-\eta)\sqrt{\frac{L+1}{L}}
\boldsymbol{Y}_{LM}^{L+1}(\boldsymbol{n}_{r})\right]
\cdot\boldsymbol{j}(\boldsymbol{r}),
\label{D.8}
\end{equation}
which may be interchangeably used as definitions of components of the
tensor $\boldsymbol{\mathsf{U}}_{L}$.

Two particular choices of $\eta$ are worth to be analyzed here.
Thus, for
\begin{equation}
\eta=\frac{2L-1}{2L+1},
\label{D.9}
\end{equation}
by virtue of the identity (\ref{C.12}), we obtain
\begin{equation}
\mathcal{U}_{LM}=-\frac{1}{L}\sqrt{\frac{4\pi}{2L+1}}
\int_{\mathbb{R}^{3}}\mathrm{d}^{3}\boldsymbol{r}\:r^{-L-1}
Y_{LM}(\boldsymbol{n}_{r})
\boldsymbol{r}\cdot\boldsymbol{j}(\boldsymbol{r}),
\label{D.10}
\end{equation}
while for
\begin{equation}
\eta=\frac{(L+1)(2L-1)}{(L-1)(2L+1)}
\qquad (L\neq1),
\label{D.11}
\end{equation}
after exploiting the relations (\ref{C.15}) and (\ref{C.17}), we find
\begin{equation}
\mathcal{U}_{LM}=\frac{\mathrm{i}}{L(L-1)}\sqrt{\frac{4\pi}{2L+1}}
\int_{\mathbb{R}^{3}}\mathrm{d}^{3}\boldsymbol{r}\:r^{-L-1}
Y_{LM}(\boldsymbol{n}_{r})\boldsymbol{\Lambda}
\cdot[\boldsymbol{r}\times\boldsymbol{j}(\boldsymbol{r})]
\qquad (L\neq1).
\label{D.12}
\end{equation}
If $L=1$ were admitted in Eq.\ (\ref{D.12}), the integral appearing
therein would vanish (see the next paragraph), and consequently on
the right-hand side we would have a $0/0$-type expression. To prepare
the ground for the use of the L'Hospital's rule, we set $L=1$ in the
spherical harmonic and $L=1+\varepsilon$ at other places in the above
formula. Hence, if we let $\varepsilon$ tend to zero, after
exploiting the aforementioned rule, we obtain
\begin{equation}
\mathcal{U}_{1M}=-\mathrm{i}\,\sqrt{\frac{4\pi}{3}}
\int_{\mathbb{R}^{3}}\mathrm{d}^{3}\boldsymbol{r}\:
r^{-2}\ln(r/r_{0})Y_{1M}(\boldsymbol{n}_{r})
\boldsymbol{\Lambda}
\cdot[\boldsymbol{r}\times\boldsymbol{j}(\boldsymbol{r})],
\label{D.13}
\end{equation}
where $r_{0}$, of the physical dimension $\mathsf{L}$ and such that
$r_{0}>0$ but otherwise arbitrary, has been introduced merely to make
the argument of the logarithm physically dimensionless.

We still owe the reader a proof that $I_{1M}=0$, where
\begin{equation}
I_{1M}=\int_{\mathbb{R}^{3}}\mathrm{d}^{3}\boldsymbol{r}\:
r^{-2}Y_{1M}(\boldsymbol{n}_{r})\boldsymbol{\Lambda}
\cdot[\boldsymbol{r}\times\boldsymbol{j}(\boldsymbol{r})].
\label{D.14}
\end{equation}
To show this, we observe that with the use of the relation
\begin{equation}
Y_{1M}(\boldsymbol{n}_{r})=\sqrt{\frac{3}{4\pi}}\,
\boldsymbol{e}_{M}\cdot\boldsymbol{n}_{r}
\label{D.15}
\end{equation}
the integral in Eq.\ (\ref{D.14}) may be transformed into
\begin{equation}
I_{1M}=\sqrt{\frac{3}{4\pi}}
\int_{\mathbb{R}^{3}}\mathrm{d}^{3}\boldsymbol{r}\:
r^{-3}[(\boldsymbol{r}\times\boldsymbol{\Lambda})
(\boldsymbol{e}_{M}\cdot\boldsymbol{r})]
\cdot\boldsymbol{j}(\boldsymbol{r}).
\label{D.16}
\end{equation}
Hence, it follows that
\begin{equation}
I_{1M}=-\mathrm{i}\sqrt{\frac{3}{4\pi}}
\int_{\mathbb{R}^{3}}\mathrm{d}^{3}\boldsymbol{r}\:
\frac{\boldsymbol{n}_{r}
\times(\boldsymbol{n}_{r}\times\boldsymbol{e}_{M})}{r}
\cdot\boldsymbol{j}(\boldsymbol{r}).
\label{D.17}
\end{equation}
Now, it holds that
\begin{equation}
\frac{\boldsymbol{n}_{r}
\times(\boldsymbol{n}_{r}\times\boldsymbol{e}_{M})}{r}
=-\boldsymbol{\nabla}(\boldsymbol{e}_{M}\cdot\boldsymbol{n}_{r}),
\label{D.18}
\end{equation}
and consequently one has
\begin{equation}
I_{1M}=\mathrm{i}\sqrt{\frac{3}{4\pi}}
\int_{\mathbb{R}^{3}}\mathrm{d}^{3}\boldsymbol{r}\:
\boldsymbol{\nabla}\cdot[(\boldsymbol{e}_{M}\cdot\boldsymbol{n}_{r})
\boldsymbol{j}(\boldsymbol{r})]
-\mathrm{i}\sqrt{\frac{3}{4\pi}}
\int_{\mathbb{R}^{3}}\mathrm{d}^{3}\boldsymbol{r}\:
(\boldsymbol{e}_{M}\cdot\boldsymbol{n}_{r})
\boldsymbol{\nabla}\cdot\boldsymbol{j}(\boldsymbol{r}).
\label{D.19}
\end{equation}
The first integral on the right-hand side is zero by virtue of the
Gauss' theorem and the constraints in Eq.\ (\ref{D.6}), while in the
second one the integrand vanishes identically since, by assumption,
the current is sourceless. This completes the proof.
%
%\newpage
%
\section{Formulas for the generalized hypergeometric series
${}_{3}F_{2}(a,1,1;b,n;1)$ with $n=3,4,5$} 
\label{E}
\setcounter{equation}{0}
In the course of evaluation of the three kinds of shielding factors,
presented in Secs.\ \ref{VI} to \ref{VIII}, we have encountered the
specialized generalized hypergeometric series
${}_{3}F_{2}(a,1,1;b,n;1)$ with $n=3,4,5$. It may be found in the
literature that the one with $n=3$ may be expressed in terms of the
digamma function as \cite[Eq.\ (7.4.4.41)]{Prud03}
\begin{eqnarray}
&& \hspace*{-3em}
{}_{3}F_{2}
\left(
\begin{array}{c}
a,\:
1,\:
1 \\
b,\:
3
\end{array}
;1
\right)
=\frac{2(b-1)}{a-1}
-\frac{2(b-1)(b-a)}{(a-1)(a-2)}[\psi(b-1)-\psi(b-a+1)]
\qquad [\Real(b-a)>-1],
\nonumber \\
&&
\label{E.1}
\end{eqnarray}
which, after exploiting the relation
\begin{equation}
\psi(z+1)=\psi(z)+\frac{1}{z},
\label{E.2}
\end{equation}
may be transformed into
\begin{eqnarray}
&& {}_{3}F_{2}
\left(
\begin{array}{c}
a,\:
1,\:
1 \\
b,\:
3
\end{array}
;1
\right)
=\frac{2(b-1)}{a-2}
-\frac{2(b-1)(b-a)}{(a-1)(a-2)}[\psi(b-1)-\psi(b-a)]
\qquad [\Real(b-a)>-1].
\nonumber \\
&&
\label{E.3}
\end{eqnarray}
Below, we shall derive analogous expressions for the two remaining
${}_{3}F_{2}(1)$ functions of interest in the context of this work.

Playing with the definition
\begin{eqnarray}
{}_{3}F_{2}
\left(
\begin{array}{c}
a_{1},\:
a_{2},\:
a_{3} \\
b_{1},\:
b_{2}
\end{array}
;z
\right)
&=& \frac{\Gamma(b_{1})\Gamma(b_{2})}
{\Gamma(a_{1})\Gamma(a_{2})\Gamma(a_{3})}
\sum_{n=0}^{\infty}\frac{\Gamma(n+a_{1})\Gamma(n+a_{2})
\Gamma(n+a_{3})}{\Gamma(n+b_{1})\Gamma(n+b_{2})}\frac{z^{n}}{n!}
\nonumber \\
&& [\textrm{$|z|\leqslant1$; 
$\Real(b_{1}+b_{2}-a_{1}-a_{2}-a_{3})>0$ for $z=1$}]
\label{E.4} 
\end{eqnarray}
(the constraints on $z$, $a$'s and $b$'s will be tacitly assumed to
hold throughout the rest of this appendix), it is possible to obtain
the recurrence relation
\begin{eqnarray}
{}_{3}F_{2}
\left(
\begin{array}{c}
a_{1},\:
a_{2},\:
a_{3} \\
b_{1},\:
b_{2}
\end{array}
;z
\right)
&=& \frac{a_{3}}{a_{3}-b_{2}+1}\,
{}_{3}F_{2}
\left(
\begin{array}{c}
a_{1},\:
a_{2},\:
a_{3}+1 \\
b_{1},\:
b_{2}
\end{array}
;z
\right)
\nonumber \\
&& -\,\frac{b_{2}-1}{a_{3}-b_{2}+1}\,
{}_{3}F_{2}
\left(
\begin{array}{c}
a_{1},\:
a_{2},\:
a_{3} \\
b_{1},\:
b_{2}-1
\end{array}
;z
\right),
\label{E.5}
\end{eqnarray}
from which we deduce that
\begin{equation}
{}_{3}F_{2}
\left(
\begin{array}{c}
a,\:
1,\:
1 \\
b,\:
n
\end{array}
;1
\right)
=-\frac{1}{n-2}\,
{}_{3}F_{2}
\left(
\begin{array}{c}
a,\:
1,\:
2 \\
b,\:
n
\end{array}
;1
\right)
+\frac{n-1}{n-2}\,
{}_{3}F_{2}
\left(
\begin{array}{c}
a,\:
1,\:
1 \\
b,\:
n-1
\end{array}
;1
\right).
\label{E.6}
\end{equation}
Next, we exploit the identity
\begin{eqnarray}
{}_{3}F_{2}
\left(
\begin{array}{c}
a_{1},\:
1,\:
a_{3} \\
b_{1},\:
b_{2}
\end{array}
;z
\right)
=\frac{(b_{1}-1)(b_{2}-1)}{(a_{1}-1)(a_{3}-1)z}\,
{}_{3}F_{2}
\left(
\begin{array}{c}
a_{1}-1,\:
1,\:
a_{3}-1 \\
b_{1}-1,\:
b_{2}-1
\end{array}
;z
\right)
-\frac{(b_{1}-1)(b_{2}-1)}{(a_{1}-1)(a_{3}-1)z},
\nonumber \\
&&
\label{E.7}
\end{eqnarray}
which also may be directly inferred from the definition (\ref{E.4}).
As the particular case of that identity, we have
\begin{equation}
{}_{3}F_{2}
\left(
\begin{array}{c}
a,\:
1,\:
2 \\
b,\:
n
\end{array}
;1
\right)
=\frac{(n-1)(b-1)}{a-1}\,
{}_{3}F_{2}
\left(
\begin{array}{c}
a-1,\:
1,\:
1 \\
b-1,\:
n-1
\end{array}
;1
\right)
-\frac{(n-1)(b-1)}{a-1}.
\label{E.8}
\end{equation}
Insertion of Eq.\ (\ref{E.8}) into Eq.\ (\ref{E.6}) yields the
relationship
\begin{eqnarray}
{}_{3}F_{2}
\left(
\begin{array}{c}
a,\:
1,\:
1 \\
b,\:
n
\end{array}
;1
\right)
&=& \frac{n-1}{n-2}\,
{}_{3}F_{2}
\left(
\begin{array}{c}
a,\:
1,\:
1 \\
b,\:
n-1
\end{array}
;1
\right)
-\frac{(n-1)(b-1)}{(n-2)(a-1)}\,
{}_{3}F_{2}
\left(
\begin{array}{c}
a-1,\:
1,\:
1 \\
b-1,\:
n-1
\end{array}
;1
\right)
\nonumber \\
&& +\,\frac{(n-1)(b-1)}{(n-2)(a-1)}.
\label{E.9}
\end{eqnarray}
If in Eq.\ (\ref{E.9}) we put $n=4$, after simplifying the right-hand
side with the use of Eqs.\ (\ref{E.3}) and (\ref{E.2}), we obtain
\begin{eqnarray}
&& {}_{3}F_{2}
\left(
\begin{array}{c}
a,\:
1,\:
1 \\
b,\:
4
\end{array}
;1
\right)
=\frac{3(b-1)(3a-2b-4)}{2(a-2)(a-3)}
+\frac{3(b-1)(b-a)(b-a+1)}{(a-1)(a-2)(a-3)}[\psi(b-1)-\psi(b-a)]
\nonumber \\
&& \hspace*{25em} [\Real(b-a)>-2].
\label{E.10}
\end{eqnarray}
Employing Eq.\ (\ref{E.9}) recursively, in the similar manner we find
\begin{eqnarray}
{}_{3}F_{2}
\left(
\begin{array}{c}
a,\:
1,\:
1 \\
b,\:
5
\end{array}
;1
\right)
&=& \frac{2(b-1)(6b^{2}+24b-15ab-40a+11a^{2}+36)}{3(a-2)(a-3)(a-4)}
\nonumber \\
&& -\,\frac{4(b-1)(b-a)(b-a+1)(b-a+2)}
{(a-1)(a-2)(a-3)(a-4)}[\psi(b-1)-\psi(b-a)]
\nonumber \\
&& \hspace*{15em} [\Real(b-a)>-3].
\label{E.11}
\end{eqnarray}
For the sake of completeness, we observe that apparent singularities
at some values of the parameter $a$ in the expressions on the
right-hand sides of Eqs.\ (\ref{E.3}), (\ref{E.10}) and (\ref{E.11})
are removable with an application of the L'Hospital's rule.
%
%\newpage
%

%
%\newpage
%
%%%%%%%%%%%%%%%%%%%%%%%%%%%%%%%%%%%%%%%%%%%%%%%%%%%%%%%%%%%%%%%%%%%%%%
%                                TABLES
%%%%%%%%%%%%%%%%%%%%%%%%%%%%%%%%%%%%%%%%%%%%%%%%%%%%%%%%%%%%%%%%%%%%%%
%
%%%%%%%%%%%%%%%%%%%%%%%%%
%       TABLE 1
%%%%%%%%%%%%%%%%%%%%%%%%%
%
\begin{landscape} 
\begin{table}[t] 
\caption{Comparison of present exact values of the static electric
dipole, quadrupole, octupole and hexadecapole polarizabilities for
the hydrogen atom ($Z=1$) in the ground state with those obtained
numerically by other authors using either the $B$-spline Galerkin
method \cite{Tang12} or the Lagrange-mesh method \cite{Fili14}. The
number in brackets following the entries is the power of 10 by which
the entry is to be multiplied. The present results have been computed
from the analytical formula in Eq.\ (\ref{3.41}), using both the
currently recommended CODATA 2014 value of the inverse of the
fine-structure constant $\alpha^{-1}$ and, for the sake of making
comparison with the results from Refs.\ \cite{Tang12,Fili14} more
explicit, also its previous CODATA 2010 value.}
\label{T.1} 
\vspace*{1ex} 
{\footnotesize
\begin{center} 
\begin{tabular}{cllll} 
\hline 
\hline \\
\multicolumn{1}{c}{Source} & 
\multicolumn{1}{c}{$\alpha_{1}$ ($a_{0}^{3}$)} & 
\multicolumn{1}{c}{$\alpha_{2}$ ($a_{0}^{5}$)} &
\multicolumn{1}{c}{$\alpha_{3}$ ($a_{0}^{7}$)} &
\multicolumn{1}{c}{$\alpha_{4}$ ($a_{0}^{9}$)} \\*[1ex] 
\hline \\
\multicolumn{5}{c}{$\alpha^{-1}=137.035\:999\:139$ 
(from CODATA 2014)} \\*[1ex]
present & 
4.499\:751\:495\:177\:875\:011\:523\:552 &
1.499\:882\:982\:285\:755\:177\:470\:692\:($+1$) &
1.312\:378\:214\:478\:562\:151\:040\:415\:($+2$) &
2.126\:028\:674\:499\:338\:786\:454\:128\:($+3$) \\
(exact) &&&& \\*[1ex]
\multicolumn{5}{c}{$\alpha^{-1}=137.035\:999\:074$ 
(from CODATA 2010)} \\*[1ex]
present & 
4.499\:751\:495\:177\:639\:267\:396\:013 &
1.499\:882\:982\:285\:644\:169\:960\:840\:($+1$) &
1.312\:378\:214\:478\:446\:621\:510\:730\:($+2$) &
2.126\:028\:674\:499\:128\:831\:459\:952\:($+3$) \\
(exact) &&&& \\*[0.5ex]
Ref.\ \cite{Tang12} & 
4.499\:751\:495\:177\:639\:267\:396\:02 & 
1.499\:882\:982\:285\:644\:169\:960\:8\:($+1$) &
1.312\:378\:214\:478\:446\:621\:510\:($+2$) &
2.126\:028\:674\:499\:128\:831\:46\:($+3$) \\*[0.5ex]
Ref.\ \cite{Fili14} & 
4.499\:751\:495\:177\:639 &
1.499\:882\:982\:285\:648\:($+1$) & 
1.312\:378\:214\:478\:460\:($+2$) & 
2.126\:028\:674\:499\:147\:($+3$) \\*[1ex] 
\hline 
\hline
\end{tabular} 
\end{center} 
}
\end{table} 
\end{landscape}
%
%%%%%%%%%%%%%%%%%%%%%%%%%
%       TABLE 2
%%%%%%%%%%%%%%%%%%%%%%%%%
%
\begin{landscape}
\begin{table}[t] 
\caption{The static electric multipole polarizabilities $\alpha_{L}$
with $1\leqslant L\leqslant4$ for selected hydrogenic ions in the
ground state, computed from the analytical formula in Eq.\
(\ref{3.41}). The number in brackets following the entries is the
power of 10 by which the entry is to be multiplied. The values of the
inverse of the fine-structure constant used in calculations have been
$\alpha^{-1}=137.035\:999\:139$ (from CODATA 2014; results in the
first row for each $Z$) and $\alpha^{-1}=137.035\:999\:074$ (from
CODATA 2010; results in the second row for each $Z$).}
\label{T.2}
\vspace*{1ex}
{\footnotesize
\begin{center}
\begin{tabular}{rllll}
\hline 
\hline \\
\multicolumn{1}{c}{$Z$} & 
\multicolumn{1}{c}{$\alpha_{1}$ ($a_{0}^{3}$)} & 
\multicolumn{1}{c}{$\alpha_{2}$ ($a_{0}^{5}$)} & 
\multicolumn{1}{c}{$\alpha_{3}$ ($a_{0}^{7}$)} & 
\multicolumn{1}{c}{$\alpha_{4}$ ($a_{0}^{9}$)} \\*[1ex]
\hline \\
1 &
4.499\:751\:495\:177\:875\:011\:524\:($+$0) & 
1.499\:882\:982\:285\:755\:177\:471\:($+$1) & 
1.312\:378\:214\:478\:562\:151\:040\:($+$2) & 
2.126\:028\:674\:499\:338\:786\:454\:($+$3) \\
  &
4.499\:751\:495\:177\:639\:267\:396\:($+$0) & 
1.499\:882\:982\:285\:644\:169\:961\:($+$1) & 
1.312\:378\:214\:478\:446\:621\:511\:($+$2) & 
2.126\:028\:674\:499\:128\:831\:460\:($+$3) \\*[0.5ex]
2 &
2.811\:878\:749\:185\:621\:693\:079\:($-$1) & 
2.343\:018\:679\:358\:605\:821\:404\:($-$1) & 
5.125\:050\:375\:239\:509\:639\:311\:($-$1) & 
2.075\:551\:546\:062\:025\:167\:458\:($+$0) \\
  & 
2.811\:878\:749\:185\:032\:354\:088\:($-$1) & 
2.343\:018\:679\:357\:912\:100\:497\:($-$1) & 
5.125\:050\:375\:237\:704\:773\:127\:($-$1) & 
2.075\:551\:546\:061\:205\:188\:564\:($+$0) \\*[0.5ex]
5 & 
7.190\:061\:246\:057\:044\:497\:594\:($-$3) & 
9.581\:285\:372\:341\:791\:023\:296\:($-$4) & 
3.352\:210\:608\:794\:400\:804\:905\:($-$4) & 
2.171\:618\:426\:950\:907\:701\:171\:($-$4) \\
  &
7.190\:061\:246\:047\:617\:462\:734\:($-$3) & 
9.581\:285\:372\:324\:045\:391\:938\:($-$4) & 
3.352\:210\:608\:787\:016\:177\:224\:($-$4) & 
2.171\:618\:426\:945\:541\:124\:049\:($-$4) \\*[0.5ex]
10 & 
4.475\:164\:360\:648\:818\:455\:398\:($-$4) & 
1.488\:319\:383\:924\:471\:652\:035\:($-$5) & 
1.300\:352\:899\:799\:117\:538\:323\:($-$6) & 
2.104\:187\:645\:771\:176\:667\:402\:($-$7) \\
   &
4.475\:164\:360\:625\:272\:208\:993\:($-$4) & 
1.488\:319\:383\:913\:411\:036\:344\:($-$5) & 
1.300\:352\:899\:787\:624\:240\:146\:($-$6) & 
2.104\:187\:645\:750\:314\:281\:973\:($-$7) \\*[0.5ex]
20 &
2.750\:523\:499\:121\:230\:880\:495\:($-$5) & 
2.271\:146\:583\:119\:162\:558\:714\:($-$7) & 
4.938\:640\:072\:445\:973\:751\:899\:($-$9) & 
1.991\:062\:443\:096\:913\:712\:558\:($-$10) \\
   & 
2.750\:523\:499\:062\:579\:076\:757\:($-$5) & 
2.271\:146\:583\:050\:793\:178\:033\:($-$7) & 
4.938\:640\:072\:269\:204\:679\:633\:($-$9) & 
1.991\:062\:443\:016\:985\:982\:651\:($-$10) \\*[0.5ex]
40 &
1.604\:002\:839\:692\:739\:227\:002\:($-$6) & 
3.218\:326\:876\:777\:344\:766\:036\:($-$9) & 
1.717\:671\:116\:979\:420\:309\:765\:($-$11) & 
1.707\:067\:336\:752\:410\:513\:201\:($-$13) \\
   & 
1.604\:002\:839\:548\:263\:708\:676\:($-$6) & 
3.218\:326\:876\:368\:960\:501\:741\:($-$9) & 
1.717\:671\:116\:720\:549\:846\:487\:($-$11) & 
1.707\:067\:336\:464\:092\:957\:761\:($-$13) \\*[0.5ex]
60 &
2.797\:090\:475\:043\:278\:635\:137\:($-$7) & 
2.371\:147\:053\:789\:095\:379\:606\:($-$10) & 
5.443\:579\:082\:032\:361\:630\:978\:($-$13) & 
2.345\:208\:225\:059\:515\:662\:725\:($-$15) \\
   & 
2.797\:090\:474\:417\:353\:213\:078\:($-$7) & 
2.371\:147\:053\:044\:287\:467\:178\:($-$10) & 
5.443\:579\:080\:005\:381\:231\:668\:($-$13) & 
2.345\:208\:224\:082\:184\:652\:454\:($-$15) \\*[0.5ex]
80 &
7.256\:230\:366\:973\:582\:550\:880\:($-$8) & 
3.196\:013\:750\:479\:081\:681\:921\:($-$11) & 
3.921\:694\:890\:318\:033\:520\:828\:($-$14) & 
9.141\:669\:900\:197\:762\:990\:639\:($-$17) \\
   & 
7.256\:230\:363\:582\:213\:484\:281\:($-$8) & 
3.196\:013\:748\:393\:893\:427\:240\:($-$11) & 
3.921\:694\:887\:293\:335\:137\:309\:($-$14) & 
9.141\:669\:892\:322\:525\:362\:993\:($-$17) \\*[0.5ex]
100 & 
2.168\:647\:589\:558\:746\:030\:622\:($-$8) & 
5.405\:559\:190\:598\:306\:646\:268\:($-$12) & 
3.923\:335\:160\:173\:613\:355\:838\:($-$15) & 
5.514\:202\:255\:886\:621\:078\:935\:($-$18) \\
    &
2.168\:647\:587\:493\:674\:593\:651\:($-$8) & 
5.405\:559\:183\:469\:571\:103\:004\:($-$12) & 
3.923\:335\:154\:079\:824\:764\:316\:($-$15) & 
5.514\:202\:246\:346\:336\:408\:556\:($-$18) \\*[0.5ex]
120 &
5.962\:322\:886\:341\:222\:518\:228\:($-$9) & 
8.350\:889\:829\:402\:244\:690\:391\:($-$13) & 
3.675\:741\:125\:180\:401\:231\:657\:($-$16) & 
3.240\:357\:008\:255\:786\:619\:282\:($-$19) \\
    & 
5.962\:322\:872\:671\:608\:906\:403\:($-$9) & 
8.350\:889\:803\:174\:165\:102\:275\:($-$13) & 
3.675\:741\:111\:635\:729\:711\:889\:($-$16) & 
3.240\:356\:994\:994\:917\:555\:786\:($-$19) \\*[0.5ex]
137 &
5.748\:648\:400\:700\:781\:043\:837\:($-$10) & 
3.159\:735\:442\:958\:334\:459\:896\:($-$14) & 
6.689\:221\:528\:538\:828\:261\:530\:($-$18) & 
3.139\:717\:909\:213\:994\:303\:355\:($-$21) \\
    & 
5.748\:647\:780\:061\:286\:039\:034\:($-$10) & 
3.159\:734\:965\:973\:634\:785\:405\:($-$14) & 
6.689\:220\:281\:323\:310\:175\:469\:($-$18) & 
3.139\:717\:224\:793\:817\:462\:022\:($-$21) \\*[1ex]
\hline
\hline
\end{tabular}
\end{center}
}
\end{table}
\end{landscape}
%
%%%%%%%%%%%%%%%%%%%%%%%%%
%       TABLE 3
%%%%%%%%%%%%%%%%%%%%%%%%%
%
\begin{table}[t]
\caption{Quasi-relativistic approximations for the static electric
multipole polarizabilities $\alpha_{L}$ with $1\leqslant
L\leqslant4$ for the Dirac one-electron atom in the ground state. The
expressions have been derived from Eq.\ (\ref{3.51}).}
\label{T.3}
\begin{center}
\begin{tabular}{ccl}
\hline
\hline \\*[-1ex]
$L$ && 
\multicolumn{1}{c}{$\alpha_{L}$} \\*[1ex]
\hline \\
1 && 
$\displaystyle\frac{a_{0}^{3}}{Z^{4}}
\frac{9}{2}\left[1-\frac{28}{27}(\alpha Z)^{2}\right]$
\\*[4ex]
2 && 
$\displaystyle\frac{a_{0}^{5}}{Z^{6}}
15\left[1-\frac{293}{200}(\alpha Z)^{2}\right]$
\\*[4ex]
3 && 
$\displaystyle\frac{a_{0}^{7}}{Z^{8}}
\frac{525}{4}\left[1-\frac{5123}{2940}(\alpha Z)^{2}\right]$
\\*[4ex]
4 && 
$\displaystyle\frac{a_{0}^{9}}{Z^{10}}
\frac{8505}{4}\left[1-\frac{33251}{17010}(\alpha Z)^{2}\right]$
\\*[3ex]
\hline
\hline
\end{tabular}
\end{center}
\end{table}
%
%%%%%%%%%%%%%%%%%%%%%%%%%
%       TABLE 4
%%%%%%%%%%%%%%%%%%%%%%%%%
%
\begin{landscape}
\begin{table}[t]
\caption{The static electric-to-magnetic multipole
cross-susceptibilities $\alpha_{\mathrm{E}L\to\mathrm{M}(L-1)}$ with
$2\leqslant L\leqslant4$ for selected hydrogenic ions in the ground
state, computed from the analytical formula in Eq.\ (\ref{4.42}). The
number in brackets following the entries is the power of 10 by which
the entry is to be multiplied. The value of the inverse of the
fine-structure constant used in calculations has been
$\alpha^{-1}=137.035\:999\:139$ (from CODATA 2014).}
\label{T.4}
\vspace*{1ex}
\begin{center}
\begin{tabular}{rlll}
\hline
\hline \\
\multicolumn{1}{c}{$Z$} & 
\multicolumn{1}{c}{$\alpha_{\mathrm{E}2\to\mathrm{M}1}$ 
($\alpha_{0}^{4}$)} &
\multicolumn{1}{c}{$\alpha_{\mathrm{E}3\to\mathrm{M}2}$ 
($\alpha_{0}^{6}$)} &
\multicolumn{1}{c}{$\alpha_{\mathrm{E}4\to\mathrm{M}3}$ 
($\alpha_{0}^{8}$)} \\*[1ex]
\hline \\
1 & 
7.447\:801\:428\:671\:972\:587\:443\:($-$8) & 
7.840\:878\:107\:823\:287\:180\:845\:($-$7) & 
1.234\:923\:524\:976\:238\:056\:122\:($-$5) \\*[0.5ex]
2 & 
1.861\:763\:994\:267\:934\:300\:844\:($-$8) & 
4.899\:814\:158\:632\:429\:697\:547\:($-$8) & 
1.929\:209\:479\:771\:076\:123\:480\:($-$7) \\*[0.5ex]
5 & 
2.976\:734\:913\:049\:369\:044\:218\:($-$9) & 
1.253\:036\:091\:447\:461\:201\:593\:($-$9) & 
7.891\:765\:327\:700\:245\:051\:915\:($-$10) \\*[0.5ex]
10 & 
7.423\:191\:448\:826\:149\:314\:350\:($-$10) & 
7.802\:108\:784\:450\:419\:927\:811\:($-$11) & 
1.227\:360\:291\:299\:865\:248\:608\:($-$11) \\*[0.5ex]
20 & 
1.837\:121\:846\:172\:278\:346\:751\:($-$10) & 
4.803\:053\:653\:104\:366\:927\:699\:($-$12) & 
1.882\:114\:263\:521\:074\:101\:279\:($-$13) \\*[0.5ex]
40 & 
4.404\:779\:739\:691\:493\:136\:149\:($-$11) & 
2.820\:265\:342\:696\:907\:444\:999\:($-$13) & 
2.722\:166\:133\:092\:580\:216\:657\:($-$15) \\*[0.5ex]
60 & 
1.816\:173\:088\:080\:121\:360\:450\:($-$11) & 
4.983\:382\:447\:871\:323\:764\:845\:($-$14) & 
2.082\:589\:394\:725\:505\:508\:280\:($-$16) \\*[0.5ex]
80 & 
9.069\:233\:476\:744\:296\:004\:781\:($-$12) & 
1.323\:207\:519\:235\:624\:648\:194\:($-$14) & 
2.987\:634\:353\:754\:512\:204\:114\:($-$17) \\*[0.5ex]
100 & 
4.810\:527\:179\:192\:603\:390\:954\:($-$12) & 
4.126\:084\:143\:007\:186\:502\:312\:($-$15) & 
5.611\:810\:738\:654\:085\:031\:945\:($-$18) \\*[0.5ex]
120 & 
2.398\:905\:441\:473\:722\:523\:318\:($-$12) & 
1.243\:760\:509\:438\:331\:764\:772\:($-$15) & 
1.064\:754\:960\:931\:800\:535\:247\:($-$18) \\*[0.5ex]
137 & 
6.535\:200\:353\:860\:103\:739\:288\:($-$13) & 
1.835\:801\:306\:142\:691\:499\:392\:($-$16) & 
9.459\:904\:274\:972\:434\:692\:572\:($-$20) \\*[1ex]
\hline
\hline
\end{tabular}
\end{center}
\end{table}
\end{landscape}
%
%%%%%%%%%%%%%%%%%%%%%%%%%
%       TABLE 5
%%%%%%%%%%%%%%%%%%%%%%%%%
%
\begin{landscape} 
\begin{table}[t] 
\caption{The static electric-to-magnetic multipole
cross-susceptibilities $\alpha_{\mathrm{E}L\to\mathrm{M}(L+1)}$ with
$1\leqslant L\leqslant4$ for selected hydrogenic ions in the ground
state, computed from the analytical formula in Eq.\ (\ref{4.43}). The
number in brackets following the entries is the power of 10 by which
the entry is to be multiplied. The value of the inverse of the
fine-structure constant used in calculations has been
$\alpha^{-1}=137.035\:999\:139$ (from CODATA 2014).}
\label{T.5}
\vspace*{1ex} 
{\footnotesize
\begin{center} 
\begin{tabular}{rllll}
\hline 
\hline \\ 
\multicolumn{1}{c}{$Z$} &
\multicolumn{1}{c}{$\alpha_{\mathrm{E}1\to\mathrm{M}2}$ 
($a_{0}^{4}$)} &
\multicolumn{1}{c}{$\alpha_{\mathrm{E}2\to\mathrm{M}3}$
($a_{0}^{6}$)} &
\multicolumn{1}{c}{$\alpha_{\mathrm{E}3\to\mathrm{M}4}$
($a_{0}^{8}$)} &
\multicolumn{1}{c}{$\alpha_{\mathrm{E}4\to\mathrm{M}5}$
($a_{0}^{10}$)} \\*[1ex] 
\hline \\ 
1  & 
3.283\:609\:988\:211\:738\:758\:318\:($-$2) & 
1.641\:779\:910\:277\:485\:809\:376\:($-$1) & 
1.915\:387\:218\:714\:341\:770\:148\:($+$0) & 
3.878\:621\:704\:266\:776\:216\:353\:($+$1) \\*[0.5ex] 
2  & 
2.051\:883\:757\:574\:382\:373\:522\:($-$3) & 
2.564\:697\:947\:344\:304\:870\:041\:($-$3) & 
7.480\:014\:760\:822\:452\:355\:077\:($-$3) & 
3.786\:611\:359\:323\:821\:745\:607\:($-$2) \\*[0.5ex] 
5  & 
5.246\:149\:090\:003\:828\:601\:954\:($-$5) & 
1.048\:828\:974\:423\:698\:581\:549\:($-$5) & 
4.893\:086\:196\:104\:772\:001\:998\:($-$6) & 
3.962\:443\:793\:356\:229\:498\:171\:($-$6) \\*[0.5ex] 
10 & 
3.263\:961\:669\:540\:895\:253\:204\:($-$6) & 
1.629\:485\:276\:850\:519\:559\:523\:($-$7) & 
1.898\:813\:208\:632\:113\:003\:285\:($-$8) & 
3.841\:383\:757\:263\:664\:871\:786\:($-$9) \\*[0.5ex] 
20  & 
2.002\:912\:101\:253\:560\:623\:960\:($-$7) & 
2.488\:285\:199\:832\:601\:762\:511\:($-$9) & 
7.222\:989\:034\:894\:080\:541\:776\:($-$11) & 
3.642\:467\:546\:249\:702\:021\:127\:($-$12) \\*[0.5ex] 
40 & 
1.160\:561\:045\:666\:125\:723\:140\:($-$8) & 
3.536\:787\:823\:830\:012\:544\:234\:($-$11) & 
2.529\:227\:814\:953\:036\:617\:056\:($-$13) & 
3.150\:752\:599\:057\:498\:903\:191\:($-$15) \\*[0.5ex] 
60 & 
2.001\:899\:182\:261\:258\:129\:378\:($-$9) & 
2.622\:086\:172\:857\:569\:638\:163\:($-$12) & 
8.120\:641\:172\:461\:031\:595\:607\:($-$15) & 
4.402\:352\:754\:592\:142\:591\:412\:($-$17) \\*[0.5ex] 
80 & 
5.112\:923\:654\:383\:558\:308\:447\:($-$10) & 
3.576\:878\:765\:048\:380\:684\:672\:($-$13) & 
5.987\:127\:613\:932\:832\:469\:744\:($-$16) & 
1.767\:496\:114\:014\:112\:526\:321\:($-$18) \\*[0.5ex] 
100 & 
1.497\:342\:934\:870\:773\:796\:740\:($-$10) & 
6.199\:237\:022\:887\:862\:092\:759\:($-$14) & 
6.253\:163\:632\:340\:007\:961\:109\:($-$17) & 
1.125\:356\:310\:598\:765\:215\:249\:($-$19) \\*[0.5ex] 
120 & 
4.025\:596\:828\:302\:637\:918\:375\:($-$11) & 
1.018\:161\:258\:748\:539\:778\:861\:($-$14) & 
6.466\:767\:858\:953\:814\:026\:653\:($-$18) & 
7.469\:936\:049\:505\:921\:356\:006\:($-$21) \\*[0.5ex] 
137 & 
4.095\:281\:709\:362\:274\:791\:736\:($-$12) & 
5.745\:732\:037\:761\:833\:358\:905\:($-$16) & 
2.193\:610\:026\:613\:034\:032\:245\:($-$19) & 
1.601\:547\:888\:113\:102\:917\:378\:($-$22) \\*[1ex] 
\hline
\hline 
\end{tabular}
\end{center}
}
\end{table}
\end{landscape}
%
%%%%%%%%%%%%%%%%%%%%%%%%%
%       TABLE 6
%%%%%%%%%%%%%%%%%%%%%%%%%
%
\begin{table}[t]
\caption{Quasi-relativistic approximations for the static
electric-to-magnetic multipole cross-susceptibilities
$\alpha_{\mathrm{E}L\to\mathrm{M}(L-1)}$ with $2\leqslant
L\leqslant4$ and $\alpha_{\mathrm{E}L\to\mathrm{M}(L+1)}$ with
$1\leqslant L\leqslant4$ for the Dirac one-electron atom in the
ground state. The expressions have been derived from Eqs.\
(\ref{4.48}) and (\ref{4.49}).}
\label{T.6}
\begin{center}
\begin{tabular}{cclcl}
\hline
\hline \\*[-1ex]
$L$ && 
\multicolumn{1}{c}{$\alpha_{\mathrm{E}L\to\mathrm{M}(L-1)}$} 
&&
\multicolumn{1}{c}{$\alpha_{\mathrm{E}L\to\mathrm{M}(L+1)}$} 
\\*[1ex]
\hline \\
1 && 
&&
$\displaystyle\frac{\alpha a_{0}^{4}}{Z^{4}}
\frac{9}{2}\left[1-\frac{409}{360}(\alpha Z)^{2}\right]$
\\*[4ex]
2 && 
$\displaystyle\frac{\alpha a_{0}^{4}}{Z^{4}}
\frac{23}{120}(\alpha Z)^{2}$
&& 
$\displaystyle\frac{\alpha a_{0}^{6}}{Z^{6}}
\frac{45}{2}\left[1-\frac{1793}{1260}(\alpha Z)^{2}\right]$
\\*[4ex]
3 && 
$\displaystyle\frac{\alpha a_{0}^{6}}{Z^{6}}
\frac{113}{56}(\alpha Z)^{2}$
&& 
$\displaystyle\frac{\alpha a_{0}^{8}}{Z^{8}}
\frac{525}{2}\left[1-\frac{3317}{2016}(\alpha Z)^{2}\right]$
\\*[4ex]
4 && 
$\displaystyle\frac{\alpha a_{0}^{8}}{Z^{8}}
\frac{1017}{32}(\alpha Z)^{2}$
&&
$\displaystyle\frac{\alpha a_{0}^{10}}{Z^{10}}
\frac{42525}{8}\left[1-\frac{759449}{415800}(\alpha Z)^{2}\right]$
\\*[3ex]
\hline
\hline
\end{tabular}
\end{center}
\end{table}
%
%%%%%%%%%%%%%%%%%%%%%%%%%
%       TABLE 7
%%%%%%%%%%%%%%%%%%%%%%%%%
%
\begin{landscape}
\begin{table}[t]
\caption{The static electric-to-toroidal-magnetic multipole
cross-susceptibilities $\alpha_{\mathrm{E}L\to\mathrm{T}L}$ with
$1\leqslant L\leqslant4$ for selected hydrogenic ions in the ground
state, computed from the analytical formula in Eq.\ (\ref{5.23}).
The number in brackets following the entries is the power of 10 by
which the entry is to be multiplied. The value of the inverse of the
fine-structure constant used in calculations has been
$\alpha^{-1}=137.035\:999\:139$ (from CODATA 2014).}
\label{T.7}
\vspace*{1ex}
{\footnotesize
\begin{center}
\begin{tabular}{rllll}
\hline
\hline \\
\multicolumn{1}{c}{$Z$} & 
\multicolumn{1}{c}{$\alpha_{\mathrm{E}1\to\mathrm{T}1}$ 
($a_{0}^{4}$)} &
\multicolumn{1}{c}{$\alpha_{\mathrm{E}2\to\mathrm{T}2}$ 
($a_{0}^{6}$)} &
\multicolumn{1}{c}{$\alpha_{\mathrm{E}3\to\mathrm{T}3}$ 
($a_{0}^{8}$)} &
\multicolumn{1}{c}{$\alpha_{\mathrm{E}4\to\mathrm{T}4}$ 
($a_{0}^{10}$)} \\*[1ex]
\hline \\
1 & 
8.208\:992\:048\:775\:267\:692\:967\:($-$3) & 
1.824\:181\:749\:468\:105\:034\:635\:($-$2) & 
1.197\:103\:615\:575\:008\:327\:209\:($-$1) & 
1.551\:430\:471\:541\:743\:800\:219\:($+$0) \\*[0.5ex]
2 & 
5.129\:627\:095\:441\:285\:938\:736\:($-$4) & 
2.849\:550\:959\:876\:121\:988\:679\:($-$4) & 
4.674\:799\:949\:880\:579\:980\:738\:($-$4) & 
1.514\:573\:425\:522\:621\:881\:987\:($-$3) \\*[0.5ex]
5 & 
1.311\:405\:660\:900\:123\:528\:828\:($-$5) & 
1.165\:075\:455\:456\:423\:974\:216\:($-$6) & 
3.057\:322\:790\:698\:241\:861\:565\:($-$7) & 
1.584\:512\:136\:366\:077\:505\:749\:($-$7) \\*[0.5ex]
10 & 
8.156\:619\:781\:840\:358\:592\:750\:($-$7) & 
1.808\:732\:920\:303\:151\:270\:625\:($-$8) & 
1.185\:426\:820\:773\:881\:525\:597\:($-$9) & 
1.534\:745\:348\:089\:175\:903\:565\:($-$10) \\*[0.5ex]
20 & 
4.999\:128\:489\:397\:474\:153\:175\:($-$8) & 
2.753\:638\:596\:007\:254\:295\:135\:($-$10) & 
4.493\:949\:807\:482\:088\:856\:482\:($-$12) & 
1.450\:075\:102\:228\:029\:731\:541\:($-$13) \\*[0.5ex]
40 & 
2.881\:625\:203\:511\:699\:805\:970\:($-$9) & 
3.864\:378\:956\:524\:133\:881\:659\:($-$12) & 
1.551\:228\:291\:910\:292\:572\:158\:($-$14) & 
1.235\:602\:013\:604\:566\:784\:123\:($-$16) \\*[0.5ex]
60 & 
4.921\:478\:469\:925\:998\:029\:824\:($-$10) & 
2.797\:665\:976\:330\:397\:344\:301\:($-$13) & 
4.849\:315\:540\:872\:668\:880\:698\:($-$16) & 
1.678\:687\:380\:721\:213\:304\:038\:($-$18) \\*[0.5ex]
80 & 
1.235\:279\:761\:023\:554\:795\:070\:($-$10) & 
3.666\:029\:498\:029\:863\:171\:866\:($-$14) & 
3.417\:412\:738\:364\:817\:715\:426\:($-$17) & 
6.427\:122\:689\:305\:431\:754\:024\:($-$20) \\*[0.5ex]
100 & 
3.504\:059\:007\:754\:496\:354\:616\:($-$11) & 
5.922\:743\:695\:588\:936\:340\:089\:($-$15) & 
3.298\:079\:781\:266\:877\:967\:284\:($-$18) & 
3.764\:776\:924\:113\:921\:574\:119\:($-$21) \\*[0.5ex]
120 & 
8.770\:433\:963\:368\:894\:978\:760\:($-$12) & 
8.385\:272\:051\:823\:546\:261\:885\:($-$16) & 
2.882\:550\:881\:212\:957\:241\:036\:($-$19) & 
2.089\:625\:125\:226\:873\:612\:390\:($-$22) \\*[0.5ex]
137 & 
5.927\:356\:551\:405\:830\:197\:553\:($-$13) & 
2.011\:157\:100\:297\:711\:698\:158\:($-$17) & 
3.384\:692\:316\:101\:185\:264\:964\:($-$21) & 
1.335\:922\:787\:227\:771\:518\:934\:($-$24) \\*[1ex]
\hline
\hline
\end{tabular}
\end{center}
}
\end{table}
\end{landscape}
%
%%%%%%%%%%%%%%%%%%%%%%%%%
%       TABLE 8
%%%%%%%%%%%%%%%%%%%%%%%%%
%
\begin{table}[t]
\caption{Quasi-relativistic approximations for the static
electric-to-toroidal-magnetic multipole cross-susceptibilities
$\alpha_{\mathrm{E}L\to\mathrm{T}L}$ with $1\leqslant L\leqslant4$
for the Dirac one-electron atom in the ground state. The expressions
have been derived from Eq.\ (\ref{5.29}).}
\label{T.8}
\begin{center}
\begin{tabular}{ccl}
\hline
\hline \\*[-1ex]
$L$ && 
\multicolumn{1}{c}{$\alpha_{\mathrm{E}L\to\mathrm{T}L}$} \\*[1ex]
\hline \\
1 && 
$\displaystyle\frac{\alpha a_{0}^{4}}{Z^{4}}
\frac{9}{8}\left[1-\frac{785}{648}(\alpha Z)^{2}\right]$
\\*[4ex]
2 && 
$\displaystyle\frac{\alpha a_{0}^{6}}{Z^{6}}
\frac{5}{2}\left[1-\frac{11591}{7200}(\alpha Z)^{2}\right]$
\\*[4ex]
3 && 
$\displaystyle\frac{\alpha a_{0}^{8}}{Z^{8}}
\frac{525}{32}\left[1-\frac{654611}{352800}(\alpha Z)^{2}\right]$
\\*[4ex]
4 && 
$\displaystyle\frac{\alpha a_{0}^{10}}{Z^{10}}
\frac{1701}{8}\left[1-\frac{8356217}{4082400}(\alpha Z)^{2}\right]$
\\*[3ex]
\hline
\hline
\end{tabular}
\end{center}
\end{table}
%
%%%%%%%%%%%%%%%%%%%%%%%%%
%       TABLE 9
%%%%%%%%%%%%%%%%%%%%%%%%%
%
\begin{landscape}
\begin{table}[t]
\caption{Exact analytical expressions for the static electric
multipole nuclear shielding constants
$\sigma_{\mathrm{E}L\to\mathrm{E}L}$ with $1\leqslant L\leqslant4$
for the Dirac one-electron atom in the ground state. The expressions
have been derived from Eq.\ (\ref{6.19}).}
\label{T.9}
\begin{center}
\begin{tabular}{cclcc}
\hline
\hline \\*[-1ex]
$L$ && 
\multicolumn{1}{c}{$\sigma_{\mathrm{E}L\to\mathrm{E}L}$} &&
constraint on $Z$ \\*[1ex]
\hline \\
1 && 
$\displaystyle\frac{1}{Z}$ &&
$Z<\alpha^{-1}$ \\*[4ex]
2 && 
$\displaystyle\frac{1}{Z}\frac{104\gamma_{1}^{2}+110\gamma_{1}-79}
{15(2\gamma_{1}+7)(4\gamma_{1}-1)}$ &&
$\displaystyle Z<\alpha^{-1}\frac{\sqrt{15}}{4}$ \\*[4ex]
3 && 
$\displaystyle\frac{1}{Z}
\frac{2064\gamma_{1}^{4}+14764\gamma_{1}^{3}+30968\gamma_{1}^{2}
+7181\gamma_{1}-17177}
{42(\gamma_{1}+7)(2\gamma_{1}+7)(4\gamma_{1}+11)(6\gamma_{1}-1)}$ &&
$\displaystyle Z<\alpha^{-1}\frac{\sqrt{35}}{6}$ \\*[4ex]
4 && 
$\displaystyle\frac{1}{Z}
\frac{14208\gamma_{1}^{6}+251184\gamma_{1}^{5}+1662556\gamma_{1}^{4}
+4813404\gamma_{1}^{3}+5195413\gamma_{1}^{2}-862740\gamma_{1}-3136025}
{90(\gamma_{1}+5)(\gamma_{1}+7)(2\gamma_{1}+5)(2\gamma_{1}+23)
(4\gamma_{1}+11)(8\gamma_{1}-1)}$ &&
$\displaystyle Z<\alpha^{-1}\frac{3\sqrt{7}}{8}$ \\*[3ex]
\hline
\hline
\end{tabular}
\end{center}
\end{table}
\end{landscape}
%
%%%%%%%%%%%%%%%%%%%%%%%%%
%       TABLE 10
%%%%%%%%%%%%%%%%%%%%%%%%%
%
\begin{table}[t]
\caption{Quasi-relativistic approximations for static electric
multipole shielding constants $\sigma_{\mathrm{E}L\to\mathrm{E}L}$
with $1\leqslant L\leqslant4$ for the Dirac one-electron atom in the
ground state. The expressions have been derived from Eq.\
(\ref{6.27}).}
\label{T.10}
\begin{center}
\begin{tabular}{ccl}
\hline
\hline \\*[-1ex]
$L$ && 
\multicolumn{1}{c}{$\sigma_{\mathrm{E}L\to\mathrm{E}L}$}
\\*[1ex]
\hline \\
1 && 
$\displaystyle\frac{1}{Z}$ \quad (exact)
\\*[4ex]
2 && 
$\displaystyle\frac{1}{Z}
\frac{1}{3}\left[1-\frac{2}{5}(\alpha Z)^{2}\right]$ 
\\*[4ex]
3 && 
$\displaystyle\frac{1}{Z}
\frac{1}{6}\left[1-\frac{59}{84}(\alpha Z)^{2}\right]$
\\*[4ex]
4 && 
$\displaystyle\frac{1}{Z}
\frac{1}{10}\left[1-\frac{529}{540}(\alpha Z)^{2}\right]$
\\*[3ex]
\hline
\hline
\end{tabular}
\end{center}
\end{table}
%
%%%%%%%%%%%%%%%%%%%%%%%%%
%       TABLE 11
%%%%%%%%%%%%%%%%%%%%%%%%%
%
\begin{landscape}
\begin{table}[t]
\caption{Exact analytical expressions for the near-nucleus static
electric-to-magnetic multipole cross-susceptibilities of the Dirac
one-electron atom in the ground state. The expressions for
$\sigma_{\mathrm{E}L\to\mathrm{M}(L-1)}$ with $2\leqslant
L\leqslant4$, derived from Eq.\ (\ref{7.16}) and given in the second
column, are valid provided that $Z<\alpha^{-1}$. The last column
displays constraints on the nuclear charge number $Z$ under which the
expressions for $\sigma_{\mathrm{E}L\to\mathrm{M}(L+1)}$ with
$1\leqslant L\leqslant4$, obtained from Eq.\ (\ref{7.17}) and given
in the third column, remain valid.}
\label{T.11}
\begin{center}
\begin{tabular}{cclclcc}
\hline
\hline \\*[-1ex]
$L$ && 
\multicolumn{1}{c}{$\sigma_{\mathrm{E}L\to\mathrm{M}(L-1)}$} &&
\multicolumn{1}{c}{$\sigma_{\mathrm{E}L\to\mathrm{M}(L+1)}$} && 
constraint on $Z$ \\
&&&&&& in $\sigma_{\mathrm{E}L\to\mathrm{M}(L+1)}$ \\*[1ex]
\hline \\
1 && 
&&
$\displaystyle-\frac{\alpha Z}{a_{0}}
\frac{6(3\gamma_{1}+1)}{5\gamma_{1}(\gamma_{1}+1)(4\gamma_{1}-1)}$
&& $\displaystyle Z<\alpha^{-1}\frac{\sqrt{15}}{4}$ \\*[4ex]
2 && 
$\displaystyle-\frac{\alpha a_{0}}{Z}\frac{2\gamma_{1}+1}{9}$ 
&& 
$\displaystyle\phantom{-}\frac{\alpha Z}{a_{0}}
\frac{16(6\gamma_{1}^{3}-\gamma_{1}^{2}-36\gamma_{1}-14)}
{35\gamma_{1}(\gamma_{1}+1)(2\gamma_{1}+7)(6\gamma_{1}-1)}$ 
&& $\displaystyle Z<\alpha^{-1}\frac{\sqrt{35}}{6}$ \\*[4ex]
3 && 
$\displaystyle-\frac{\alpha a_{0}}{Z}
\frac{3(2\gamma_{1}+1)(2\gamma_{1}+5)}{28(2\gamma_{1}+7)}$ 
&& 
$\displaystyle\phantom{-}\frac{\alpha Z}{a_{0}}
\frac{50(32\gamma_{1}^{4}+164\gamma_{1}^{3}-53\gamma_{1}^{2}
-584\gamma_{1}-231)}{189\gamma_{1}(\gamma_{1}+1)(\gamma_{1}+7)
(4\gamma_{1}+11)(8\gamma_{1}-1)}$ 
&& $\displaystyle Z<\alpha^{-1}\frac{3\sqrt{7}}{8}$ \\*[4ex]
4 && 
$\displaystyle-\frac{\alpha a_{0}}{Z}
\frac{2(2\gamma_{1}+1)(68\gamma_{1}^{2}+483\gamma_{1}+709)}
{315(\gamma_{1}+7)(4\gamma_{1}+11)}$ 
&&
$\displaystyle\phantom{-}\frac{\alpha Z}{a_{0}}
\frac{2(60\gamma_{1}^{5}+744\gamma_{1}^{4}+2065\gamma_{1}^{3}
-964\gamma_{1}^{2}-5905\gamma_{1}-2300)}
{11\gamma_{1}(\gamma_{1}+1)(\gamma_{1}+5)(2\gamma_{1}+5)
(2\gamma_{1}+23)(10\gamma_{1}-1)}$ 
&& $\displaystyle Z<\alpha^{-1}\frac{3\sqrt{11}}{10}$ \\*[3ex]
\hline
\hline
\end{tabular}
\end{center}
\end{table}
\end{landscape}
%
%%%%%%%%%%%%%%%%%%%%%%%%%
%       TABLE 12
%%%%%%%%%%%%%%%%%%%%%%%%%
%
\begin{table}[t]
\caption{Quasi-relativistic approximations for the near-nucleus
static electric-to-magnetic multipole cross-susceptibilities
$\sigma_{\mathrm{E}L\to\mathrm{M}(L-1)}$ with $2\leqslant
L\leqslant4$ and $\sigma_{\mathrm{E}L\to\mathrm{M}(L+1)}$ with
$1\leqslant L\leqslant4$ for the Dirac one-electron atom in the
ground state. The expressions have been derived from Eqs.\
(\ref{7.24}) and (\ref{7.25}).}
\label{T.12}
\begin{center}
\begin{tabular}{cclcl}
\hline
\hline \\*[-1ex]
$L$ && 
\multicolumn{1}{c}{$\sigma_{\mathrm{E}L\to\mathrm{M}(L-1)}$} 
&&
\multicolumn{1}{c}{$\sigma_{\mathrm{E}L\to\mathrm{M}(L+1)}$} 
\\*[1ex]
\hline \\
1 && 
&&
$\displaystyle-\frac{\alpha a_{0}}{Z}
\frac{4}{5}\left[1+\frac{25}{24}(\alpha Z)^{2}\right]$ 
\\*[4ex]
2 && 
$\displaystyle-\frac{\alpha Z}{a_{0}}
\frac{1}{3}\left[1-\frac{1}{3}(\alpha Z)^{2}\right]$ 
&&
$\displaystyle-\frac{\alpha a_{0}}{Z}
\frac{8}{35}\left[1+\frac{223}{180}(\alpha Z)^{2}\right]$ 
\\*[4ex]
3 && 
$\displaystyle-\frac{\alpha Z}{a_{0}}
\frac{1}{4}\left[1-\frac{23}{63}(\alpha Z)^{2}\right]$ 
&&
$\displaystyle-\frac{\alpha a_{0}}{Z}
\frac{20}{189}\left[1+\frac{1641}{1120}(\alpha Z)^{2}\right]$ 
\\*[4ex]
4 && 
$\displaystyle-\frac{\alpha Z}{a_{0}}
\frac{1}{5}\left[1-\frac{1931}{5040}(\alpha Z)^{2}\right]$ 
&&
$\displaystyle-\frac{\alpha a_{0}}{Z}
\frac{2}{33}\left[1+\frac{10721}{6300}(\alpha Z)^{2}\right]$ 
\\*[3ex]
\hline
\hline
\end{tabular}
\end{center}
\end{table}
%
%%%%%%%%%%%%%%%%%%%%%%%%%
%       TABLE 13
%%%%%%%%%%%%%%%%%%%%%%%%%
%
\begin{table}[t]
\caption{Exact analytical expressions for the near-nucleus static
electric-to-magnetic-toroidal multipole cross-susceptibilities
$\sigma_{\mathrm{E}L\to\mathrm{T}L}$ with $1\leqslant L\leqslant4$
for the Dirac one-electron atom in the ground state. The formulas
have been derived from Eq.\ (\ref{8.16}) and are valid under the
constraint $Z<\alpha^{-1}$.}
\label{T.13}
\begin{center}
\begin{tabular}{ccl}
\hline
\hline \\*[-1ex]
$L$ && 
\multicolumn{1}{c}{$\sigma_{\mathrm{E}L\to\mathrm{T}L}$} 
\\*[1ex]
\hline \\
1 && 
$\displaystyle-\frac{\alpha a_{0}}{Z}
\frac{1}{9}\left[\frac{(\gamma_{1}+1)(4\gamma_{1}+1)}{2\gamma_{1}}
+\frac{(\gamma_{1}-2)(2\gamma_{1}+1)}{\gamma_{2}+\gamma_{1}+3}\,
{}_{3}F_{2}\left(
\begin{array}{c}
1,\:
1,\:
\gamma_{2}-\gamma_{1}-1 \\
\gamma_{2}-\gamma_{1}+2,\:
\gamma_{2}+\gamma_{1}+1
\end{array}
;1
\right)
\right]$
\\*[4ex]
2 && 
$\displaystyle-\frac{\alpha a_{0}}{Z}
\frac{(2\gamma_{1}+1)(2\gamma_{1}+3)}{20(2\gamma_{1}+7)}$
\\*[4ex]
3 && 
$\displaystyle-\frac{\alpha a_{0}}{Z}
\frac{(\gamma_{1}+1)(2\gamma_{1}+1)(4\gamma_{1}^{2}+34\gamma_{1}+67)}
{21(\gamma_{1}+7)(2\gamma_{1}+7)(4\gamma_{1}+11)}$
\\*[4ex]
4 && 
$\displaystyle-\frac{\alpha a_{0}}{Z}
\frac{(2\gamma_{1}+1)(48\gamma_{1}^{5}+1092\gamma_{1}^{4}
+8856\gamma_{1}^{3}+31625\gamma_{1}^{2}+48384\gamma_{1}+23395)}
{216(\gamma_{1}+5)(\gamma_{1}+7)(2\gamma_{1}+5)(2\gamma_{1}+23)
(4\gamma_{1}+11)}$
\\*[3ex]
\hline
\hline
\end{tabular}
\end{center}
\end{table}
%
%%%%%%%%%%%%%%%%%%%%%%%%%
%       TABLE 14
%%%%%%%%%%%%%%%%%%%%%%%%%
%
\begin{table}[t]
\caption{The near-nucleus static electric-to-toroidal-magnetic dipole
cross-susceptibilities $\sigma_{\mathrm{E}1\to\mathrm{T}1}$ for
selected hydrogenic ions in the ground state, computed from Eq.\
(\ref{8.16}) with $L=1$. The number in brackets following the
entries is the power of 10 by which the entry is to be multiplied.
The value of the inverse of the fine-structure constant used in
calculations has been $\alpha^{-1}=137.035\:999\:139$ (from CODATA
2014).}
\label{T.14}
\vspace*{1ex}
\begin{center}
\begin{tabular}{rllll}
\hline
\hline \\
\multicolumn{1}{c}{$Z$} & 
\multicolumn{1}{c}{$\sigma_{\mathrm{E}1\to\mathrm{T}1}$ ($a_{0}$)}
\\*[1ex]
\hline \\
1 & 
$-$3.648\:637\:095\:595\:566\:551\:586\:($-$3) \\*[0.5ex]
2 & 
$-$1.824\:259\:765\:869\:242\:452\:438\:($-$3) \\*[0.5ex]
5 & 
$-$7.295\:393\:076\:288\:932\:910\:227\:($-$4) \\*[0.5ex]
10 & 
$-$3.644\:756\:614\:282\:248\:624\:473\:($-$4) \\*[0.5ex]
20 & 
$-$1.816\:493\:443\:607\:184\:794\:164\:($-$4) \\*[0.5ex]
40 & 
$-$8.964\:422\:142\:806\:221\:802\:427\:($-$5) \\*[0.5ex]
60 & 
$-$5.844\:669\:492\:463\:283\:200\:203\:($-$5) \\*[0.5ex]
80 & 
$-$4.246\:284\:894\:885\:672\:771\:322\:($-$5) \\*[0.5ex]
100 & 
$-$3.264\:717\:305\:265\:228\:591\:534\:($-$5) \\*[0.5ex]
120 & 
$-$2.646\:549\:207\:670\:313\:520\:251\:($-$5) \\*[0.5ex]
137 & 
$-$1.413\:712\:752\:786\:307\:100\:801\:($-$4) \\*[1ex]
\hline
\hline
\end{tabular}
\end{center}
\end{table}
%
%%%%%%%%%%%%%%%%%%%%%%%%%
%       TABLE 15
%%%%%%%%%%%%%%%%%%%%%%%%%
%
\begin{table}[t]
\caption{Quasi-relativistic approximations for the near-nucleus
static electric-to-toroidal-magnetic multipole cross-susceptibilities
$\sigma_{\mathrm{E}L\to\mathrm{T}L}$ with $1\leqslant L\leqslant4$
for the Dirac one-electron atom in the ground state. The expression
for $L=1$ is the one displayed in Eq.\ (\ref{8.23}), while these for
$2\leqslant L\leqslant4$ have been derived from Eq.\ (\ref{8.21}).}
\label{T.15}
\begin{center}
\begin{tabular}{ccl}
\hline
\hline \\*[-1ex]
$L$ && 
\multicolumn{1}{c}{$\sigma_{\mathrm{E}L\to\mathrm{T}L}$} \\*[1ex]
\hline \\
1 && 
$\displaystyle-\frac{\alpha a_{0}}{Z}
\frac{1}{2}\left[1-\left(\frac{3}{4}-\frac{\pi^{2}}{18}\right)
(\alpha Z)^{2}\right]$
\\*[4ex]
2 && 
$\displaystyle-\frac{\alpha a_{0}}{Z}
\frac{1}{12}\left[1-\frac{19}{45}(\alpha Z)^{2}\right]$
\\*[4ex]
3 && 
$\displaystyle-\frac{\alpha a_{0}}{Z}
\frac{1}{36}\left[1-\frac{343}{720}(\alpha Z)^{2}\right]$
\\*[4ex]
4 && 
$\displaystyle-\frac{\alpha a_{0}}{Z}
\frac{1}{80}\left[1-\frac{113623}{226800}(\alpha Z)^{2}\right]$
\\*[3ex]
\hline
\hline
\end{tabular}
\end{center}
\end{table}
%
%%%%%%%%%%%%%%%%%%%%%%%%%
%       TABLE 16
%%%%%%%%%%%%%%%%%%%%%%%%%
%
\begin{table}[t] 
\caption{The collection of formulas defining the multipole
susceptibilities considered in the present paper.}
\label{T.16} 
\vspace*{1ex} 
\begin{center} 
\begin{tabular}{lclcl}
\hline 
\hline \\
\multicolumn{1}{c}{susceptibility} &&
\multicolumn{1}{c}{related induced moment} &&
\multicolumn{1}{c}{constraints} \\*[1ex]
\hline \\
\multicolumn{5}{c}{far-field zone} 
\\*[1ex]
$\alpha_{\mathrm{E}L\to\mathrm{E}L}$ 
&& $\boldsymbol{\mathsf{Q}}_{L}^{(1)}=(4\pi\epsilon_{0})
\alpha_{\mathrm{E}L\to\mathrm{E}L}\,
\boldsymbol{\mathsf{C}}_{L}^{(1)}$
&& \\*[3ex]
$\alpha_{\mathrm{E}L\to\mathrm{M}\lambda}$
&& $\displaystyle\boldsymbol{\mathsf{M}}_{\lambda}^{(1)}
=(4\pi\epsilon_{0})c\alpha_{\mathrm{E}L\to\mathrm{M}\lambda}\,
\frac{\big\{\boldsymbol{\nu}\otimes\boldsymbol{\mathsf{C}}_{L}^{(1)}
\big\}_{\lambda}}{\langle10L0\big|\lambda0\rangle}$
&& $\lambda=
\left\{
\begin{array}{ll}
2 & \textrm{for $L=1$} \\
L\mp1 & \textrm{for $L\geqslant2$}
\end{array}
\right.$ \\*[3ex]
$\alpha_{\mathrm{E}L\to\mathrm{T}L}$ 
&& $\displaystyle\boldsymbol{\mathsf{T}}_{L}^{(1)}
=\mathrm{i}(4\pi\epsilon_{0})c\alpha_{\mathrm{E}L\to\mathrm{T}L}\,
\sqrt{L(L+1)}\,\big\{\boldsymbol{\nu}
\otimes\boldsymbol{\mathsf{C}}_{L}^{(1)}\big\}_{L}$
&& \\*[3ex]
\multicolumn{5}{c}{near-nucleus zone} 
\\*[1ex]
$\sigma_{\mathrm{E}L\to\mathrm{E}L}$ 
&& $\boldsymbol{\mathsf{R}}_{L}^{(1)}=(4\pi\epsilon_{0})
\sigma_{\mathrm{E}L\to\mathrm{E}L}\,
\boldsymbol{\mathsf{C}}_{L}^{(1)}$
&& \\*[3ex]
$\sigma_{\mathrm{E}L\to\mathrm{M}\lambda}$
&& $\displaystyle\boldsymbol{\mathsf{N}}_{\lambda}^{(1)}
=(4\pi\epsilon_{0})c\sigma_{\mathrm{E}L\to\mathrm{M}\lambda}\,
\frac{\big\{\boldsymbol{\nu}\otimes\boldsymbol{\mathsf{C}}_{L}^{(1)}
\big\}_{\lambda}}{\langle10L0\big|\lambda0\rangle}$
&& $\lambda=
\left\{
\begin{array}{ll}
2 & \textrm{for $L=1$} \\
L\mp1 & \textrm{for $L\geqslant2$}
\end{array}
\right.$ \\*[3ex]
$\sigma_{\mathrm{E}L\to\mathrm{T}L}$ 
&& $\displaystyle\boldsymbol{\mathsf{U}}_{L}^{(1)}
=\mathrm{i}(4\pi\epsilon_{0})c\sigma_{\mathrm{E}L\to\mathrm{T}L}\,
\sqrt{L(L+1)}\,\big\{\boldsymbol{\nu}
\otimes\boldsymbol{\mathsf{C}}_{L}^{(1)}\big\}_{L}$
&& \\*[3ex]
\mbox{} &&&& \\
\hline 
\hline
\end{tabular} 
\end{center} 
\end{table}
%
%%%%%%%%%%%%%%%%%%%%%%%%%
%       TABLE 17
%%%%%%%%%%%%%%%%%%%%%%%%%
%
\begin{landscape} 
\begin{table}[t] 
\caption{The table shows how the susceptibilities studied in the
present paper enter the near- and far-zone asymptotic representations
of static electric $\boldsymbol{E}^{(1)}(\boldsymbol{r})$ and
magnetic $\boldsymbol{B}^{(1)}(\boldsymbol{r})$ fields, and of their
potentials: scalar $\phi^{(1)}(\boldsymbol{r})$ and vector
$\boldsymbol{A}^{(1)}(\boldsymbol{r})$, which are due to the
first-order charge and current densities induced in the ground state
of the hydrogenic atom by an external $2^{L}$-pole ($L\geqslant1$)
electric field $\boldsymbol{\mathcal{E}}_{L}^{(1)}(\boldsymbol{r})$
derivable from the scalar potential
$\varphi_{L}^{(1)}(\boldsymbol{r})$ defined in Eq.\ (\ref{2.1}).}
\label{T.17} 
\vspace*{1ex} 
{\footnotesize
\begin{center}
\begin{tabular}{lclcl}
\hline 
\hline \\
\multicolumn{1}{c}{induced field} &&
\multicolumn{1}{c}{near-zone representation} &&
\multicolumn{1}{c}{far-zone representation} \\*[1ex]
\hline \\
$\phi^{(1)}(\boldsymbol{r})$ 
&& $\displaystyle\sigma_{\mathrm{E}L\to\mathrm{E}L}
\sqrt{\frac{4\pi}{2L+1}}\,r^{L}\sum_{M=-L}^{L}\mathcal{C}_{LM}^{(1)}
Y_{LM}^{*}(\boldsymbol{n}_{r})=-\sigma_{\mathrm{E}L\to\mathrm{E}L}
\varphi_{L}^{(1)}(\boldsymbol{r})$
&& $\displaystyle\alpha_{\mathrm{E}L\to\mathrm{E}L}
\sqrt{\frac{4\pi}{2L+1}}\,r^{-L-1}
\sum_{M=-L}^{L}\mathcal{C}_{LM}^{(1)}Y_{LM}^{*}(\boldsymbol{n}_{r})$
\\*[3ex]
$\boldsymbol{E}^{(1)}(\boldsymbol{r})$ 
&& $\displaystyle-\sigma_{\mathrm{E}L\to\mathrm{E}L}
\sqrt{4\pi L}\,r^{L-1}\sum_{M=-L}^{L}\mathcal{C}_{LM}^{(1)}
\boldsymbol{Y}_{LM}^{L-1*}(\boldsymbol{n}_{r})
=-\sigma_{\mathrm{E}L\to\mathrm{E}L}
\boldsymbol{\mathcal{E}}_{L}^{(1)}(\boldsymbol{r})$
&& $\displaystyle-\alpha_{\mathrm{E}L\to\mathrm{E}L}
\sqrt{4\pi(L+1)}\,r^{-L-2}\sum_{M=-L}^{L}\mathcal{C}_{LM}^{(1)}
\boldsymbol{Y}_{LM}^{L+1*}(\boldsymbol{n}_{r})$
\\*[3ex]
$\boldsymbol{A}^{(1)}(\boldsymbol{r})$
&& $\displaystyle-\sum_{\lambda=L\mp1}\mathrm{i}c^{-1}
\sigma_{\mathrm{E}L\to\mathrm{M}\lambda}(1-\delta_{\lambda0})
\sqrt{\frac{4\pi\lambda}{(\lambda+1)(2\lambda+1)}}\,r^{\lambda}
\sum_{\mu=-\lambda}^{\lambda}\frac{\big\{\boldsymbol{\nu}
\otimes\boldsymbol{\mathsf{C}}_{L}^{(1)}\big\}_{\lambda\mu}}
{\langle10L0\big|\lambda0\rangle}
\boldsymbol{Y}_{\lambda\mu}^{\lambda*}(\boldsymbol{n}_{r})$ 
&& $\displaystyle\sum_{\lambda=L\mp1}\mathrm{i}c^{-1}
\alpha_{\mathrm{E}L\to\mathrm{M}\lambda}(1-\delta_{\lambda0})
\sqrt{\frac{4\pi(\lambda+1)}{\lambda(2\lambda+1)}}\,r^{-\lambda-1}
\sum_{\mu=-\lambda}^{\lambda}\frac{\big\{\boldsymbol{\nu}
\otimes\boldsymbol{\mathsf{C}}_{L}^{(1)}\big\}_{\lambda\mu}}
{\langle10L0\big|\lambda0\rangle}
\boldsymbol{Y}_{\lambda\mu}^{\lambda*}(\boldsymbol{n}_{r})$ \\
&& $\displaystyle-\,\mathrm{i}c^{-1}\sigma_{\mathrm{E}L\to\mathrm{T}L}
\sqrt{4\pi(L+1)}\,Lr^{L-1}\sum_{M=-L}^{L}\big\{\boldsymbol{\nu}
\otimes\boldsymbol{\mathsf{C}}_{L}^{(1)}\big\}_{LM}
\boldsymbol{Y}_{LM}^{L-1*}(\boldsymbol{n}_{r})$
&& $\displaystyle-\,\mathrm{i}c^{-1}\alpha_{\mathrm{E}L\to\mathrm{T}L}
\sqrt{4\pi L}\,(L+1)r^{-L-2}\sum_{M=-L}^{L}\big\{\boldsymbol{\nu}
\otimes\boldsymbol{\mathsf{C}}_{L}^{(1)}\big\}_{LM}
\boldsymbol{Y}_{LM}^{L+1*}(\boldsymbol{n}_{r})$
\\*[3ex]
$\boldsymbol{B}^{(1)}(\boldsymbol{r})$ 
&& $-\displaystyle\sum_{\lambda=L\mp1}c^{-1}
\sigma_{\mathrm{E}L\to\mathrm{M}\lambda}(1-\delta_{\lambda0})
\sqrt{4\pi\lambda}\,r^{\lambda-1}
\sum_{\mu=-\lambda}^{\lambda}\frac{\big\{\boldsymbol{\nu}
\otimes\boldsymbol{\mathsf{C}}_{L}^{(1)}\big\}_{\lambda\mu}}
{\langle10L0\big|\lambda0\rangle}
\boldsymbol{Y}_{\lambda\mu}^{\lambda-1*}(\boldsymbol{n}_{r})$
&& $\displaystyle-\sum_{\lambda=L\mp1}c^{-1}
\alpha_{\mathrm{E}L\to\mathrm{M}\lambda}(1-\delta_{\lambda0})
\sqrt{4\pi(\lambda+1)}\,r^{-\lambda-2}\sum_{\mu=-\lambda}^{\lambda}
\frac{\big\{\boldsymbol{\nu}
\otimes\boldsymbol{\mathsf{C}}_{L}^{(1)}\big\}_{\lambda\mu}}
{\langle10L0\big|\lambda0\rangle}
\boldsymbol{Y}_{\lambda\mu}^{\lambda+1*}(\boldsymbol{n}_{r})$
\\*[3ex]
\mbox{} && \\
\hline 
\hline
\end{tabular} 
\end{center}
}
\end{table}
\end{landscape}
%
%%%%%%%%%%%%%%%%%%%%%%%%%
%       TABLE 18
%%%%%%%%%%%%%%%%%%%%%%%%%
%
\begin{landscape} 
\begin{table}[t] 
\caption{The collection of exact analytical expressions for the
far-field static electric multipole susceptibilities for the Dirac
one-electron atom in the ground state: the polarizability
$\alpha_{\mathrm{E}L\to\mathrm{E}L}$ ($\equiv\alpha_{L}$), the
electric-to-magnetic cross-susceptibilities
$\alpha_{\mathrm{E}L\to\mathrm{M}(L\mp1)}$ and the
electric-to-toroidal-magnetic cross-susceptibility
$\alpha_{\mathrm{E}L\to\mathrm{T}L}$. All the formulas are valid for
$L\geqslant1$ and under the constraint $Z<\alpha^{-1}$.}
\label{T.18} 
\vspace*{1ex} 
{\footnotesize
\begin{center} 
\begin{tabular}{l} 
\hline 
\hline \\
\multicolumn{1}{c}{susceptibility} \\*[1ex]
\hline \\
$\begin{array}{lcl}
\displaystyle 
\alpha_{\mathrm{E}L\to\mathrm{E}L} & 
\displaystyle = & 
\displaystyle
\frac{a_{0}^{2L+1}}{Z^{2L+2}}\frac{\Gamma(2\gamma_{1}+2L+2)}
{2^{2L}L(L+1)(2L+1)\Gamma(2\gamma_{1}+1)}
\\*[3ex]
&&
\displaystyle
\times\,\bigg\{1+\frac{L^{2}(L+1)^{2}(\gamma_{1}+1)^{2}
\Gamma^{2}(\gamma_{L}+\gamma_{1}+L+1)}
{(2L+1)(\gamma_{L}-\gamma_{1}+1)\Gamma(2\gamma_{1}+2L+2)
\Gamma(2\gamma_{L}+1)}\,
{}_{3}F_{2}
\left(
\begin{array}{c}
\gamma_{L}-\gamma_{1}-L,\:
\gamma_{L}-\gamma_{1}-L,\:
\gamma_{L}-\gamma_{1}+1 \\
\gamma_{L}-\gamma_{1}+2,\:
2\gamma_{L}+1
\end{array}
;1
\right) 
\\*[3ex]
&& 
\displaystyle
\quad -\,\frac{(L+1)^{2}[L(\gamma_{1}-1)-1]^{2}
\Gamma^{2}(\gamma_{L+1}+\gamma_{1}+L+1)}
{(2L+1)(\gamma_{L+1}-\gamma_{1}+1)\Gamma(2\gamma_{1}+2L+2)
\Gamma(2\gamma_{L+1}+1)}\,
{}_{3}F_{2}
\left(
\begin{array}{c}
\gamma_{L+1}-\gamma_{1}-L,\:
\gamma_{L+1}-\gamma_{1}-L,\:
\gamma_{L+1}-\gamma_{1}+1 \\
\gamma_{L+1}-\gamma_{1}+2,\:
2\gamma_{L+1}+1
\end{array}
;1
\right)
\bigg\}
\end{array}
$
\\
\mbox{} \\
$\begin{array}{lcl}
\displaystyle
\alpha_{\mathrm{E}L\to\mathrm{M}(L-1)} & 
\displaystyle = & 
\displaystyle
\frac{\alpha a_{0}^{2L}}{Z^{2L}}
\frac{(L-1)\Gamma(2\gamma_{1}+2L+1)}
{2^{2L-1}(L+1)(4L^{2}-1)\Gamma(2\gamma_{1}+1)} 
\\*[3ex]
&& 
\displaystyle
\times\,\bigg\{1-\frac{L(L+1)(\gamma_{1}+1)
\Gamma(\gamma_{L}+\gamma_{1}+L)\Gamma(\gamma_{L}+\gamma_{1}+L+1)}
{(\gamma_{L}-\gamma_{1}+1)\Gamma(2\gamma_{1}+2L+1)
\Gamma(2\gamma_{L}+1)}\,
{}_{3}F_{2}
\left(
\begin{array}{c}
\gamma_{L}-\gamma_{1}-L,\:
\gamma_{L}-\gamma_{1}-L+1,\:
\gamma_{L}-\gamma_{1}+1 \\
\gamma_{L}-\gamma_{1}+2,\:
2\gamma_{L}+1
\end{array}
;1
\right)
\bigg\}
\end{array}
$
\\
\mbox{} \\
$\begin{array}{lcl}
\displaystyle
\alpha_{\mathrm{E}L\to\mathrm{M}(L+1)} &
\displaystyle = & 
\displaystyle
\frac{\alpha a_{0}^{2L+2}}{Z^{2L+2}}
\frac{(L+1)\Gamma(2\gamma_{1}+2L+3)}
{2^{2L+1}L(2L+1)(2L+3)\Gamma(2\gamma_{1}+1)}
\\*[3ex]
&&
\displaystyle
\times\,\bigg\{1+\frac{(L+2)[L(\gamma_{1}-1)-1]
\Gamma(\gamma_{L+1}+\gamma_{1}+L+1)
\Gamma(\gamma_{L+1}+\gamma_{1}+L+2)}
{(\gamma_{L+1}-\gamma_{1}+1)\Gamma(2\gamma_{1}+2L+3)
\Gamma(2\gamma_{L+1}+1)}\,
{}_{3}F_{2}
\left(
\begin{array}{c}
\gamma_{L+1}-\gamma_{1}-L-1,\:
\gamma_{L+1}-\gamma_{1}-L,\:
\gamma_{L+1}-\gamma_{1}+1 \\
\gamma_{L+1}-\gamma_{1}+2,\:
2\gamma_{L+1}+1
\end{array}
;1
\right)
\bigg\}
\end{array}
$
\\
\mbox{} \\
$\begin{array}{lcl}
\displaystyle 
\alpha_{\mathrm{E}L\to\mathrm{T}L} &
\displaystyle = & 
\displaystyle 
\frac{\alpha a_{0}^{2L+2}}{Z^{2L+2}}
\frac{1}{2^{2L+1}(2L+1)^{2}\Gamma(2\gamma_{1}+1)}
\bigg\{\frac{(\gamma_{1}+1)\Gamma(\gamma_{L}+\gamma_{1}+L+1)
\Gamma(\gamma_{L}+\gamma_{1}+L+2)}
{(\gamma_{L}-\gamma_{1}+1)\Gamma(2\gamma_{L}+1)}\,
{}_{3}F_{2}
\left(
\begin{array}{c}
\gamma_{L}-\gamma_{1}-L-1,\:
\gamma_{L}-\gamma_{1}-L,\:
\gamma_{L}-\gamma_{1}+1 \\
\gamma_{L}-\gamma_{1}+2,\:
2\gamma_{L}+1
\end{array}
;1
\right)
\\*[3ex]
&& 
\displaystyle
+\,\frac{[L(\gamma_{1}-1)-1]\Gamma(\gamma_{L+1}+\gamma_{1}+L+1)
\Gamma(\gamma_{L+1}+\gamma_{1}+L+2)}
{(L+1)(\gamma_{L+1}-\gamma_{1}+1)\Gamma(2\gamma_{L+1}+1)}\,
{}_{3}F_{2}
\left(
\begin{array}{c}
\gamma_{L+1}-\gamma_{1}-L-1,\:
\gamma_{L+1}-\gamma_{1}-L,\:
\gamma_{L+1}-\gamma_{1}+1 \\
\gamma_{L+1}-\gamma_{1}+2,\:
2\gamma_{L+1}+1
\end{array}
;1
\right)
\bigg\}
\end{array}
$
\\
\mbox{} \\
\hline 
\hline
\end{tabular} 
\end{center} 
}
\end{table}
\end{landscape}
%
%%%%%%%%%%%%%%%%%%%%%%%%%
%       TABLE 19
%%%%%%%%%%%%%%%%%%%%%%%%%
%
\begin{landscape} 
\begin{table}[t] 
\caption{The collection of exact analytical expressions for the
near-nucleus static electric multipole susceptibilities for the Dirac
one-electron atom in the ground state: the electric nuclear shielding
constant $\sigma_{\mathrm{E}L\to\mathrm{E}L}$, the
electric-to-magnetic cross-susceptibilities
$\sigma_{\mathrm{E}L\to\mathrm{M}(L\mp1)}$ and the
electric-to-toroidal-magnetic cross-susceptibility
$\sigma_{\mathrm{E}L\to\mathrm{T}L}$. All the formulas hold for
$L\geqslant1$, except for the one for
$\sigma_{\mathrm{E}L\to\mathrm{M}(L-1)}$, which makes physical sense
only for $L\geqslant2$.}
\label{T.19} 
\vspace*{1ex} 
{\footnotesize
\begin{center} 
\begin{tabular}{lcl} 
\hline 
\hline \\
\multicolumn{1}{c}{susceptibility} &&
\multicolumn{1}{c}{constraints} \\*[1ex]
\hline \\
$\begin{array}{lcl}
\displaystyle
\sigma_{\mathrm{E}L\to\mathrm{E}L} & 
\displaystyle = &
\displaystyle
\frac{2}{ZL(L+1)(2L+1)}
\bigg\{1+\frac{L^{2}(L+1)(\gamma_{1}+1)[\gamma_{1}(L+1)-L]}
{(2L+1)(\gamma_{L}-\gamma_{1}+1)(\gamma_{L}+\gamma_{1}-L)}\,
{}_{3}F_{2}
\left(
\begin{array}{c}
-L+1,\:
1,\:
\gamma_{L}-\gamma_{1}-L \\
\gamma_{L}-\gamma_{1}+2,\:
\gamma_{L}+\gamma_{1}-L+1
\end{array}
;1
\right)
\\*[3ex]
&& 
\displaystyle
-\,\frac{L(L+1)^{2}(\gamma_{1}+1)[L(\gamma_{1}-1)-1]}
{(2L+1)(\gamma_{L+1}-\gamma_{1}+1)(\gamma_{L+1}+\gamma_{1}-L)}\,
{}_{3}F_{2}
\left(
\begin{array}{c}
-L+1,\:
1,\:
\gamma_{L+1}-\gamma_{1}-L \\
\gamma_{L+1}-\gamma_{1}+2,\:
\gamma_{L+1}+\gamma_{1}-L+1
\end{array}
;1
\right)
\bigg\}
\end{array}
$
&& 
$Z<
\left\{
\begin{array}{ll}
\alpha^{-1} & \textrm{for $L=1$} \\*[1ex]
\displaystyle
\alpha^{-1}\frac{\sqrt{4L^{2}-1}}{2L} 
& \textrm{for $L\geqslant2$}
\end{array}
\right.
$ \\
\mbox{} && \\
$\begin{array}{lcl}
\displaystyle
\sigma_{\mathrm{E}L\to\mathrm{M}(L-1)} & 
\displaystyle = & 
\displaystyle
-\frac{\alpha a_{0}}{Z}
\frac{L(2\gamma_{1}+1)}{(L+1)(4L^{2}-1)} 
\bigg\{1+\frac{(L^{2}-1)(\gamma_{1}+1)}
{(\gamma_{L}-\gamma_{1}+1)(\gamma_{L}+\gamma_{1}-L+1)}\,
{}_{3}F_{2}
\left(
\begin{array}{c}
-L+2,\:
1,\:
\gamma_{L}-\gamma_{1}-L \\
\gamma_{L}-\gamma_{1}+2,\:
\gamma_{L}+\gamma_{1}-L+2 
\end{array}
;1
\right)
\bigg\}
\end{array}
$
&& $\textrm{$L\geqslant2$; $Z<\alpha^{-1}$ }$ \\
\mbox{} \\
$\begin{array}{lcl}
\displaystyle
\sigma_{\mathrm{E}L\to\mathrm{M}(L+1)} &
\displaystyle = & 
\displaystyle
-\frac{\alpha Z}{a_{0}}\frac{2(L+2)}{\gamma_{1}L(2L+1)(2L+3)}
\bigg\{1-\frac{(L+1)[L(\gamma_{1}-1)-1]}
{(\gamma_{L+1}-\gamma_{1}+1)(\gamma_{L+1}+\gamma_{1}-L-1)}\,
{}_{3}F_{2}
\left(
\begin{array}{c}
-L,\:
1,\:
\gamma_{L+1}-\gamma_{1}-L \\
\gamma_{L+1}-\gamma_{1}+2,\:
\gamma_{L+1}+\gamma_{1}-L 
\end{array}
;1
\right)
\bigg\}
\end{array}
$
&& 
$\displaystyle 
Z<\alpha^{-1}\frac{\sqrt{(2L+1)(2L+3)}}{2(L+1)}$ \\
\mbox{} && \\
$\begin{array}{lcl}
\displaystyle 
\sigma_{\mathrm{E}L\to\mathrm{T}L} &
\displaystyle = & 
\displaystyle 
-\frac{\alpha a_{0}}{Z}\frac{(L+1)(2\gamma_{1}+1)}{L(2L+1)^{2}}
\bigg\{\frac{\gamma_{1}+1}{(\gamma_{L}-\gamma_{1}+1)
(\gamma_{L}+\gamma_{1}-L+1)}\,
{}_{3}F_{2}
\left(
\begin{array}{c}
-L+2,\:
1,\:
\gamma_{L}-\gamma_{1}-L \\
\gamma_{L}-\gamma_{1}+2,\:
\gamma_{L}+\gamma_{1}-L+2
\end{array}
;1
\right) 
\\*[3ex]
&&
\displaystyle
+\,\frac{L(\gamma_{1}-1)-1}{(L+1)(\gamma_{L+1}-\gamma_{1}+1)
(\gamma_{L+1}+\gamma_{1}-L+1)}\,
{}_{3}F_{2}
\left(
\begin{array}{c}
-L+2,\:
1,\:
\gamma_{L+1}-\gamma_{1}-L \\
\gamma_{L+1}-\gamma_{1}+2,\:
\gamma_{L+1}+\gamma_{1}-L+2
\end{array}
;1
\right)
\bigg\}
\end{array}
$
&& $Z<\alpha^{-1}$ \\
\mbox{} && \\
\hline 
\hline
\end{tabular} 
\end{center} 
}
\end{table}
\end{landscape}

\begin{thebibliography}{99}
\bibitem{Zon72}
   B.\ A.\ Zon, N.\ L.\ Manakov, L.\ P.\ Rapoport,
   Coulomb Green function in the $x$-representation and the
   relativistic polarizability of hydrogen atom,
   Yad.\ Fiz.\ 15 (1972) 508 
   [English translation: Sov.\ J.\ Nucl.\ Phys.\ 15 (1972) 282]
\bibitem{Labz73a}
   L.\ N.\ Labzowsky,
   Polarizability of a hydrogenlike ion with an arbitrary nuclear
   charge,
   Opt.\ Spektrosk.\ 35 (1973) 561
\bibitem{Labz73b}
   L.\ N.\ Labzowsky,
   Multicharged ions in external electric field,
   Vest.\ Leningr.\ Univ.\ Ser.\ Fiz.\ Khim.\ (2) (1973) 19
\bibitem{Shes74}
   A.\ F.\ Shestakov, S.\ V.\ Khristenko,
   Polarizability of a relativistic hydrogen atom,
   Opt.\ Spektrosk.\ 36 (1974) 635
\bibitem{Labz93}
   L.\ N.\ Labzowsky, G.\ L.\ Klimchitskaya, Yu.\ Yu.\ Dmitriev,
   Relativistic Effects in the Spectra of Atomic Systems,
   Institute of Physics, Bristol, 1993, Chap.\ 6
\bibitem{LeAn94}
   Le Anh Thu, Le Van Hoang, L.\ I.\ Komarov, T.\ S.\ Romanova,
   Operator representation of the Dirac Coulomb Green function and
   relativistic polarizability of hydrogen-like atoms,
   J.\ Phys.\ B 27 (1994) 4083
\bibitem{Szmy97}
   R.\ Szmytkowski,
   The Dirac--Coulomb Sturmians and the series expansion of the
   Dirac--Coulomb Green function: application to the relativistic
   polarizability of the hydrogen-like atom,
   J.\ Phys.\ B 30 (1997) 825
   [erratum: J.\ Phys.\ B 30 (1997) 2747; 
   addendum: arXiv:physics/9902050]
\bibitem{Yakh03}
   V.\ Yakhontov,
   Relativistic linear response wave functions and dynamic scattering
   tensor for the $n$s$_{1/2}$ states in hydrogenlike atoms,
   Phys.\ Rev.\ Lett.\ 91 (2003) 093001
\bibitem{Szmy04}
   R.\ Szmytkowski, K.\ Mielewczyk,
   Gordon decomposition of the static dipole polarizability of the
   relativistic hydrogen-like atom: application of the Sturmian
   expansion of the first-order Dirac--Coulomb Green function,
   J.\ Phys.\ B 37 (2004) 3961
\bibitem{Bart69}
   M.\ L.\ Bartlett, E.\ A.\ Power,
   Relativistic corrections to $S_{-2}$ for atomic hydrogen,
   J.\ Phys.\ A 2 (1969) 419
\bibitem{Rutk90}
   A.\ Rutkowski, W.\ H.\ E.\ Schwarz,
   Relativistic perturbation theory of chemical properties,
   Theor.\ Chim.\ Acta 76 (1990) 391
\bibitem{Turs01}
   P.\ Turski, A.\ J.\ Sadlej, 
   The change of picture contribution to relativistic corrections to
   second-order properties,
   Chem.\ Phys.\ Lett.\ 338 (2001) 345
\bibitem{Balu95}
   K.\ L.\ Baluja,
   Relativistic correction to the dipole polarizability of a
   hydrogenic ion,
   Pramana 45 (1995) 533
\bibitem{Mana74}
   N.\ L.\ Manakov, L.\ P.\ Rapoport, S.\ A.\ Zapryagaev,
   Relativistic electromagnetic susceptibilities of hydrogen-like
   atoms,
   J.\ Phys.\ B 7 (1974) 1076
\bibitem{Zapr81}
   S.\ A.\ Zapryagaev, N.\ L.\ Manakov,
   Application of the Green's function for the Dirac equation to
   study of relativistic and correlation effects in multicharged ions,
   Izv.\ Akad.\ Nauk SSSR Ser.\ Fiz.\ 45 (1981) 2336
\bibitem{Zapr85}
   S.\ A.\ Zapryagaev, N.\ L.\ Manakov, V.\ G.\ Palchikov,
   Theory of Multi-charged Ions with One and Two Electrons,
   Energoatomizdat, Moscow, 1985 (in Russian)
\bibitem{Kane77}
   S.\ Kaneko,
   Relativistic corrections to the electric multipole polarisability
   and the shielding factor of a hydrogen-like ion,
   J.\ Phys.\ B 10 (1977) 3347 
   [erratum: J.\ Phys.\ B 11 (1978) 1879]
\bibitem{Drac85}
   R.\ J.\ Drachman,
   Rydberg states of helium: Relativistic and second-order
   corrections,
   Phys.\ Rev.\ A 31 (1985) 1253
   [erratum: Phys.\ Rev.\ A 38 (1988) 1659]
\bibitem{Gold89}
   S.\ P.\ Goldman,
   Gauge-invariance method for accurate atomic-physics calculations:
   Application to relativistic polarizabilities,
   Phys.\ Rev.\ A 39 (1989) 976
\bibitem{Zhan12}
   Y.-H.\ Zhang, L.-Y.\ Tang, X.-Z.\ Zhang, T.-Y.\ Shi, J.\ Mitroy,
   Relativistic quadrupole polarizability for the ground state of
   hydrogen-like ions,
   Chin.\ Phys.\ Lett.\ 29 (2012) 063101
\bibitem{Tang12}
   L.-Y.\ Tang, Y.-H.\ Zhang, X.-Z.\ Zhang, J.\ Jiang, J.\ Mitroy,
   Computational investigation of static multipole polarizabilities
   and sum rules for ground-state hydrogenlike ions,
   Phys.\ Rev.\ A 86 (2012) 012505
\bibitem{Fili14}
   L.\ Filippin, M.\ Godefroid, D.\ Baye,
   Relativistic polarizabilities with the Lagrange-mesh method,
   Phys.\ Rev.\ A 90 (2014) 052520
\bibitem{Szmy14}
   R.\ Szmytkowski, P.\ Stefa{\'n}ska,
   Electric-field-induced magnetic quadrupole moment in the ground
   state of the relativistic hydrogenlike atom: Application of the
   Sturmian expansion of the generalized Dirac--Coulomb Green
   function,
   Phys.\ Rev.\ A 89 (2014) 012501
\bibitem{Lewi95}
   R.\ R.\ Lewis, S.\ M.\ Blinder,
   Stark-induced anapole magnetic fields in atoms,
   Phys.\ Rev.\ A 52 (1995) 4439
\bibitem{Miel06}
   K.\ Mielewczyk, R.\ Szmytkowski,
   Stark-induced magnetic anapole moment in the ground state of the
   relativistic hydrogenlike atom: Application of the Sturmian 
   expansion of the generalized Dirac--Coulomb Green function,
   Phys.\ Rev.\ A 73 (2006) 022511
   [erratum: Phys.\ Rev.\ A 73 (2006) 039908]
\bibitem{Zapr74}
   S.\ A.\ Zapryagaev, N.\ L.\ Manakov, L.\ P.\ Rapoport,
   Multipole screening of nuclei of hydrogen-like atoms,
   Yad.\ Fiz.\ 19 (1974) 1136
   [English translation: Sov.\ J.\ Nucl.\ Phys.\ 19 (1974) 582]
\bibitem{Szmy01}
   R.\ Szmytkowski,
   Dynamic polarizability of the relativistic hydrogenlike atom:
   Application of the Sturmian expansion of the Dirac--Coulomb Green
   function,
   Phys.\ Rev.\ A 65 (2001) 012503
\bibitem{Szmy02}
   R.\ Szmytkowski,
   Magnetizability of the relativistic hydrogen-like atom:
   application of the Sturmian expansion of the first-order
   Dirac--Coulomb Green function,
   J.\ Phys.\ B 35 (2002) 1379
\bibitem{Szmy11}
   R.\ Szmytkowski, P.\ Stefa{\'n}ska,
   Comment on ``Four-component relativistic theory for NMR parameters:
   Unified formulation and numerical assessment of different
   approaches'' [J.\ Chem.\ Phys.\ 130, 144102 (2009)],
   arXiv:1102.1811
\bibitem{Stef12}
   P.\ Stefa{\'n}ska, R.\ Szmytkowski,
   Electric and magnetic dipole shielding constants for the ground
   state of the relativistic hydrogen-like atom: Application of the
   Sturmian expansion of the generalized Dirac--Coulomb Green
   function,
   Int.\ J.\ Quantum Chem.\ 112 (2012) 1363
\bibitem{Szmy12}
   R.\ Szmytkowski, P.\ Stefa{\'n}ska, 
   Magnetic-field-induced electric quadrupole moment in the ground
   state of the relativistic hydrogenlike atom: Application of the
   Sturmian expansion of the generalized Dirac--Coulomb Green
   function, 
   Phys.\ Rev.\ A 85 (2012) 042502
\bibitem{Stef15a}
   P.\ Stefa{\'n}ska,
   Magnetizability of the relativistic hydrogenlike atom in an
   arbitrary discrete energy eigenstate: Application of the Sturmian 
   expansion of the generalized Dirac--Coulomb Green function,
   Phys.\ Rev.\ A 92 (2015) 032504
\bibitem{Stef15b}
   P.\ Stefa{\'n}ska,
   Magnetizabilities of relativistic hydrogenlike atoms in some
   arbitrary discrete energy eigenstates,
   At.\ Data Nucl.\ Data Tables, doi:\/~10.1016/j.adt.2015.09.001
\bibitem{Stef15c}
   P.\ Stefa{\'n}ska,
   Magnetic-field-induced electric quadrupole moment in an arbitrary
   discrete energy eigenstate of the relativistic hydrogenlike atom:
   Application of the Sturmian expansion of the generalized
   Dirac--Coulomb Green function,
   submitted for publication
\bibitem{Szmy07}
   R.\ Szmytkowski,
   Recurrence and differential relations for spherical spinors,
   J.\ Math.\ Chem.\ 42 (2007) 397
\bibitem{Magn66}
   W.\ Magnus, F.\ Oberhettinger, R.\ P.\ Soni,
   Formulas and Theorems for the Special Functions of Mathematical
   Physics, 3rd ed.,
   Springer, Berlin, 1966
\bibitem{Grad07}
   I.\ S.\ Gradshteyn, I.\ M.\ Ryzhik,
   Table of Integrals, Series, and Products, 7th ed.,
   Elsevier, Amsterdam, 2007
\bibitem{Vars75}
   D.\ A.\ Varshalovich, A.\ N.\ Moskalev, V.\ K.\ Khersonskii,
   Quantum Theory of Angular Momentum,
   Nauka, Leningrad, 1975 (in Russian)
   [English translation: World Scientific, Singapore, 1988]
\bibitem{Agre11}
   M.\ Ya.\ Agre,
   Multipole expansions in magnetostatics,
   Phys.\ Usp.\ 54 (2) (2011) 167
\bibitem{Dubo90}
   V.\ M.\ Dubovik, V.\ V.\ Tugushev,
   Toroid moments in electrodynamics and solid-state physics,
   Phys.\ Rep.\ 187 (1990) 145
\bibitem{Prud03}
   A.\ P.\ Prudnikov, Yu.\ A.\ Brychkov, O.\ I.\ Marichev,
   Integrals and Series. Vol.\ 3: Special Functions. Supplementary
   Chapters, 2nd ed.,
   Fizmatlit, Moscow, 2003 (in Russian)
\end{thebibliography}
\end{document}